\documentclass[preprint,12pt,authoryear]{elsarticle}

\usepackage{amsmath}
\usepackage{amsfonts}
\usepackage{amssymb}
\usepackage{bm}
\usepackage{graphicx}
\usepackage{upgreek}
\usepackage{hyperref}

\newcommand{\pdv}[2]{\frac{\partial #1}{\partial #2}}
\newcommand{\dprime}{{\prime\prime}}

\begin{document}

\begin{frontmatter}

\title{Multiscale retinal flow on a spherical cap of varying aperture}

\author[aff1]{Chang Lin}
\author[aff2]{Zilong Song}
\author[aff3]{Bob Eisenberg}
\author[aff4]{Shixin Xu}
\author[aff4]{Huaxiong Huang\corref{cor1}}
\ead{huaxiong.huang@dukekunshan.edu.cn}
\cortext[cor1]{Corresponding author}

\affiliation[aff1]{organization={School of Mathematical Sciences, Beijing Normal University},
                    city={Beijing},
                    country={China}}
\affiliation[aff2]{organization={Department of Mathematics and Statistics, Utah State University},
                    city={Logan},
                    state={Utah},
                    country={USA}}
\affiliation[aff3]{organization={Department of Applied Mathematics, Illinois Institute of Technology},
                    city={Chicago},
                    state={Illinois},
                    country={USA}}
\affiliation[aff4]{organization={Zu Chongzhi Centre, Duke Kunshan University},
                    city={Kunshan},
                    country={China}}

\begin{abstract}
Modelling retinal haemodynamics is crucial for understanding retinal microcirculation but is computationally demanding because it involves coupling between the vasculature and surrounding tissue across multiple scales.
This computational burden has been substantially alleviated by a recent analytic solution on the planar disc that enables lumping the capillary bed and surrounding tissue into an effective resistor.
However, that formulation treats the retina as a flat surface, whereas the retina is a curved surface with a finite anterior aperture.
In this work, we develop a nontrivial and physiologically necessary extension to spherical-cap tissue domains with varying apertures, where surface curvature and finite-aperture boundaries complicate solving coupled Darcy equations on a curved manifold.
Using a stereographic projection and a decoupling transformation, we derive an analytic solution for the capillary-tissue system on the spherical cap that represents flow in both the capillary bed and interstitial tissue more realistically while retaining the efficient resistor formulation, a key advantage of the planar-disc formulation.
This solution is coupled to one-dimensional (1D) arteriolar and venular flows to obtain a multiscale description of retinal haemodynamics.
Using a vasculature model designed to capture retinal vascular features, we show that the multiscale model's predictions are consistent with experimental data.
We further explore aperture effects using both a fixed hemispherical vasculature and aperture-dependent vasculature.
The aperture affects retinal haemodynamics mainly through changes in the constructed vasculature itself, whereas the surface-averaged pressures and relative terminal flow distributions remain nearly unchanged.
This framework provides a foundation for studying retinal pathophysiology on more anatomically realistic domains.
\end{abstract}

\begin{keyword}
biological fluid dynamics \sep biomedical flows \sep retinal haemodynamics
\end{keyword}

\end{frontmatter}

\section{Introduction}
Retinal haemodynamics is essential to visual function, and its disruption can compromise the supply of oxygen and nutrients, leading to several retinopathies \citep{kornfield_measurement_2015,garhofer_response_2004,caprioli_blood_2010,chng_ocular_2021,osborne_retinal_2004}.
While experimental studies have extensively explored retinal microcirculation, mathematical modelling provides a complementary framework for integrating the physiological mechanisms that are challenging to capture \textit{in vivo} \citep{ruffini_mathematical_2024,siggers_fluid_2012}.

Several models have been developed to explore retinal haemodynamics, including zero-dimensional (0D) lumped models \citep{flores_novel_2016,ganesan_analysis_2010,fry_predicting_2018,sala_eye2heart_2026}, one-dimensional (1D) models \citep{julien_one-dimensional_2023} and three-dimensional (3D) models \citep{liu_computational_2009,rebhan_computational_2019,malek_computational_2015}.
For example, 0D models have been used to interpret clinical data \citep{guidoboni_intraocular_2014}, while 1D and 3D models have been used to study retinal transport and perform patient-specific computations \citep{liu_computational_2009,causin_blood_2016,julien_one-dimensional_2023,rebhan_computational_2019,malek_computational_2015,tripathy_image-based_2023}.
These works have provided crucial physiological and pathological insights into retinal haemodynamics; however, they usually rely on significant simplifications of the components in retinal microcirculation.
Specifically, arterioles, venules, capillaries, and interstitial tissue are not treated with equal fidelity, and one or more of these components are typically modelled as effective resistors.
A more balanced approach, adapted from models of other organs and tissues \citep{desposito_computational_2018,fritz_1d0d3d_2022,sweeney_threedimensional_2024,shipley_hybrid_2019}, describes capillary-bed and tissue flow using Darcy equations and couples them to the arterial and venous trees through point sources and sinks.
Although this formulation provides necessary details for each component, it is computationally demanding, as the Darcy equations must be solved simultaneously with arterial and venous flows \citep{qohar_nonlinear_2021,fritz_1d0d3d_2022}.
To address this, some works describe the flow in the capillary bed using an analytic solution \citep{shipley_hybrid_2019,sweeney_threedimensional_2024,desposito_computational_2018}, while the flow in the tissue is treated as a far-field boundary condition.
This description allows the capillary bed to be lumped into an effective resistor that couples arteries and veins in the computation.
Moreover, thanks to the point source formulation, it also enables the full pressure field to be obtained given the flow at sources \citep{sweeney_threedimensional_2024}.
However, treating the tissue as a far-field boundary condition neglects its local fluid exchange with the capillary bed, despite the physiological interaction between these compartments \citep{fritz_1d0d3d_2022,eisenberg_structural_2023}.

Given the regular geometry of the ocular globe, the retinal surface has usually been modelled as various idealised domains in previous studies, such as a hemisphere \citep{dziubek_effect_2016,rebhan_computational_2019,siggers_fluid_2012,tripathy_image-based_2023}, a circular domain with radial cuts \citep{fry_predicting_2018,ganesan_analysis_2010}, and a rectangle with the same area as the retinal surface \citep{causin_blood_2016}.
Such regular domains facilitate analytic solutions that simplify the capillary bed and surrounding tissue into an effective resistor, enabling efficient computation.
For example, previous work modelled the retinal surface as a circular disc and derived an analytic solution for the capillary-tissue system described by coupled Darcy equations \citep{lin_novel_2026}.
The resulting analytic solution, expressed in terms of modified Bessel functions, provided a more complete interpretation of the multiscale haemodynamics in the retina, along with an efficient coupling across multiple scales.
However, the real retina is a curved surface with a finite anterior aperture.
The planar disc thus serves as a useful first simplification, and a more realistic description requires a domain that better reflects the physiological geometry of the retina.

Our work advances previous multiscale models of retinal haemodynamics \citep{fry_predicting_2018,ganesan_analysis_2010,causin_blood_2016}, particularly the planar circular-disc formulation \citep{lin_novel_2026}, in two closely connected respects.
First, the retinal surface is represented as a spherical-cap domain with a finite anterior aperture, providing a more physiological description of the retinal surface.
Accordingly, the arterioles and venules are modelled as 1D elastic segments governed by a 1D flow model, while the capillary bed and interstitial tissue are modelled as two interacting porous media described by coupled surface Darcy equations.
The surface curvature and finite-aperture boundary require a new analytic solution for the capillary-tissue system.
Second, the arteriolar and venular trees are constructed directly on spherical-cap domains of varying aperture using a vasculature model designed to account for the corresponding retinal geometry.
This vasculature model is a modification of the constrained constructive optimization (CCO) algorithm \citep{kerautret_opencco_2023,karch_three-dimensional_1999-1,bao_constrained_2024,hernandez_linking_2024}, in which the vasculature is grown through optimisation.
Following the common modelling practice, the arteriolar and venular trees are coupled to the capillary-tissue system via the point sources and sinks located at the capillary bed.
The novel aspect of our work is the derivation of the analytic solution for the capillary-tissue system on the spherical-cap domain and its subsequent implementation in retinal haemodynamics.
To the best of our knowledge, few studies focused on coupled Darcy flow problems on spherical-cap domains with varying apertures, largely due to the difficulty introduced by the finite-aperture boundary.
While one related study considered a potential problem on a spherical-cap domain \citep{kidambi_point_2000-1}, other studies solved related problems on domains with stronger symmetry, such as the hemisphere \citep{munthe-kaas_boundary_2008} and the full sphere \citep{iglewska-nowak_wavelet_2023}.
Our work thus provides the first analytic solution for the coupled capillary-tissue system on a spherical-cap domain and enables the efficient computation of multiscale retinal flow.
The proposed multiscale model is validated against experimental data \citep{kornfield_measurement_2015,riva_blood_1985,werkmeister_measurement_2012} and results from the circular-disc analytic solution \citep{lin_novel_2026}, and the pulsatile flow is also simulated.
Moreover, the sensitivity of the model to aperture parameter is analysed.

The rest of the paper is organised as follows.
The models are described in Section$~$\ref{ModelHierarchy}, including the vasculature model for arterioles and venules, and the flow models for arterioles, venules, capillary bed and interstitial tissue.
The analytic solution for the capillary-tissue system is derived in Section$~$\ref{AnalyticSolutionCapillaryTissueFlow}.
In Section$~$\ref{NumericalMethods}, we describe the numerical methods for our multiscale model.
The proposed model is validated against experimental data in Section$~$\ref{ValidationMultiscaleModel}, where pulsatile flow is also computed.
\section{Model Hierarchy}
\label{ModelHierarchy}
\subsection{Physiological basis and modelling simplifications for the retinal surface}
\begin{figure}
    \centering
    \includegraphics[width=\textwidth]{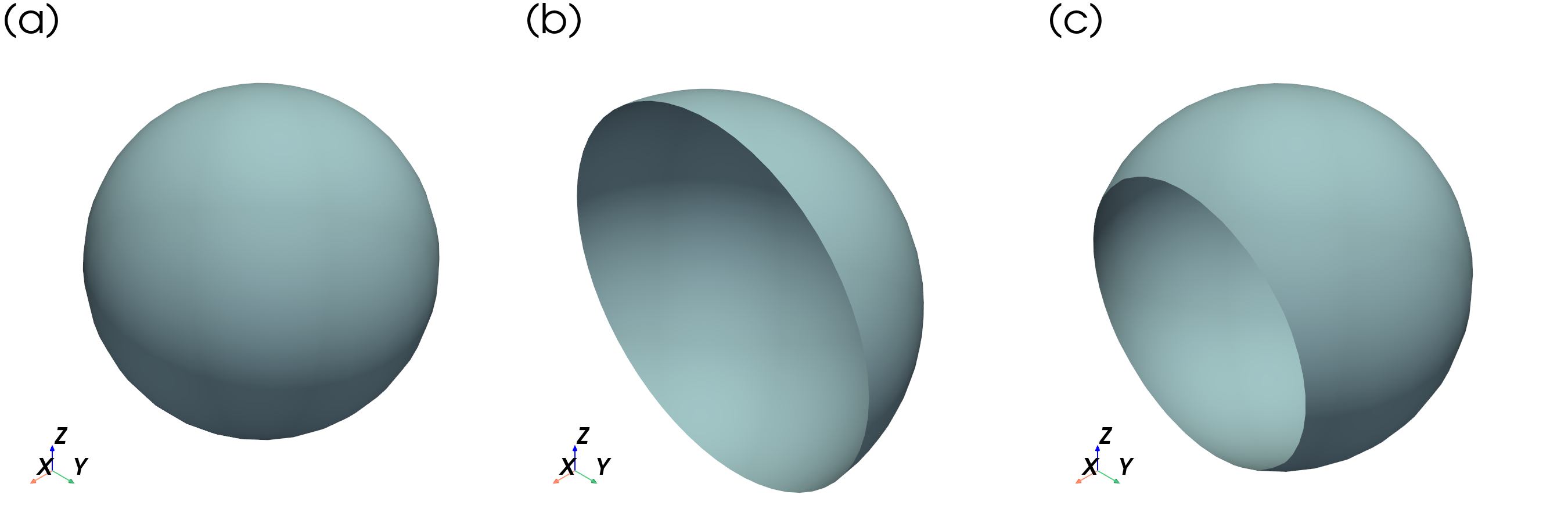}
    \caption{Shapes of the ocular globe and the retinal surface. (a) The ocular globe. (b) A hemispherical domain for the retinal surface. (c) A spherical-cap domain for the retinal surface. The $X$-axis is aligned with the visual axis.}
    \label{shape_spheres}
\end{figure}

We begin by defining the geometry of the retinal surface, which is determined by the shape of the ocular globe.
This globe is typically modelled as an ellipsoid \citep{dziubek_effect_2016,rebhan_computational_2019}:
\begin{eqnarray}
    \mathcal{E}=\left\{(X,Y,Z) \left\vert\frac{X^2}{a^2}+\frac{Y^2}{b^2}+\frac{Z^2}{c^2}\leq 1\right.\right\},
\end{eqnarray}
where $a$, $b$ and $c$ are the semi-axis lengths, and the $X$-axis is aligned with the visual axis.
In this model, the ocular globe is approximated as a sphere by setting $a=b=c=R_t$, where $R_t$ is the radius of the ocular globe, as illustrated in figure$~$\ref{shape_spheres}(a).
Because its thickness is much smaller than the radius of ocular globe, the retina is represented as a two-dimensional (2D) surface.
More precisely, the retinal surface is modelled as a spherical-cap domain with an anterior cutting plane that is perpendicular to the visual axis:
\begin{eqnarray}
    \mathcal{S}=\left\{(X,Y,Z) \left\vert X^2+Y^2+Z^2=R^2_t\right.,-R_t\leq X\leq \beta_{oc}R_t\right\},
\end{eqnarray}
where $1>\beta_{oc}\geq 0$ is the aperture parameter.
The aperture parameter $\beta_{oc}$ determines the position of the anterior cutting plane, which corresponds to the ora serrata.
The value $\beta_{oc}=0$ corresponds to a hemispherical retinal domain, as shown in figure$~$\ref{shape_spheres}(b), while increasing $\beta_{oc}$ extends the domain beyond the equator, as shown in figure$~$\ref{shape_spheres}(c).
A point on the retinal surface is denoted by ${\bm x}=(X,Y,Z)$.
\subsection{Tree-based computation for arterioles and venules}
\subsubsection{One-dimensional model}
The retinal arteriolar and venular trees are idealised as networks composed of 1D elastic segments.
The flow in each segment is described by the 1D model, which enforces mass and momentum balance and is widely adopted in haemodynamic modelling.
At any time $t>0$, the flow in each vessel is characterised by flow rate $q(s,t)$ and cross-sectional area $A(s,t)$, where $s$ denotes the axial coordinate along the vessel.
Their evolutions are described by \citep{julien_one-dimensional_2023, wang_fluid_2016}
\begin{flalign}
    \left\{
    \begin{aligned}
        &\pdv{A}{t}+\pdv{q}{s}=0, \\
        &\pdv{q}{t}+\pdv{}{s}\left(\hat{\alpha}\frac{q^2}{A}\right)+\frac{A}{\rho}\pdv{p}{s}=C_f,
        \label{1D_model}
    \end{aligned}\right.
\end{flalign}
where $\hat{\alpha}$ is the momentum flux correction factor that depends on the velocity profile within retinal vessels \citep{muller_global_2014}, $p$ is the pressure, $\rho$ is the blood density and $C_f$ is the friction term.
The coefficient $\hat{\alpha}$ is set to $\hat{\alpha}=4/3$, which corresponds to the parabolic velocity profile, consistent with the experimental observations in the retina \citep{song_combined_2014,causin_blood_2016,kornfield_measurement_2015}.
The 1D model is identical to that used in previous planar-disc formulation \citep{lin_novel_2026}, except that vessel lengths are computed as geodesic arc lengths on the spherical surface.
The blood density is fixed at $\rho=1~{\rm{g}/\rm{cm}^3}$.
The explicit form of the friction term is given by \citep{muller_global_2014,audebert_kinetic_2017}
\begin{eqnarray}
    C_f=\left\{\begin{aligned}
        &-8\pi\frac{\eta}{\rho}\frac{q}{A},\quad \text{for arterioles},\\
        &-8\pi\sqrt{\frac{A}{A_0}}\frac{\eta}{\rho}\frac{q}{A},\quad \text{for venules},
    \end{aligned}\right.
\end{eqnarray}
where $\eta=\eta(r)$ is the apparent viscosity, modelled as a function of the vessel radius $r$ to account for the rheological properties of blood.
We use an empirical relation developed by Pries \textit{et al.} to characterise the rheological properties of blood \citep{pries_resistance_1994,secomb_blood_2013,bappoo_viscosity_2017}:
\begin{eqnarray}
\eta(r)=\eta_p\left[1+(\eta_{45}-1)\left(\frac{r}{r-0.55}\right)^2\right]\left(\frac{r}{r-0.55}\right)^2,
\end{eqnarray}
where $r$ is the vessel radius (in $\rm{\upmu m}$), $\eta_p=1.2~\rm{cP}$ is the plasma viscosity and $\eta_{45}$ is the relative apparent blood viscosity for a discharge haematocrit of $0.45$ given by
\begin{eqnarray}
    \eta_{45}=6\exp(-0.17r)+3.2-2.44\exp\left[-0.06(2r)^{0.645}\right].
\end{eqnarray}

The constitutive behaviour of the vessel wall is described by the following elastic model \citep{liu_multicore_2024, muller_global_2014}:
\begin{eqnarray}
    p-p_{ext}=K\left[\left(\frac{A}{A_0}\right)^{\beta_1}-\left(\frac{A}{A_0}\right)^{\beta_2}\right],
    \label{elastic_model}
\end{eqnarray}
where $p_{ext}$ is the surrounding tissue pressure, $K$ is the stiffness modulus, $\beta_1$ and $\beta_2$ are two exponents characterising the vessel wall elasticity, and $A_0$ is the neutral area of the vessel.
All of these parameters, including $K$, $\beta_1$, $\beta_2$ and $A_0$, vary with the vessel position and diameter.
While $A_0$ is an intrinsic property of each vessel, the specific forms of $K$, $\beta_1$ and $\beta_2$ are prescribed distinctly for arterioles and venules.
The stiffness modulus is given by
\begin{eqnarray}
    K=\left\{\begin{aligned}
        &\frac{4}{3}\frac{\pi^{1/2}Eh}{A^{1/2}_0}, \quad \text{for arterioles},\\
        &\frac{1}{9}\frac{\pi^{3/2}Eh^3}{A^{3/2}_0}, \quad \text{for venules}.
    \end{aligned}\right.
\end{eqnarray}
where $E$ is the Young's modulus and $h$ is the vessel wall thickness.
The exponents $\beta_1$ and $\beta_2$ are
\begin{eqnarray}
    \beta_1=\left\{\begin{aligned}
        &1/2,\quad \text{for arterioles},\\
        &10,\quad \text{for venules},
    \end{aligned}\right.
    \quad
    \beta_2=\left\{\begin{aligned}
        &0,\quad \text{for arterioles},\\
        &-3/2,\quad \text{for venules}.
    \end{aligned}\right.
\end{eqnarray}
The expression for $K$ and the values of $\beta_1$ and $\beta_2$ for arterioles are adopted from established models of retinal arteriolar circulation, whereas the venular counterparts are adopted from the model of veins in other tissues.
The wall thickness $h$ in each vessel is proportional to the neutral vessel radius $r_0$, and is given by
\begin{eqnarray}
    h=\rm{WLR}\times 2r_0,
\end{eqnarray}
where $\rm{WLR}$ denotes the wall-to-lumen ratio.
Its value for arterioles is based on experimental observations \citep{arichika_effects_2015}, while a smaller value is employed for venules to account for their thinner walls.
Moreover, we assume a uniform value of $E$ for both arterioles and venules as a reasonable approximation.
Note that the external pressure for arterioles and venules is approximately the intraocular pressure (IOP) \citep{kiel_ocular_2011,guidoboni_intraocular_2014}.
Thus, we set $p_{ext}=\rm{IOP}$.
\subsubsection{Vasculature model}
We next describe the vasculature model for the retinal vascular trees, which is a modification of the CCO algorithm \citep{kerautret_opencco_2023,karch_three-dimensional_1999-1,bao_constrained_2024,hernandez_linking_2024}.
The superior and inferior vascular trees are grown separately \citep{hernandez_linking_2024} and then combined.
The superior retinal domain is given by
\begin{eqnarray}
    \mathcal{S}_{sup}=\left\{(X,Y,Z)\in \mathcal{S}\vert Z\geq 0\right\},
\end{eqnarray}
while the inferior retinal domain is given by
\begin{eqnarray}
    \mathcal{S}_{inf}=\left\{(X,Y,Z)\in \mathcal{S}\vert Z\leq 0\right\},
\end{eqnarray}
so $Z=0$ is the plane that separates the superior and inferior trees.
We next describe the growth of the arteriolar and venular trees on the superior domain, whereas growth on the inferior domain follows a similar procedure.
\begin{figure}
    \centering
    \includegraphics[width=0.76\textwidth]{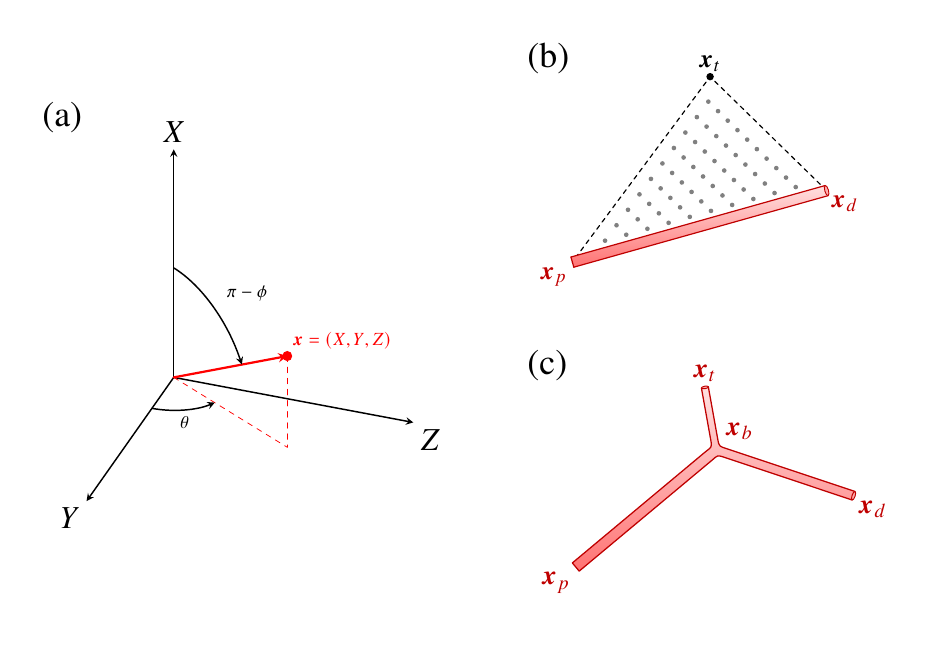}
    \caption{Construction of a candidate vascular tree on the spherical cap.
    (a) Spherical coordinates centred on the $X$-axis, where $\phi$ is measured from the foveal direction $-X$ and $\theta$ is the azimuthal angle in the $YZ$-plane.
    For illustration, the $X$-axis is vertical and the angle from the positive $X$-axis is labelled $\pi-\phi$.
    (b) Discrete candidate junction positions (grey markers) in the spherical barycentric triangle defined by ${\bm x}_p$, ${\bm x}_d$ and ${\bm x}_t$.
    The red cylinder represents the initial vessel from ${\bm x}_p$ to ${\bm x}_d$.
    (c) Candidate tree formed at ${\bm x}_b$, with a mother segment from ${\bm x}_p$ to ${\bm x}_b$ and daughter segments from ${\bm x}_b$ to ${\bm x}_d$ and ${\bm x}_t$.}
    \label{schematic_spherical_coordinates_barycentric}
\end{figure}

\begin{figure}
    \centering
    \includegraphics[width=\textwidth]{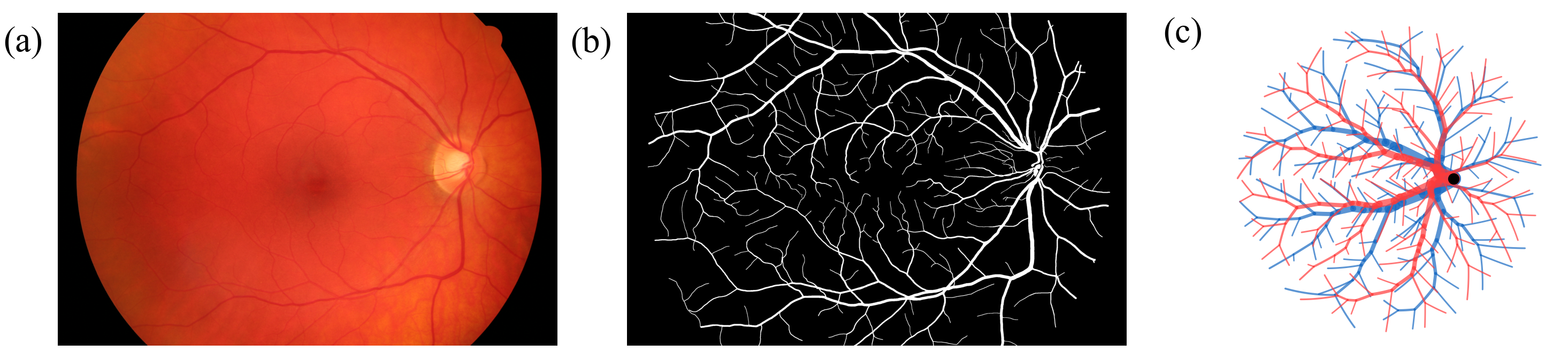}
    \caption{Comparison of real and constructed retinal vasculature.
    (a) Fundus image from the high-resolution fundus (HRF) image dataset \citep{budai_robust_2013}.
    (b) Corresponding manually segmented vasculature.
    (c) Constructed arteriolar (red) and venular (blue) trees projected onto the fovea-centred plane of the hemisphere.
    The black circle denotes the optic disc where the CRA and the CRV enter the retina.}
    \label{spherical_fundus}
\end{figure}

The foveal centre is placed at ${\bm x}_{fovea}=(-R_t,0,0)$, and the optic disc centre is ${\bm x}_{od}=(-R_t\cos\phi_{od},R_t\sin\phi_{od},0)$ with $\phi_{od}=\pi/4$.
This prescribed offset between the fovea and optic disc is consistent with physiological observations, as shown by the dark and yellow regions in figure$~$\ref{spherical_fundus}(a).
The root of each tree is initialised at the optic disc centre, and the tree is then grown by adding terminal vessels until the number of termini reaches $N_{term}$, while minimising its total volume with a prescribed inlet radius.
The number of termini is set to $N_{term}=60$, which yields a minimum radius of $9~\rm{\upmu m}$ in the constructed arteriolar trees.
Vessels of this radius are consistent with the smallest explicitly resolved vessels and provide the connections to the capillary bed \citep{fritz_1d0d3d_2022}.
The total volume is given by
\begin{eqnarray}
    V=\sum_{k=1}^{n_v} \pi r_k^2 L_k,
    \label{volume_cost}
\end{eqnarray}
where $n_v$ is the number of vessels in the candidate tree satisfying $1\leq n_v\leq 2N_{term}-1$, and $L_k$ is the geodesic length of vessel $k$.
We assume equal terminal flow rates $q_{term}$ and equal terminal pressures.
Thus, if a vessel is supplied by a subtree that contains $m$ terminal vessels, its flow rate is $mq_{term}$.

To distribute terminal vessels progressively over the retinal surface, candidate terminal positions are selected uniformly over the spherical surface area.
Since the $X$-axis is parallel to the visual axis, it is convenient to use spherical coordinates centred on the $X$-axis:
\begin{eqnarray}
    {\bm x}=(X,Y,Z)=(-R_t\cos\phi, R_t\sin\phi\cos\theta, R_t\sin\phi\sin\theta),
    \label{spherical_coordinates_spherical_cap}
\end{eqnarray}
where $\phi=\cos^{-1}(-X/R_t)$ is the polar angle measured from the foveal direction and $\theta=\tan^{-1}(Z/Y)$ is the azimuthal angle in the $YZ$-plane, measured from the positive $Y$-axis toward the positive $Z$-axis.
The quadrant of $\theta$ is determined from the signs of $Y$ and $Z$ such that $\theta\in[0,2\pi)$.
This coordinate convention is shown in figure$~$\ref{schematic_spherical_coordinates_barycentric}(a).
The superior domain becomes
\begin{eqnarray}
    \mathcal{S}_{sup}=\left\{(\phi,\theta)\vert0\leq\phi\leq\phi_{max},0\leq\theta\leq\pi\right\},\quad \phi_{max}=\cos^{-1}(-\beta_{oc}),
\end{eqnarray}
and in this convention, the foveal centre is placed at $\phi=0$.
A candidate terminal point is selected uniformly over the surface by sampling $\theta\sim\mathcal{U}(0,\pi)$ and $\cos\phi\sim\mathcal{U}(\cos\phi_{max},1)$, where $\mathcal{U}(v_1,v_2)$ denotes the uniform distribution.
After a terminal position ${\bm x}_t$ is selected, the nearest existing segments are identified using the geodesic distance from the terminal point to each great-circle arc.
For each potential parent segment with proximal and distal points ${\bm x}_p$ and ${\bm x}_d$, candidate bifurcation positions are generated inside the corresponding spherical triangle.
A candidate position for the bifurcation boundary is computed using spherical barycentric coordinates as
\begin{eqnarray}
    {\bm x}_b=R_t\frac{w_p{\bm x}_p+w_d{\bm x}_d+w_t{\bm x}_t}{\lVert w_p{\bm x}_p+w_d{\bm x}_d+w_t{\bm x}_t\rVert},
    \label{bifurcation_boundary_position}
\end{eqnarray}
where $w_p+w_d+w_t=1$ and all weights are positive.
A finite set of candidate junction points is generated using discretised barycentric weights in equation$~$(\ref{bifurcation_boundary_position}).
The candidates for which any connecting minor great-circle arc leaves the retinal domain are rejected.
The total volume is computed for each candidate point.
The minimum angle between any two vessel segments connected at the candidate junction is prescribed as $\pi/3$ \citep{hernandez_linking_2024}.
The most favourable candidates are then locally refined on the retinal surface, and the position that minimises the volume of the vascular tree is selected.
The candidate junction positions are shown in figure$~$\ref{schematic_spherical_coordinates_barycentric}(b).
The grey markers denote these positions inside the barycentric triangle formed by ${\bm x}_p$, ${\bm x}_d$ and ${\bm x}_t$.
For each admissible bifurcation position, a candidate tree is formed by splitting an existing vessel into one mother segment and two daughter segments, as illustrated in figure$~$\ref{schematic_spherical_coordinates_barycentric}(c).
The length of segment $k$ is computed as the arc distance between its proximal point ${\bm x}^k_p$ and distal point ${\bm x}^k_d$:
\begin{eqnarray}
    L_k=R_t\cos^{-1}\left(\frac{{\bm x}^k_p\cdot{\bm x}^k_d}{R^2_t}\right),
\end{eqnarray}
while the radii of all vessels must be updated before evaluating the total volume.
Given the parent radius $r_k$, two daughter radii $r_m$ and $r_n$ are determined by
\begin{eqnarray}
    r_m=\nu_mr_k, \quad r_n=\nu_nr_k,
\end{eqnarray}
where $\nu_m=r_m/r_k$ and $\nu_n=r_n/r_k$ are the daughter-to-parent radius ratios.
Thus, once $\nu_m$ and $\nu_n$ at each bifurcation boundary are obtained, the radii of all vessels can be determined downstream from the prescribed inlet radius.
Since the radii at each bifurcation satisfy
\begin{equation}
    r_k^\beta = r_m^\beta+r_n^\beta,
    \label{branching_relation}
\end{equation}
where $\beta$ is the branching exponent, the ratios $\nu_m$ and $\nu_n$ can be related to the daughter-to-daughter ratio $\nu_{mn}=r_m/r_n$ by
\begin{eqnarray}
    \nu_m=(1+\nu^{-\beta}_{mn})^{-1/\beta}, \quad \nu_n=(1+\nu^\beta_{mn})^{-1/\beta}.
\end{eqnarray}
Moreover, because the two subtrees connecting to two daughter vessels share the pressure at the bifurcation boundary and have equal terminal pressures, their pressure differences are equal, yielding
\begin{eqnarray}
    \nu_{mn}=\left(\frac{q_m\mathcal{R}_m}{q_n\mathcal{R}_n}\right)^{1/4},
\end{eqnarray}
where $q_m$ and $q_n$ are the flow rates in the two daughter vessels.
The quantities $\mathcal{R}_m$ and $\mathcal{R}_n$ are the truncated resistances of the subtrees rooted at daughter vessels $m$ and $n$, and are given by \citep{karch_three-dimensional_1999-1}
\begin{eqnarray}
    \mathcal{R}_k=\left\{
    \begin{aligned}
        &\frac{8\eta_v L_k}{\pi}, \quad \text{if vessel $k$ is a terminal vessel}, \\
        &\frac{8\eta_v L_k}{\pi}+\left(\frac{\nu_m^4}{\mathcal{R}_m}+\frac{\nu_n^4}{\mathcal{R}_n}\right)^{-1}, \quad \text{if vessel $k$ branches into vessels $m$ and $n$},
    \end{aligned}\right.
\end{eqnarray}
where $\eta_v$ is the constant viscosity for vasculature growth and $L_k$ is the length of vessel $k$.
The truncated resistance of non-terminal vessels depends on $\nu_m$ and $\nu_n$, and is thus evaluated recursively from the terminal vessels toward the inlet.
Specifically, for each candidate tree, the truncated resistances are computed progressively from the terminal vessels, level by level, until the truncated resistances of all vessels are determined.
Note that the value of $\eta_v$ does not affect the constructed vascular geometry, because it acts as a common scaling factor for all truncated resistances.
This vasculature model can also be used to construct the arteriolar and venular trees on the planar circular disc, where the vessel lengths are computed as Euclidean distances.

Figure$~$\ref{spherical_fundus} compares the constructed vasculatures with the real retinal vasculature from the high-resolution fundus (HRF) image dataset and its manual segmentation \citep{budai_robust_2013}.
The aperture is set to $\beta_{oc}=0$, corresponding to a hemispherical retinal surface.
Other parameters for the vasculature are summarised in table$~$\ref{parameters_multiscale_model}.
As shown in figure$~$\ref{spherical_fundus}(c), the constructed vasculature captures two essential features of the retinal vasculature: the large-scale curvature of the main branches and the separation into superior and inferior vascular territories.
\subsubsection{Boundary conditions}
Pressure boundary conditions are imposed at the CRA inlet and CRV outlet and are denoted by $p_{in,CRA}$ and $p_{out,CRV}$, respectively.
The corresponding areas $A_{in,CRA}$ and $A_{out,CRV}$ are determined from the elastic model:
\begin{eqnarray}
    p_{in,CRA}-p_{ext}=K\left[\left(\frac{A_{in,CRA}}{A_0}\right)^{\beta_1}-\left(\frac{A_{in,CRA}}{A_0}\right)^{\beta_2}\right], \\
    p_{out,CRV}-p_{ext}=K\left[\left(\frac{A_{out,CRV}}{A_0}\right)^{\beta_1}-\left(\frac{A_{out,CRV}}{A_0}\right)^{\beta_2}\right],
\end{eqnarray}
respectively.
While $p_{in,CRA}$ can be constant or time-dependent, $p_{out,CRV}$ is constant and thus the derivative $\partial A/\partial t=0$, which implies the boundary condition for the flow rate $\partial q/\partial s=0$.
At a bifurcation boundary, where a mother vessel $M$ connects to two daughter vessels $d_1$ and $d_2$, the bifurcation boundary conditions are given by \citep{koppl_dimension_2023}
\begin{eqnarray}
    \left\{
    \begin{aligned}
        &q_M-q_{d_1}-q_{d_2}=0,\\
        &p_M-p_{d_1}=0,\\
        &p_M-p_{d_2}=0,\\
        &\pdv{A_M}{t}+\pdv{q_M}{s_M}=0,\\
        &\pdv{A_{d_1}}{t}+\pdv{q_{d_1}}{s_{d_1}}=0,\\
        &\pdv{A_{d_2}}{t}+\pdv{q_{d_2}}{s_{d_2}}=0,
    \end{aligned}\right.
    \label{bifurct_boundary_conditions}
\end{eqnarray}
where variables with subscripts $M$, $d_1$ and $d_2$ correspond to the mother vessel and the two daughter vessels, respectively.
The arteriolar outlets and venular inlets couple the vasculature to the capillary-tissue system, as described in Section$~$\ref{CouplingVasculatureCapillaryTissueContinuum}.
\subsection{Capillary-tissue continuum on the spherical cap}
The capillary bed, which consists of the superficial, intermediate, and deep plexuses, is modelled as a porous medium.
The surrounding tissue, composed of cells, fibres and interstitial fluid, is similarly treated as a porous medium.
These two porous media are coupled through fluid exchange, akin to the coupling of electrical systems in models of syncytial tissue \citep{eisenberg_structural_2023}, as described below.

The flow in both the capillary bed and the surrounding tissue is described by Darcy's law, which states that the flow in a porous medium is driven by the pressure gradient \citep{sweeney_threedimensional_2024,shipley_hybrid_2019,qohar_nonlinear_2021,fritz_1d0d3d_2022}.
The Darcy flux, which is the volumetric flow rate per unit area ($\rm{cm\cdot s^{-1}}$), is given by
\begin{eqnarray}
    {\bm u}=-\frac{1}{\mu}{\bm k}\cdot\nabla_{\mathcal{S}} p,
    \label{Darcy_flow}
\end{eqnarray}
where $\nabla_{\mathcal{S}}$ is the surface gradient operator (the tangential component of the full 3D gradient on $\mathcal{S}$), ${\bm k}$ is the permeability tensor (${\rm{\upmu m}^2}$), $\mu$ is the viscosity of the corresponding fluid, and $p$ is the pressure of the fluid.
Using the convention from equation$~$(\ref{spherical_coordinates_spherical_cap}), $\nabla_{\mathcal{S}}$ on a function $f(\phi,\theta)$ is given by
\begin{eqnarray}
    \nabla_{\mathcal{S}}f=\frac{1}{R_t}\frac{\partial f}{\partial \phi}\hat{{\bm \phi}}+\frac{1}{R_t\sin\phi}\frac{\partial f}{\partial \theta}\hat{{\bm \theta}},
\end{eqnarray}
where $\hat{{\bm \phi}}$ is the unit vector in the increasing $\phi$ direction and $\hat{{\bm \theta}}$ is the unit vector in the increasing $\theta$ direction.
We assume that the capillary bed and the surrounding tissue are isotropic and homogeneous \citep{qohar_nonlinear_2021,vidotto_hybrid_2019}; consequently, the permeability tensor simplifies to ${\bm k}=k{\bm I}$, where $k$ is the permeability constant and ${\bm I}$ is the identity matrix.
Moreover, the blood and the interstitial fluid are incompressible, so conservation of mass yields
\begin{eqnarray}
    \nabla_{\mathcal{S}}\cdot{\bm u}=\psi,
    \label{Darcy_mass_conservation}
\end{eqnarray}
where $\psi$ is the source term (${\rm{s}^{-1}}$).
Then, combining the equations$~$(\ref{Darcy_flow}) and (\ref{Darcy_mass_conservation}) yields the generic equation for the pressure:
\begin{eqnarray}
    -\nabla_{\mathcal{S}} \cdot\left(\frac{k}{\mu}\nabla_{\mathcal{S}}p\right)=\psi.
\end{eqnarray}
Applying this equation separately to the capillary pressure $p_{cap}$ and interstitial pressure $p_t$ yields the coupled system on the surface $\mathcal{S}$:
\begin{eqnarray}
    \left\{
    \begin{aligned}
        &-\nabla_{\mathcal{S}}\cdot\left(\frac{k_{cap}}{\mu_{cap}}\nabla_{\mathcal{S}}p_{cap}\right)=\psi_{cap}({\bm x}) \quad \text{in $\mathcal{S}$},\\
        &-\nabla_{\mathcal{S}}\cdot\left(\frac{k_t}{\mu_t}\nabla_{\mathcal{S}}p_t\right)=\psi_t({\bm x}) \quad \text{in $\mathcal{S}$},
    \end{aligned}\right.
    \label{capillary_tissue_coupled_model}
\end{eqnarray}
where $k_{cap}$ and $k_t$ are the permeabilities, $\mu_{cap}$ and $\mu_t$ are the viscosities of blood and interstitial fluid, respectively, and $\psi_{cap}$ and $\psi_t$ are the source terms.
The capillary source $\psi_{cap}$ is given by
\begin{eqnarray}
    \psi_{cap}({\bm x})=\psi_{ac}+\psi_{cv}-\psi_{ct},
    \label{capillary_source}
\end{eqnarray}
where $\psi_{ac}$, $\psi_{cv}$ and $\psi_{ct}$ denote arteriolar inflow, venular drainage and fluid exchange between the capillary bed and the interstitial tissue, respectively.
The arteriolar and venular contributions are represented as a sum of point sources:
\begin{eqnarray}
    \psi_{ac}({\bm x})+\psi_{cv}({\bm x})=\sum^{N_{cap}}_{n=1}q_n\delta_{\mathcal{S}}({\bm x}-{\bm x}_n),
\end{eqnarray}
where $N_{cap}$ is the total number of point sources, $q_n$ is the flow rate at location ${\bm x}_n$, and $\delta_{\mathcal{S}}(\cdot)$ is the Dirac delta function on the surface $\mathcal{S}$.
Note that, under physiological conditions, $q_n>0$ corresponds to an arteriolar source supplying blood to the capillary bed, whereas $q_n<0$ corresponds to a venular source draining blood from the capillary bed.
The exchange $\psi_{ct}$ follows Starling's filtration principle \citep{xu_osmosis_2018,fritz_1d0d3d_2022,vidotto_hybrid_2019}:
\begin{eqnarray}
    \psi_{ct}({\bm x})=L_h S_{cap} [(p_{cap}({\bm x})-p_t({\bm x}))-\alpha_r(\pi_{cap}({\bm x})-\pi_t({\bm x}))],
\end{eqnarray}
where $L_h$ is the hydraulic conductivity of the capillary wall, $S_{cap}$ is the capillary surface area per unit volume of tissue, $\pi_{cap}$ is the capillary plasma osmotic pressure, $\pi_t$ is the interstitial fluid osmotic pressure, and $\alpha_r$ is the osmotic reflection coefficient of the capillary wall for proteins.
The capillary wall is largely impermeable to plasma proteins, so $\alpha_r\approx 1$ \citep{truskey_transport_2009}.
Moreover, we assume minimal protein leakage and insignificant spatial variation in protein concentrations within the capillary bed, and thus the osmotic pressure difference $\Delta \pi=\pi_{cap}-\pi_t$ is treated as a constant.
The exchange term $\psi_{ct}$ then simplifies to
\begin{eqnarray}
    \psi_{ct}({\bm x})=\alpha_{exch}(p_{cap}({\bm x})-p_t({\bm x})-\alpha_r\Delta\pi),
\end{eqnarray}
where $\alpha_{exch}=L_h S_{cap}$ is the fluid exchange rate (${\rm{\upmu m}\cdot\rm{s}/\rm{g}}$), and then the source $\psi_{cap}$ is given by
\begin{eqnarray}
     \psi_{cap}=\sum^{N_{cap}}_{n=1}q_n\delta_{\mathcal{S}}({\bm x}-{\bm x}_n)-\alpha_{exch}(p_{cap}-p_t-\alpha_r\Delta \pi).
\end{eqnarray}
Conservation of mass requires that fluid leaving the capillary bed enters the tissue, which yields
\begin{eqnarray}
    \psi_t=\psi_{ct}=\alpha_{exch}(p_{cap}-p_t-\alpha_r\Delta \pi).
\end{eqnarray}
Neither blood nor interstitial fluid crosses the boundary $\partial \mathcal{S}$.
Their fluxes, described by $-(k_{cap}/\mu_{cap})\nabla_{\mathcal{S}}p_{cap}\cdot {\bm n}_\mathcal{S}$ and $-(k_t/\mu_t)\nabla_{\mathcal{S}}p_t\cdot {\bm n}_\mathcal{S}$, must be zero, which yields the boundary conditions:
\begin{eqnarray}
    \left\{
    \begin{aligned}
        &-\frac{k_{cap}}{\mu_{cap}}\nabla_{\mathcal{S}}p_{cap}\cdot {\bm n}_{\mathcal{S}}=0 \quad \text{on $\partial \mathcal{S}$}, \\
        &-\frac{k_t}{\mu_t}\nabla_{\mathcal{S}}p_t\cdot {\bm n}_{\mathcal{S}}=0 \quad \text{on $\partial \mathcal{S}$},
    \end{aligned}\right.
\end{eqnarray}
where ${\bm n}_\mathcal{S}$ is the normal vector on $\partial \mathcal{S}$ that points outward.
Since both the permeability and viscosity are positive, the boundary conditions simplify to
\begin{eqnarray}
    \left\{
    \begin{aligned}
        &-\nabla_{\mathcal{S}}p_{cap}\cdot {\bm n}_{\mathcal{S}}=0 \quad \text{on $\partial \mathcal{S}$}, \\
        &-\nabla_{\mathcal{S}}p_t\cdot {\bm n}_{\mathcal{S}}=0 \quad \text{on $\partial \mathcal{S}$}.
    \end{aligned}\right.
\end{eqnarray}
\begin{figure}
    \centering
    \includegraphics[width=\textwidth]{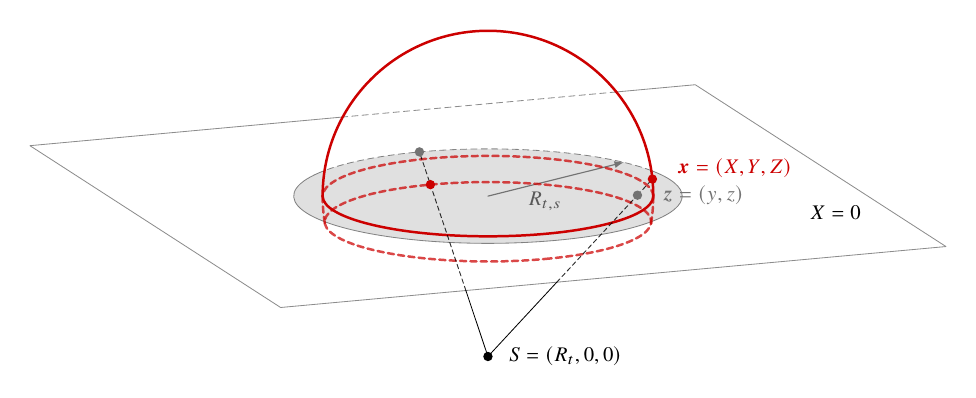}
    \caption{Schematic of the stereographic projection of the retinal surface onto the plane $X=0$.
    For illustration, the spherical cap is oriented with its aperture facing downward and is shown in red.
    A point ${\bm x}=(X,Y,Z)$ on the spherical cap is projected from $S=(R_t,0,0)$ to the point ${\bm z}=(y,z)$, where $(y,z)=(R_tY/(R_t-X),R_tZ/(R_t-X))$.
    The grey region represents the resulting projected disc, and the aperture rim is mapped onto its boundary.
    Dashed curves and lines indicate geometrically obscured portions in the selected viewing direction.}
    \label{schematic_stereographic_projection}
\end{figure}

We denote the capillary and tissue conductivities as $\kappa_{cap}=k_{cap}/\mu_{cap}$ and $\kappa_t=k_t/\mu_t$, respectively.
Then, with transformed capillary pressure $\tilde{p}_{cap}=p_{cap}-\alpha_r\Delta\pi$, the coupled system can be expressed as
\begin{eqnarray}
    \left\{
    \begin{aligned}
        &-\nabla_{\mathcal{S}} \cdot(\kappa_{cap}\nabla_{\mathcal{S}} \tilde{p}_{cap})=\sum^{N_{cap}}_{n=1}q_n\delta_{\mathcal{S}}({\bm x}-{\bm x}_n)-\alpha_{exch}(\tilde{p}_{cap}-p_t) \quad \text{in $\mathcal{S}$} \\
        &-\nabla_{\mathcal{S}} \cdot(\kappa_t\nabla_{\mathcal{S}} p_t)=\alpha_{exch}(\tilde{p}_{cap}-p_t) \quad \text{in $\mathcal{S}$}.
    \end{aligned}\right.
\end{eqnarray}
The boundary conditions are given by
\begin{eqnarray}
    \left\{
    \begin{aligned}
        &-\nabla_{\mathcal{S}} \tilde{p}_{cap}\cdot {\bm n}_\mathcal{S}=0, \quad \text{on $\partial \mathcal{S}$}, \\
        &-\nabla_{\mathcal{S}} p_t\cdot {\bm n}_\mathcal{S}=0 \quad \text{on $\partial \mathcal{S}$},
    \end{aligned}\right.
    \label{capillary_tissue_coupled_system_bc}
\end{eqnarray}
\subsection{Coupling between the vasculature and capillary-tissue continuum}
\label{CouplingVasculatureCapillaryTissueContinuum}
The arteriolar and venular trees are coupled to the capillary-tissue system through the sources located at ${\bm x}_n$ in the capillary bed.
With the analytic solution for the capillary pressure, the pressure at each source point can be expressed as a linear combination of all source flow rates, yielding $N_{cap}$ equations:
\begin{eqnarray}
    p_{cap}({\bm x}_n)=p_{cap}(q_1,\dots,q_{N_{cap}}), \quad n=1,\dots, N_{cap},
\end{eqnarray}
where $p_{cap}(q_1,\dots,q_{N_{cap}})$ denotes the linear combination of the flow rates $q_n$ with $n=1,\dots, N_{cap}$.
Each unknown flow rate $q_n$ is governed by the mass conservation equation in the system$~$(\ref{1D_model}), which implies $N_{cap}$ additional equations involving $N_{cap}$ unknown areas.
Together with the elastic model, these unknown areas can be expressed in terms of $p_{cap}({\bm x}_n)$, provided that $p_{cap}({\bm x}_n)$ is interpreted as the pressure at the distal point of the terminal vessel that is connected to the capillary bed at ${\bm x}={\bm x}_n$.
Together, these relations form $2N_{cap}$ equations for the $2N_{cap}$ unknowns: $N_{cap}$ pressures and $N_{cap}$ flow rates.
\subsection{Summary of multiscale model}
The arteriolar and venular trees are constructed using a vasculature model that grows vessels by minimising their total volume.
The capillary bed and interstitial tissue are modelled as coupled porous media, with the flow described by the coupled Darcy equations$~$(\ref{capillary_tissue_coupled_model}).
These components are coupled via source and sink points located in the capillary bed.
The multiscale coupling conditions can be specified by interpreting each source point as the distal point of the terminal vessel.
This model hierarchy describes the flow across multiple scales and thus provides a comprehensive description of retinal haemodynamics.
\section{Analytic solution for the capillary-tissue flow}
\label{AnalyticSolutionCapillaryTissueFlow}
\subsection{Stereographic projection of the spherical cap and decoupling transformation}
We next use a stereographic projection to map the spherical-cap domain to a circular disc and then decouple the mapped capillary-tissue system into mean and exchange pressure components.
Under this projection, the spherical-cap boundary becomes a circular boundary, while the geometric effect of the curvature is retained through the spatially varying conformal factor.
We denote the point in the projected domain as ${\bm z}=(y,z)$ and introduce the polar coordinates $(r,\theta)$ through
\begin{eqnarray}
    r=\sqrt{y^2+z^2}, \quad \tan\theta=z/y.
\end{eqnarray}
Thus, the angular coordinate $\theta$ in the projected plane is identical to the azimuthal angle on the spherical cap.
\begin{figure}
    \centering
    \includegraphics[width=0.76\textwidth]{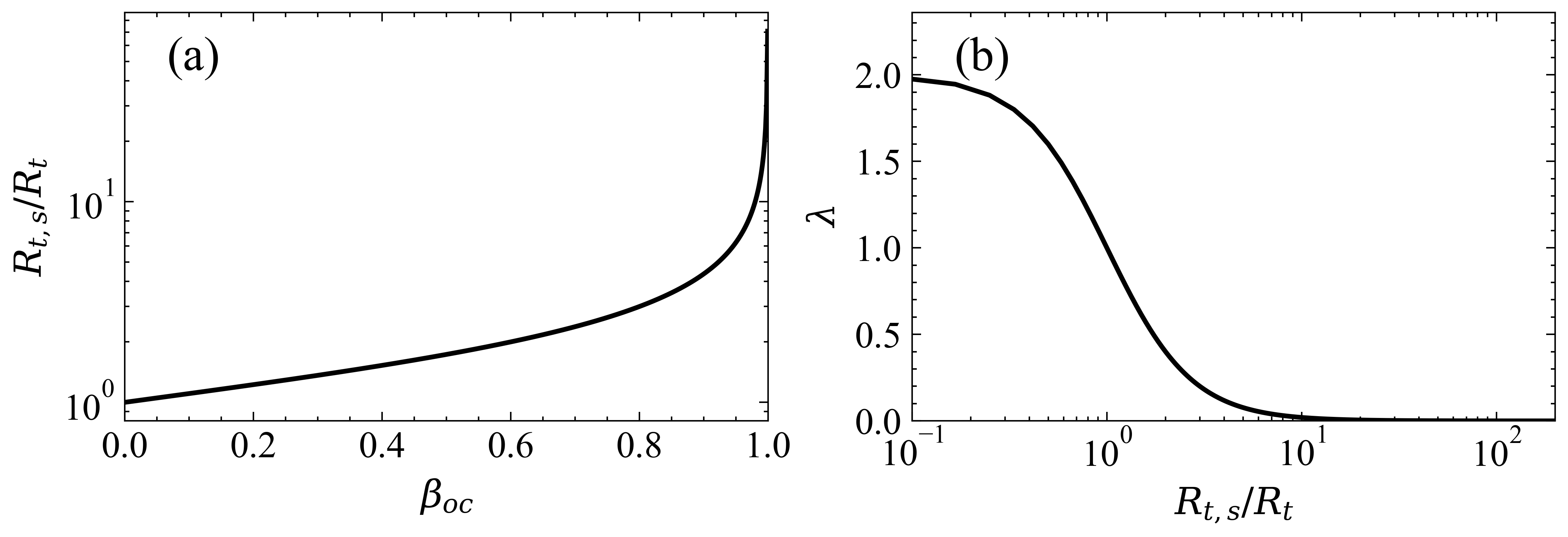}
    \caption{Projected radius and conformal factor on the spherical-cap domains with varying aperture. (a) Projected spherical-cap radius $R_{t,s}/R_t$ as a function of aperture parameter $\beta_{oc}$. (b) Conformal factor $\lambda$ as a function of $R_{t,s}/R_t$.}
    \label{conformal_factor_mapped_radius_cutting_ratio}
\end{figure}

The projection centre is placed at the pole $S=(R_t,0,0)$, and for any point ${\bm x}=(X,Y,Z)$ on the spherical cap, a line is drawn through $S$ and ${\bm x}$.
Its intersection with the plane $X=0$ yields the projected point ${\bm z}=(y,z)$.
Figure$~$\ref{schematic_stereographic_projection} illustrates this construction, with the spherical cap shown in red and the resulting projected disc shown in grey.
The red point on the aperture rim and its projection demonstrate that the aperture rim is mapped onto the boundary of the projected disc.
Dashed curves and lines denote the obscured parts.
The forward mapping, written in $(X,Y,Z)$, is given by
\begin{eqnarray}
    {\bm z}=(y,z)=\left(\frac{R_tY}{R_t-X},\frac{R_tZ}{R_t-X}\right),
    \label{stereographic_mapping}
\end{eqnarray}
and the inverse mapping is given by
\begin{eqnarray}
    {\bm x}=(X,Y,Z)=\left(R_t\frac{r^2-R^2_t}{R^2_t+r^2},\frac{2R^2_ty}{R^2_t+r^2},\frac{2R^2_tz}{R^2_t+r^2}\right),
    \label{stereographic_mapping_inverse}
\end{eqnarray}
where $r$ is the radial coordinate on the $yz$-plane.
The resulting projected domain is given by
\begin{eqnarray}
    \mathcal{D}=\{(y,z)\left\vert y^2+z^2\leq R^2_{t,s}\right.\}, 
\end{eqnarray}
where the stereographic projected radius $R_{t,s}$ is given by
\begin{eqnarray}
    R_{t,s}=R_t\sqrt{\frac{1+\beta_{oc}}{1-\beta_{oc}}}.
\end{eqnarray}
The dependence of $R_{t,s}/R_t$ on $\beta_{oc}$ is shown in figure$~$\ref{conformal_factor_mapped_radius_cutting_ratio}(a).
For a full sphere ($\beta_{oc}=1$), the projected radius satisfies $R_{t,s}\to \infty$, which corresponds to the entire plane; for a hemisphere ($\beta_{oc}=0$), the projected radius is $R_{t,s}=R_t$.
Thus, the projected radius grows as the cap extends in the direction opposite to the fovea.
Using spherical coordinates$~$(\ref{spherical_coordinates_spherical_cap}), the metric on the spherical cap is given by
\begin{eqnarray}
    ds^2_{\mathcal{S}}=R^2_t(d\phi^2+\sin^2\phi d\theta^2), 
    \label{spherical_cap_domain_metric}
\end{eqnarray}
which, in the projected coordinates $(y,z)$, becomes
\begin{eqnarray}
    ds^2_{\mathcal{S}}=\frac{4R^4_t}{(R^2_t+r^2)^2}(dy^2+dz^2).
\end{eqnarray}
Since the spherical metric is a scalar multiple of the Euclidean metric $dy^2+dz^2$, the stereographic projection$~$(\ref{stereographic_mapping}) is conformal.
The corresponding conformal factor $\lambda(y,z)$ is given by
\begin{eqnarray}
    \lambda(y,z)=\frac{2R^2_t}{R^2_t+y^2+z^2}.
\end{eqnarray}
and the derivation is provided in Appendix$~$\ref{StereographicProjectionConformalFactor}.
In polar coordinates $(r,\theta)$, the conformal factor can be expressed as
\begin{eqnarray}
    \lambda(r)=\frac{2R^2_t}{R^2_t+r^2}.
    \label{conformal_factor_radial_coordinate}
\end{eqnarray}
The value of $\lambda$ at the boundary is shown in figure$~$\ref{conformal_factor_mapped_radius_cutting_ratio}(b).
As $\beta_{oc}$ increases, the radius of projected domain increases and the value of $\lambda$ decreases.
Under the stereographic projection$~$(\ref{stereographic_mapping}), the divergence of surface Darcy flux becomes \citep{dritschel_conformal_2023,reuter_laplacebeltrami_2006}
\begin{eqnarray}
    -\nabla_{\mathcal{S}} \cdot(\kappa_{cap}\nabla_{\mathcal{S}} \tilde{p}_{cap})=-\lambda^{-2}({\bm z})\nabla\cdot(\kappa_{cap}\nabla \tilde{p}_{cap}), \\
    -\nabla_{\mathcal{S}} \cdot(\kappa_t\nabla_{\mathcal{S}} p_t)=-\lambda^{-2}({\bm z})\nabla \cdot(\kappa_t\nabla p_t).
\end{eqnarray}
Moreover, since the surface element satisfies
\begin{eqnarray}
    dA=\lambda^2({\bm z})d{\bm z},
\end{eqnarray}
the surface delta function becomes
\begin{eqnarray}
    \delta_{\mathcal{S}}({\bm x}({\bm z})-{\bm x}_n)=\frac{1}{\lambda^2({\bm z}_n)}\delta({\bm z}-{\bm z}_n),
\end{eqnarray}
where ${\bm z}_n={\bm z}({\bm x}_n)$ and $\delta({\bm z}-{\bm z}_n)$ is the two-dimensional delta function in the projected domain.
The mapped coupled system is thus given by
\begin{eqnarray}
    \left\{
    \begin{aligned}
        &-\lambda^{-2}({\bm z})\nabla\cdot(\kappa_{cap}\nabla \tilde{p}_{cap})=\sum^{N_{cap}}_{n=1}\frac{q_n}{\lambda^2({\bm z}_n)}\delta({\bm z}-{\bm z}_n)-\alpha_{exch}(\tilde{p}_{cap}-p_t) \quad \text{in $\mathcal{D}$},\\
        &-\lambda^{-2}({\bm z})\nabla \cdot(\kappa_t\nabla p_t)=\alpha_{exch}(\tilde{p}_{cap}-p_t) \quad \text{in $\mathcal{D}$},
    \end{aligned}\right.
\end{eqnarray}
and the boundary conditions are
\begin{eqnarray}
    \left\{
    \begin{aligned}
        &-\nabla \tilde{p}_{cap}\cdot {\bm n}=0, \quad \text{on $\partial \mathcal{D}$}, \\
        &-\nabla p_t\cdot {\bm n}=0 \quad \text{on $\partial \mathcal{D}$}.
    \end{aligned}\right.
\end{eqnarray}
Multiplying the mapped equations by $\lambda^2({\bm z})$ and using the identity
\begin{eqnarray}
    \lambda^2({\bm z})\delta({\bm z}-{\bm z}_n)=\lambda^2({\bm z}_n)\delta({\bm z}-{\bm z}_n),
\end{eqnarray}
the mapped coupled system becomes
\begin{eqnarray}
    \left\{
    \begin{aligned}
        &-\kappa_{cap}\nabla^2 \tilde{p}_{cap}=\sum^{N_{cap}}_{n=1}q_n\delta({\bm z}-{\bm z}_n)-\lambda^2({\bm z})\alpha_{exch}(\tilde{p}_{cap}-p_t) \quad \text{in $\mathcal{D}$},\\
        &-\kappa_t\nabla^2 p_t=\lambda^2({\bm z})\alpha_{exch}(\tilde{p}_{cap}-p_t) \quad \text{in $\mathcal{D}$}.
    \end{aligned}\right.
    \label{mapped_capillary_tissue_coupled_system}
\end{eqnarray}

The system$~$(\ref{mapped_capillary_tissue_coupled_system}) is similar to that in the circular-disc formulation \citep{lin_novel_2026}, except for the conformal factor $\lambda({\bm z})$ in the exchange term.
For completeness, the analytic solution for the capillary-tissue system on the circular disc is summarised in Appendix$~$\ref{CircularDiscCapillaryTissueSolution}.
We use the same transformation to decouple the mapped capillary-tissue system:
\begin{eqnarray}
    \begin{bmatrix}
        p_{mean} \\
        p_{exch}
    \end{bmatrix}={\bm T}
    \begin{bmatrix}
        \tilde{p}_{cap} \\
        p_t
    \end{bmatrix},
\end{eqnarray}
where the transformation matrix ${\bm T}$ is given by
\begin{eqnarray}
    {\bm T}=\frac{\kappa_{cap}}{\kappa_{cap}+\kappa_t}
    \begin{bmatrix}
        1 & \kappa_t/\kappa_{cap} \\
        -1 & 1
    \end{bmatrix}.
\end{eqnarray}
The new pressure $p_{mean}$ is the conductivity-weighted average pressure, while $p_{exch}$ is the exchange pressure, proportional to the pressure difference between the capillary bed and interstitial tissue.
Adding the two mapped equations implies
\begin{eqnarray}
    -\kappa_{cap}\nabla^2\tilde{p}_{cap}-\kappa_t\nabla^2p_t=\sum^{N_{cap}}_{n=1}q_n\delta({\bm z}-{\bm z}_n).
\end{eqnarray}
Then, we obtain
\begin{eqnarray}
    -\nabla^2 p_{mean}=\frac{1}{\kappa_{cap}+\kappa_t}\sum^{N_{cap}}_{n=1}q_n\delta({\bm z}-{\bm z}_n).
\end{eqnarray}
Subtracting the tissue equation divided by $\kappa_t$ from the capillary equation divided by $\kappa_{cap}$ gives
\begin{eqnarray}
   -\nabla^2\tilde{p}_{cap}+\nabla^2p_t=\frac{1}{\kappa_{cap}}\sum^{N_{cap}}_{n=1}q_n\delta({\bm z}-{\bm z}_n)-\gamma^2({\bm z})(\tilde{p}_{cap}-p_t),
\end{eqnarray}
where $\gamma({\bm z})=\gamma_0\lambda({\bm z})$ is the inverse screening length on the projected circular disc, and $\gamma_0$ is the constant inverse screening length on the spherical cap, given by
\begin{eqnarray}
    \gamma_0=\sqrt{\alpha_{exch}(1/\kappa_{cap}+1/\kappa_t)}.
\end{eqnarray}
Then, the equation for $p_{exch}$ is given by
\begin{eqnarray}
    -\nabla^2 p_{exch}+\lambda^2({\bm z})\alpha_{exch}\left(\frac{1}{\kappa_{cap}}+\frac{1}{\kappa_t}\right)p_{exch}=-\frac{1}{\kappa_{cap}+\kappa_t}\sum^{N_{cap}}_{n=1}q_n\delta({\bm z}-{\bm z}_n).
\end{eqnarray}
The decoupled system is then given by
\begin{eqnarray}
    \left\{
    \begin{aligned}
        &-\nabla^2 p_{mean}=\frac{1}{\kappa_{cap}+\kappa_t}\sum^{N_{cap}}_{n=1}q_n\delta({\bm z}-{\bm z}_n) \quad \text{in $\mathcal{D}$},\\
        &-\nabla^2 p_{exch}+\gamma^2({\bm z})p_{exch}=-\frac{1}{\kappa_{cap}+\kappa_t}\sum^{N_{cap}}_{n=1}q_n\delta({\bm z}-{\bm z}_n) \quad \text{in $\mathcal{D}$}.
    \end{aligned}\right.
\end{eqnarray}
The Green's function for the mean pressure satisfies \citep{kim_proof_1993,riley_mathematical_2006}
\begin{eqnarray}
    \left\{
    \begin{aligned}
        &{\nabla^\prime}^2 f_{mean}({\bm z},{\bm z}^\prime)=\delta({\bm z}-{\bm z}^\prime)\quad &\text{in $\mathcal{D}$}, \\
        &\nabla^\prime f_{mean}({\bm z},{\bm z}^\prime)\cdot {\bm n}^\prime=1/S\quad &\text{on $\partial\mathcal{D}$},
        \label{p_mean_unit_source}
    \end{aligned}\right.
\end{eqnarray}
where $\nabla^\prime$ denotes the gradient operator with respect to ${\bm z}^\prime$ and $S=2\pi R_{t,s}$ is the length of the boundary $\partial\mathcal{D}$. The Green's function for the exchange pressure satisfies
\begin{eqnarray}
    \left\{\begin{aligned}
        &\nabla^2 f_{exch}-\gamma^2({\bm z})f_{exch}=\delta({\bm z}-{\bm z}_n) \quad \text{in $\mathcal{D}$}, \\
        &\nabla f_{exch}\cdot {\bm n}=0 \quad \text{on $\partial \mathcal{D}$}.
        \label{f_exch_equation_boundary_condition}
    \end{aligned}\right.
\end{eqnarray}
Once $f_{mean}({\bm z},{\bm z}^\prime)$ is determined, the mean pressure is obtained via the following convolution \citep{riley_mathematical_2006,kim_proof_1993}
\begin{eqnarray}
    p_{mean}({\bm z})&=&\int_{\mathcal D}f_{mean}{\nabla^\prime}^2 p_{mean}({\bm z}^\prime)d{\bm z}^\prime \nonumber\\
    &+&\int_{\partial {\mathcal D}}\left[p_{mean}({\bm z}^\prime)\nabla^\prime f_{mean}-f_{mean}\nabla^\prime p_{mean}({\bm z}^\prime)\right]\cdot {\bm n}^\prime ds^\prime,
\end{eqnarray}
and using the boundary conditions for $p_{mean}$ and $f_{mean}$ yields
\begin{eqnarray}
    p_{mean}({\bm z})=-\frac{1}{\kappa_{cap}+\kappa_t}\sum^{N_{cap}}_{n=1}q_nf_{mean}({\bm z},{\bm z}_n)+\overline{p_{mean}},
\end{eqnarray}
where $\overline{p_{mean}}$ is the boundary average of mean pressure given by
\begin{eqnarray}
    \overline{p_{mean}}=\frac{1}{S}\int_{\partial {\mathcal D}}p_{mean}({\bm z}^\prime) ds^\prime.
\end{eqnarray}
Moreover, the exchange pressure can be constructed by using the following superposition of $f_{exch}({\bm z},{\bm z}_n)$:
\begin{eqnarray}
    p_{exch}({\bm z})=\frac{1}{\kappa_{cap}+\kappa_t}\sum^{N_{cap}}_{n=1}q_nf_{exch}({\bm z},{\bm z}_n).
\end{eqnarray}
Note that the value of $\overline{p_{mean}}$ is not fixed by the capillary-tissue system; it is determined by coupling to the arteriolar and venular trees, which provide the absolute pressure level through specified inlet and outlet pressures.
The no-flux boundary condition further requires the compatibility condition
\begin{eqnarray}
    \sum^{N_{cap}}_{n=1}q_n=0,
    \label{compatibility_condition}
\end{eqnarray}
which expresses the conservation of mass.
\subsection{Green's functions for mean and exchange pressures}
\begin{figure}
    \centering
    \includegraphics[width=0.76\textwidth]{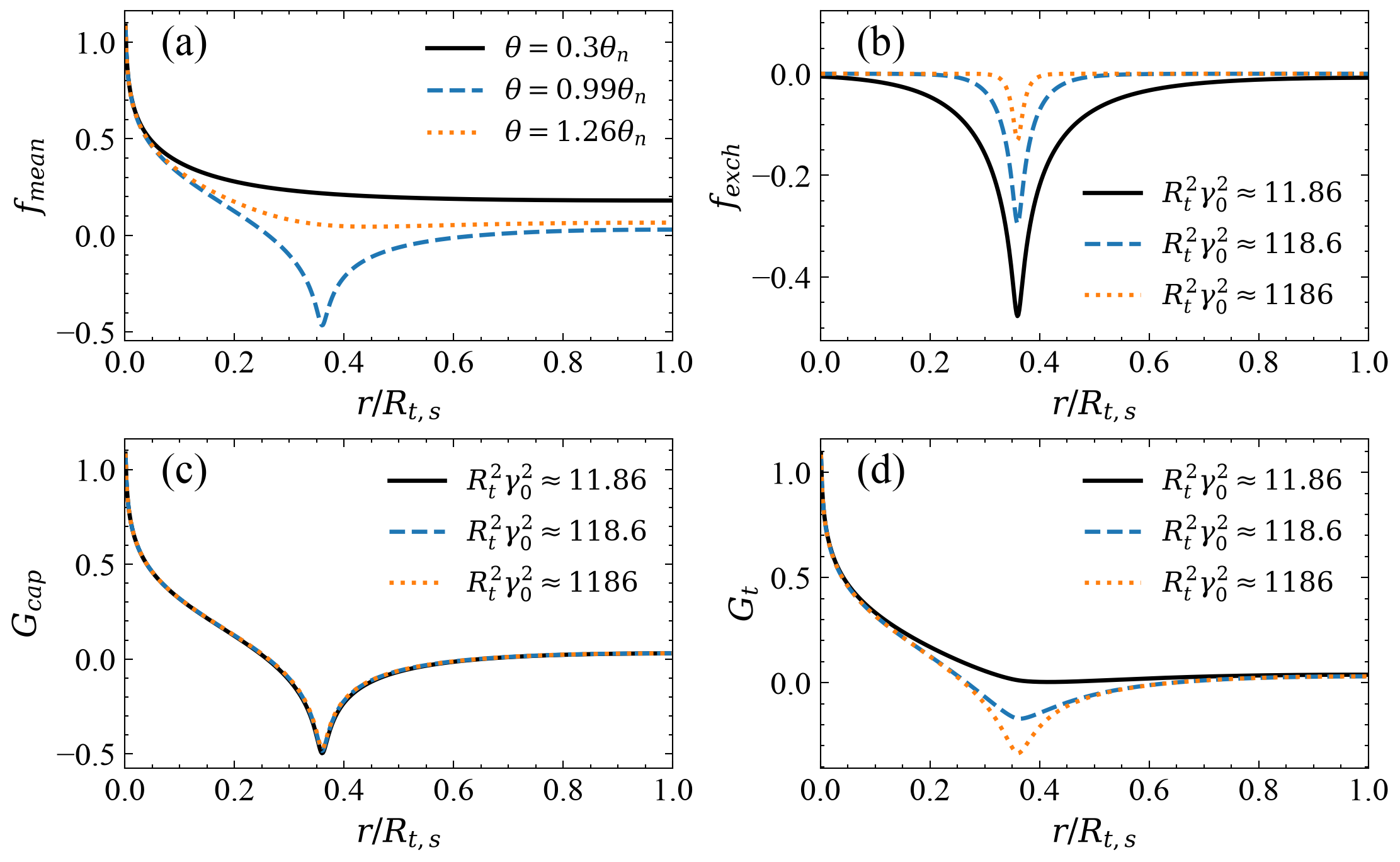}
    \caption{Radial profiles of the Green's functions for a single source at $(r_n,\theta_n)=(0.36R_{t,s},2\pi/3)$ with $R_t=1.26~\rm{cm}$. (a) Mean-pressure Green's function $f_{mean}$ for $\theta=0.3\theta_n$, $0.99\theta_n$ and $1.26\theta_n$. (b) Exchange-pressure Green's function $f_{exch}$. (c) Capillary-pressure Green's function $G_{cap}$. (d) Tissue-pressure Green's function $G_t$. In panels$~$(b)--(d), $\theta=0.99\theta_n$, and the black solid, blue dashed and orange dotted curves denote $R_t^2\gamma_0^2\approx11.86$, $R_t^2\gamma_0^2\approx118.6$ and $R_t^2\gamma_0^2\approx1186$, respectively. The middle value is computed using the physiologically typical parameter values in table$~$\ref{parameters_multiscale_model} and corresponds to $\gamma_0\approx8.643\times10^{-4}~\rm{\upmu m^{-1}}$.}
    \label{f_mean_f_exch_g_cap_g_t_variation}
\end{figure}

It is standard to solve $f_{mean}({\bm z},{\bm z}^\prime)$ by using the method of images.
Since ${\bm z}^\prime$ is a variable for the operator $\nabla^\prime$, we place an image source at ${\bm z}^\dprime=(R^2_{t,s}/\vert {\bm z}\vert^2){\bm z}$, which ensures that $\nabla^\prime f_{mean}\cdot {\bm n}^\prime=1/S$ on $\partial \mathcal{D}$.
The Green's function for the mean pressure is then given by
\begin{eqnarray}
    f_{mean}({\bm z},{\bm z}^\prime)=\frac{1}{4\pi}\left(\log\vert {\bm z}^\prime-{\bm z}\vert^2+\log\left\vert {\bm z}^\prime-\frac{R^2_{t,s}}{\vert {\bm z}\vert^2}{\bm z}\right\vert^2\right),
\end{eqnarray}
and substituting ${\bm z}^\prime={\bm z}_n$ into $f_{mean}({\bm z},{\bm z}^\prime)$ yields
\begin{eqnarray}
    f_{mean}({\bm z},{\bm z}_n)=\frac{1}{4\pi}\left(\log\vert {\bm z}_n-{\bm z}\vert^2+\log\left\vert {\bm z}_n-\frac{R^2_{t,s}}{\vert {\bm z}\vert^2}{\bm z}\right\vert^2\right).
\end{eqnarray}
The Green's function $f_{mean}$ in polar coordinates $(r,\theta)$ is given by
\begin{eqnarray}
   f_{mean}(r,\theta;r_n,\theta_n)&=&\frac{1}{4\pi}\log(r^2_n+r^2-2r_nr\cos(\theta_n-\theta)) \nonumber\\
    &&+\frac{1}{4\pi}\log\left(r^2_n+\frac{R^4_{t,s}}{r^2}-2r_n\frac{R^2_{t,s}}{r}\cos(\theta_n-\theta)\right),
    \label{f_mean_solution}
\end{eqnarray}
where $(r_n,\theta_n)$ are the polar coordinates of source point ${\bm z}_n=(r_n\cos\theta_n,r_n\sin\theta_n)$.
Figure$~$\ref{f_mean_f_exch_g_cap_g_t_variation}(a) shows the Green's function $f_{mean}$ as a function of $r/R_{t,s}$ for a source at $(r_n,\theta_n)=(0.36R_{t,s},2\pi/3)$.
Curves with different colours correspond to various values of $\theta$.
The profile of $f_{mean}$ becomes increasingly sharp near $r=r_n$ as $\theta$ approaches $\theta_n$.

The equation for $f_{exch}$ can be written in polar coordinates $(r,\theta)$ as
\begin{eqnarray}
    \frac{1}{r}\frac{\partial}{\partial r}\left(r\frac{\partial f_{exch}}{\partial r}\right)+\frac{1}{r^2}\frac{\partial^2 f_{exch}}{\partial \theta^2}-\gamma^2(r)f_{exch}=\frac{\delta(r-r_n)\delta(\theta-\theta_n)}{r},
\end{eqnarray}
where $\gamma(r)=\gamma_0\lambda(r)$.
We assume the following trial solution for $f_{exch}$:
\begin{eqnarray}
    f_{exch}=\sum^\infty_{m=-\infty}f^m_{exch}(r)\exp(im(\theta-\theta_n)),
\end{eqnarray}
where $m\in\mathbb{Z}$.
Employing the following expansion of $\delta(\theta-\theta_n)$:
\begin{eqnarray}
    \delta(\theta-\theta_n)=\frac{1}{2\pi}\sum^\infty_{m=-\infty}\exp(im(\theta-\theta_n)),
\end{eqnarray}
the equation for $f^m_{exch}(r)$ becomes
\begin{eqnarray}
    \frac{d^2f^m_{exch}}{dr^2}+\frac{1}{r}\frac{df^m_{exch}}{dr}-\left(\gamma^2(r)+\frac{m^2}{r^2}\right)f^m_{exch}=\frac{1}{2\pi r}\delta(r-r_n).
    \label{exchange_pressure_radial_component_radial_coordinate}
\end{eqnarray}
Multiplying both sides by $r^2$ yields
\begin{eqnarray}
    r^2\frac{d^2f^m_{exch}}{dr^2}+r\frac{df^m_{exch}}{dr}-(\gamma^2_0r^2\lambda^2(r)+m^2)f^m_{exch}=\frac{r}{2\pi}\delta(r-r_n).
\end{eqnarray}
The boundary condition for the radial component $f^m_{exch}$ is
\begin{eqnarray}
    \left. \frac{df^m_{exch}}{dr}\right\vert_{r=R_{t,s}}=0,
\end{eqnarray}
and $f^m_{exch}$ must be finite at $r=0$, which implies the other boundary condition.
To solve for the radial component $f^m_{exch}$, we introduce a new variable
\begin{eqnarray}
    w=\frac{R^2_t-r^2}{R^2_t+r^2},
    \label{variable_w}
\end{eqnarray}
which satisfies $-\beta_{oc}\leq w\leq 1$, since $0\leq r\leq R_t\sqrt{(1+\beta_{oc})/(1-\beta_{oc})}$.
The aperture parameter enters the analytic solution only through the lower boundary value $w=-\beta_{oc}$, which determines the boundary condition for the associated Legendre functions.
From the equation$~$(\ref{variable_w}), we obtain
\begin{eqnarray}
    r\frac{dw}{dr}=-\frac{4R^2_tr^2}{(R^2_t+r^2)^2}=-(1-w^2), \label{derivative_w_conformal_factor} \\
    r^2\left(\frac{dw}{dr}\right)^2=\frac{16R^4_tr^4}{(R^2_t+r^2)^4}=(1-w^2)^2, \\
    r^2\frac{d^2w}{dr^2}=\frac{4R^2_tr^2(3r^2-R^2_t)}{(R^2_t+r^2)^3}=(1-w^2)(1-2w).
\end{eqnarray}
The first derivative term $rdf^m_{exch}/dr$ becomes
\begin{eqnarray}
    r\frac{df^m_{exch}}{dr}=r\frac{df^m_{exch}}{dw}\frac{dw}{dr}=-(1-w^2)\frac{df^m_{exch}}{dw},
\end{eqnarray}
while the second derivative term $r^2d^2f^m_{exch}/dr^2$ is
\begin{eqnarray}
    r^2\frac{d^2f^m_{exch}}{dr^2}&=&r^2\frac{d^2f^m_{exch}}{dw^2}\left(\frac{dw}{dr}\right)^2+r^2\frac{df^m_{exch}}{dw}\frac{d^2w}{dr^2} \\
    &=&(1-w^2)^2\frac{d^2f^m_{exch}}{dw^2}+(1-w^2)(1-2w)\frac{df^m_{exch}}{dw}.
\end{eqnarray}
Thus, the combination in the radial equation is
\begin{eqnarray}
    r^2\frac{d^2f^m_{exch}}{dr^2}+r\frac{df^m_{exch}}{dr}=(1-w^2)^2\frac{d^2f^m_{exch}}{dw^2}-2w(1-w^2)\frac{df^m_{exch}}{dw}.
\end{eqnarray}
Moreover, from the equations$~$(\ref{derivative_w_conformal_factor}) and (\ref{conformal_factor_radial_coordinate}), we obtain
\begin{eqnarray}
    r^2\lambda^2(r)=R^2_t(1-w^2),
\end{eqnarray}
while the function $\delta(r-r_n)$ can be written as
\begin{eqnarray}
    \delta(r-r_n)=\left\vert \frac{dw}{dr}\right\vert_{r=r_n}\delta(w-w_n)=\frac{1-w^2_n}{r_n}\delta(w-w_n).
\end{eqnarray}
Thus,
\begin{eqnarray}
    \frac{r}{2\pi}\delta(r-r_n)=\frac{r_n}{2\pi}\delta(r-r_n)=\frac{1-w^2_n}{2\pi}\delta(w-w_n),
\end{eqnarray}
and the equation for the radial component becomes
\begin{eqnarray}
    \frac{d}{dw}\left((1-w^2)\frac{df^m_{exch}}{dw}\right)-\left(R^2_t\gamma^2_0+\frac{m^2}{1-w^2}\right)f^m_{exch}=\frac{1}{2\pi}\delta(w-w_n).
    \label{exchange_pressure_radial_component}
\end{eqnarray}
Since the equation has a singularity at $w=w_n$, we construct the solution piecewise as
\begin{eqnarray}
    f^m_{exch}(w,w_n)=\left\{
    \begin{aligned}
        &C_mf^m_1(w)f^m_2(w_n), \quad w<w_n, \\
        &C_mf^m_1(w_n)f^m_2(w), \quad w>w_n,
    \end{aligned}\right.
\end{eqnarray}
where $f^m_1$ is the solution satisfying the boundary conditions at $w=-\beta_{oc}$ and $f^m_2$ is the other solution that is finite at $w=1$.
The constant $C_m$ is determined by imposing the jump condition induced by the singular source at $w=w_n$.
The solution can equivalently be written as
\begin{eqnarray}
    f^m_{exch}(w,w_n)=C_mf^m_1(w_<)f^m_2(w_>),
    \label{f_m_exch_small_large}
\end{eqnarray}
where $w_<=\min\{w,w_n\}$, $w_>=\max\{w,w_n\}$, and this form makes $f^m_{exch}$ continuous at $w=w_n$.
The equation$~$(\ref{exchange_pressure_radial_component}) is the associated Legendre equation \citep[\S14.2(ii)]{olver_nist_2010-1} with order $\vert m\vert$ and degree $\nu$, where $\nu$ is determined from $\nu(\nu+1)=-R^2_t\gamma^2_0$ and given by
\begin{eqnarray}
    \nu=-\frac{1}{2}+i\tau, \quad \tau=\sqrt{R^2_t\gamma^2_0-1/4},
\end{eqnarray}
where $\tau$ is complex-valued when $R^2_t\gamma^2_0< 1/4$ and real-valued when $R^2_t\gamma^2_0>1/4$.
This determines whether the Legendre functions oscillate or decay, corresponding physically to whether the screening length is larger or smaller than the characteristic size of the tissue domain.
Note that the variable $w$ has a direct geometric interpretation on the spherical cap.

The two solutions $f^m_1$ and $f^m_2$ are constructed from combinations of associated Legendre functions of first and second kind, $P^{\vert m\vert}_\nu(w)$ and $Q^{\vert m\vert}_\nu(w)$, respectively.
Specifically, $f^m_1$ is constructed to satisfy the boundary condition at $w=-\beta_{oc}$, while $f^m_2$ is constructed to remain finite at $w=1$.
The coefficient $C_m$ is then obtained from the jump condition.
The detailed derivation is provided in Appendix$~$\ref{SolutionRadialComponentGreensFunctionExchangePressureSimplification}.
The solution for the radial component $f^m_{exch}$ is thus given by
\begin{eqnarray}
    f^m_{exch}&=&\frac{\Gamma(\nu-\vert m\vert+1)}{2\pi\Gamma(\nu+\vert m\vert+1)}\frac{Q^{\vert m\vert\prime}_\nu(-\beta_{oc})}{P^{\vert m\vert\prime}_\nu(-\beta_{oc})}P^{\vert m\vert}_\nu(w_<)P^{\vert m\vert}_\nu(w_>) \nonumber\\
    &&-\frac{\Gamma(\nu-\vert m\vert+1)}{2\pi\Gamma(\nu+\vert m\vert+1)}Q^{\vert m\vert}_\nu(w_<)P^{\vert m\vert}_\nu(w_>),
    \label{f_m_exch_Q_P}
\end{eqnarray}
where $Q^{\vert m\vert\prime}_\nu(w)$ and $P^{\vert m\vert\prime}_\nu(w)$ are derivatives of functions $Q^{\vert m\vert}_\nu(w)$ and $P^{\vert m\vert}_\nu(w)$.
The Green's function for the exchange pressure is obtained by summing over all angular components.
To obtain a more compact form of $f_{exch}$, we express $Q^{\vert m\vert}_\nu(w)$ and its derivative in terms of $P^{\vert m\vert}_\nu(w)$.
The Legendre addition formula then sums one component explicitly, while the finite-aperture boundary correction remains a series.
The simplification is provided in Appendix$~$\ref{SolutionRadialComponentGreensFunctionExchangePressureSimplification}.
Then, $f_{exch}$ simplifies to
\begin{eqnarray}
    f_{exch}(r,\theta;r_n,\theta_n)&=&\frac{1}{4\sin(\pi\nu)}P^0_\nu\left(-w_nw-\sqrt{1-w^2_n}\sqrt{1-w^2}\cos(\theta-\theta_n)\right) \nonumber \\
    &&+\sum^\infty_{m=0}c_m(\nu,\beta_{oc})P^m_\nu(w_n)P^m_\nu(w)\cos(m(\theta-\theta_n)),
    \label{f_exch_solution}
\end{eqnarray}
where $w=(R^2_t-r^2)/(R^2_t+r^2)$ and
\begin{eqnarray}
    c_m(\nu,\beta_{oc})=\left\{
    \begin{aligned}
        &\frac{1}{4\sin(\pi\nu)}\frac{P^{0\prime}_\nu(\beta_{oc})}{P^{0\prime}_\nu(-\beta_{oc})}, \quad m=0, \\
        &\frac{\Gamma(\nu-m+1)}{\Gamma(\nu+ m+1)}\frac{(-1)^m}{2\sin(\pi\nu)}\frac{P^{ m\prime}_\nu(\beta_{oc})}{P^{m\prime}_\nu(-\beta_{oc})}, \quad m\geq 1.
    \end{aligned}\right.
\end{eqnarray}
Note that all components in equation$~$(\ref{f_exch_solution}) are real-valued, and the first term is the solution independent of the boundary while the series corresponds to the effects from the finite aperture.

Figure$~$\ref{f_mean_f_exch_g_cap_g_t_variation}(b) displays $f_{exch}$ as a function of $r/R_{t,s}$ for $(r_n,\theta_n)=(0.36R_{t,s},2\pi/3)$ and $\theta=0.99\theta_n$. The blue dashed curve corresponds to $R^2_t\gamma^2_0\approx118.6$, computed using the typical parameters in table$~$\ref{parameters_multiscale_model}, while the black solid and orange dotted curves correspond to $R^2_t\gamma^2_0\approx11.86$ and $R^2_t\gamma^2_0\approx1186$, respectively.
The function $f_{exch}$ has a minimum near $r=r_n$; as $R^2_t\gamma^2_0$ increases, this minimum becomes narrower and less negative.
\subsection{Solutions for capillary and tissue pressures}
We denote the Green's functions for the capillary and tissue pressures as
\begin{eqnarray}
    G_{cap}(r,\theta;r_n,\theta_n)=f_{mean}(r,\theta;r_n,\theta_n)+\frac{\kappa_t}{\kappa_{cap}}f_{exch}(r,\theta;r_n,\theta_n), 
    \label{Greens_func_cap} \\
    G_t(r,\theta;r_n,\theta_n)=f_{mean}(r,\theta;r_n,\theta_n)-f_{exch}(r,\theta;r_n,\theta_n),
    \label{Greens_func_t}
\end{eqnarray}
where the functions $f_{mean}$ and $f_{exch}$ are given by equations$~$(\ref{f_mean_solution}) and (\ref{f_exch_solution}), respectively.
Figure$~$\ref{f_mean_f_exch_g_cap_g_t_variation}(c) and (d) show the dependence of $G_{cap}$ and $G_t$ on $r/R_{t,s}$ for a source $(r_n,\theta_n)=(0.36R_{t,s},2\pi/3)$ with $\theta=0.99\theta_n$, where different colours correspond to various values of $R^2_t\gamma^2_0$.
At $r=r_n$, the function $G_{cap}$ has a logarithmic singularity, appearing as a sharp minimum, whereas $G_t$ remains finite because the singular parts of $f_{mean}$ and $f_{exch}$ are balanced.
As $R_t^2\gamma_0^2$ increases, the exchange response becomes more localised around the source.
Accordingly, the profiles of $G_{cap}$ remain nearly indistinguishable, whereas the local minimum in $G_t$ becomes narrower and more pronounced.
The pressures can then be expressed as
\begin{eqnarray}
    p_{cap}(r,\theta)=-\frac{1}{\kappa_{cap}+\kappa_t}\sum^{N_{cap}}_{n=1}q_nG_{cap}(r,\theta;r_n,\theta_n)+\overline{p_{mean}}+\alpha_r\Delta\pi, \label{capillary_pressure} \\
    p_t(r,\theta)=-\frac{1}{\kappa_{cap}+\kappa_t}\sum^{N_{cap}}_{n=1}q_nG_t(r,\theta;r_n,\theta_n)+\overline{p_{mean}}.
    \label{tissue_pressure}
\end{eqnarray}
\begin{figure}
    \centering
    \includegraphics[width=0.76\textwidth]{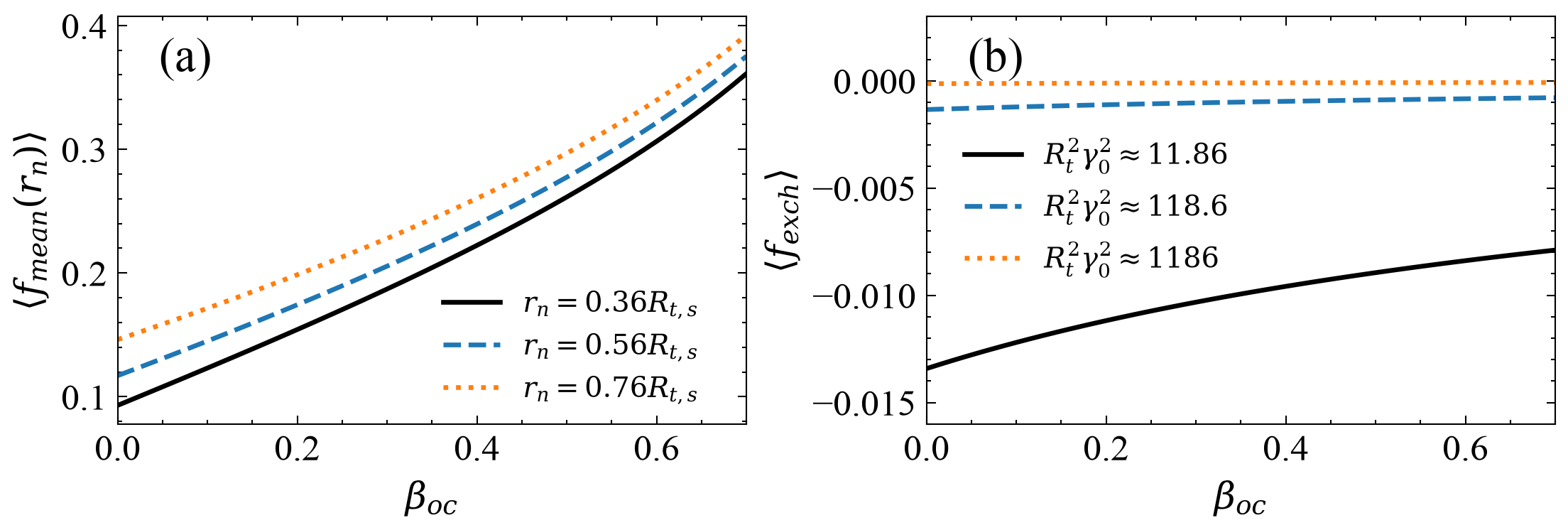}
    \caption{Dependence of the surface-averaged Green's functions on the aperture parameter $\beta_{oc}$ for $R_t=1.26~\rm{cm}$.
    (a) Surface average $\langle f_{mean}(r_n)\rangle$, where the black solid, blue dashed and orange dotted curves correspond to $r_n=0.36R_{t,s}$, $r_n=0.56R_{t,s}$ and $r_n=0.76R_{t,s}$, respectively.
    (b) Surface average $\langle f_{exch}\rangle$, where the black solid, blue dashed and orange dotted curves correspond to $R_t^2\gamma_0^2\approx11.86$, $R_t^2\gamma_0^2\approx118.6$ and $R_t^2\gamma_0^2\approx1186$, respectively. The middle value is computed using the typical parameter values in table$~$\ref{parameters_multiscale_model}.}
    \label{surface_averages_f_mean_f_exch_aperture}
\end{figure}

The surface-averaged capillary and tissue pressures provide a compact summary of the parameter dependence of the proposed model.
Define the surface average of a function $f$ as
\begin{eqnarray}
    \langle f\rangle=\frac{1}{\vert\mathcal{S}\vert}\int_{\mathcal{D}}f({\bm z})\lambda^2({\bm z})d{\bm z},
\end{eqnarray}
where the surface area $\vert\mathcal{S}\vert$ is given by
\begin{eqnarray}
    \vert\mathcal{S}\vert=\int_{\mathcal{D}}\lambda^2({\bm z})d{\bm z}.
\end{eqnarray}
The surface averages of pressures are thus given by
\begin{eqnarray}
   \langle p_{cap}\rangle=-\frac{1}{\kappa_{cap}+\kappa_t}\sum^{N_{cap}}_{n=1}q_n\left(\langle f_{mean}(r_n)\rangle+\frac{\kappa_t}{\kappa_{cap}}\langle f_{exch}\rangle\right)+\overline{p_{mean}}+\alpha_r\Delta\pi, \\
   \langle p_t\rangle=-\frac{1}{\kappa_{cap}+\kappa_t}\sum^{N_{cap}}_{n=1}q_n\left(\langle f_{mean}(r_n)\rangle-\langle f_{exch}\rangle\right)+\overline{p_{mean}},
\end{eqnarray}
where $\langle f_{exch}\rangle$ is independent of the source position, while $\langle f_{mean}(r_n)\rangle$ depends on the source position $r_n$.
The surface averages of $f_{mean}$ and $f_{exch}$ are given by
\begin{eqnarray}
    \langle f_{mean}(r_n)\rangle=\frac{1}{\pi}\log \left(R_t\sqrt{\frac{1+\beta_{oc}}{1-\beta_{oc}}}\right)+\frac{\log\left(\frac{2}{1+(R^2_t-r^2_n)/(R^2_t+r^2_n)}\right)}{2\pi(1+\beta_{oc})},\\
    \langle f_{exch}\rangle=-\frac{1}{2\pi\gamma^2_0 R^2_t(1+\beta_{oc})},
\end{eqnarray}
and the derivations of these averages are provided in Appendix$~$\ref{SurfaceBoundaryAveragesGreensFunction}.

While $\langle f_{mean}(r_n)\rangle$ varies with the source position $r_n$ and aperture parameter $\beta_{oc}$, $\langle f_{exch}\rangle$ depends on the inverse screening length $\gamma_0$ and $\beta_{oc}$.
Figure$~$\ref{surface_averages_f_mean_f_exch_aperture} shows the dependence of the surface averages $\langle f_{mean}(r_n)\rangle$ and $\langle f_{exch}\rangle$ on $\beta_{oc}$ with fixed $R_t=1.26~\rm{cm}$.
In subfigure$~$(a), the black solid, blue dashed and orange dotted curves correspond to $r_n=0.36R_{t,s}$, $r_n=0.56R_{t,s}$ and $r_n=0.76R_{t,s}$, respectively.
The surface average $\langle f_{mean}(r_n)\rangle$ increases with both $\beta_{oc}$ and $r_n/R_{t,s}$.
In subfigure$~$(b), the black solid, blue dashed and orange dotted curves correspond to $R^2_t\gamma^2_0\approx11.86$, $R^2_t\gamma^2_0\approx118.6$ and $R^2_t\gamma^2_0\approx1186$, respectively.
The surface average $\langle f_{exch}\rangle$ is negative and increases toward zero as $\beta_{oc}$ increases, while its magnitude decreases as $R^2_t\gamma^2_0$ increases.
\begin{figure}
    \centering
    \includegraphics[width=0.76\textwidth]{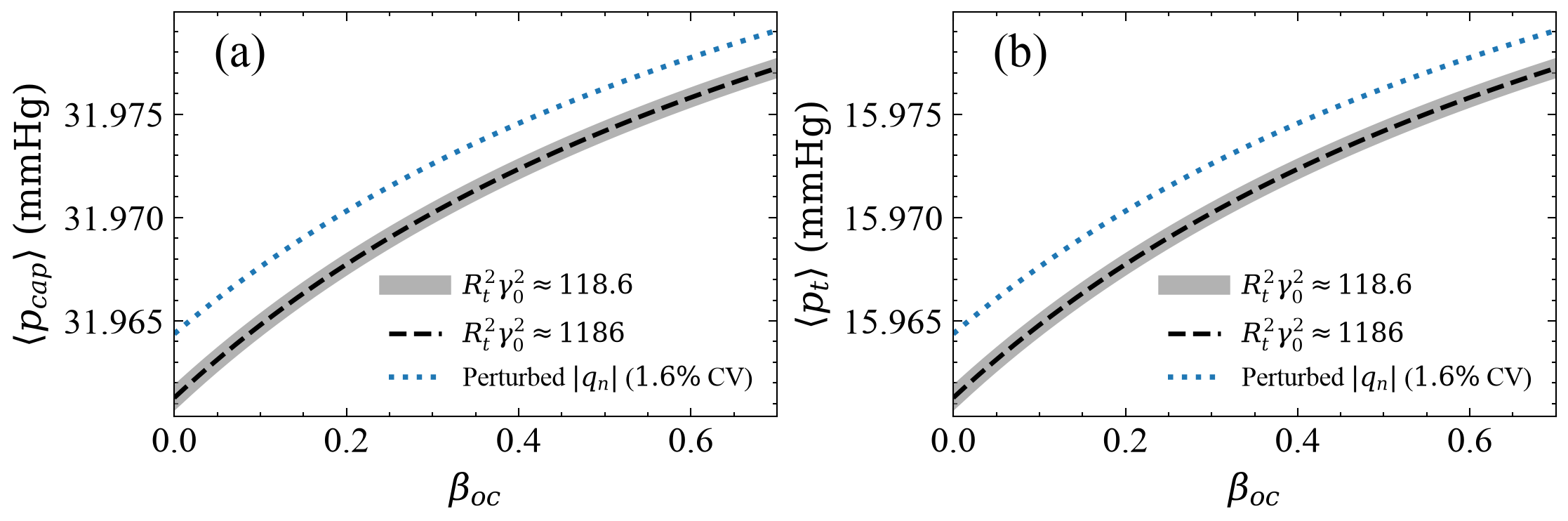}
    \caption{Surface-averaged pressures as functions of $\beta_{oc}$ for $R_t=1.26~\rm{cm}$.
    (a) Capillary pressure $\langle p_{cap}\rangle$.
    (b) Tissue pressure $\langle p_t\rangle$.
    The grey solid and black dashed curves correspond to $R_t^2\gamma_0^2\approx118.6$ and $R_t^2\gamma_0^2\approx1186$, obtained using the baseline and tenfold exchange coefficients, respectively, with fixed $\kappa_{cap}$ and $\kappa_t$.
    The hemispherical source locations in figure$~$\ref{spherical_fundus}(c) are used, with $q_{in,CRA}=16~\rm{\upmu L/min}$ distributed uniformly among $120$ arteriolar and $120$ venular sources, giving $q_n=\pm0.133333~\rm{\upmu L/min}$.
    The blue dotted curves correspond to a single nonuniform terminal-flow allocation at $R_t^2\gamma_0^2\approx118.6$, generated by applying independent lognormal noise with a target coefficient of variation of $1.6~\%$ to the arteriolar and venular terminal-flow magnitudes $\vert q_n\vert$.
    The arteriolar and venular flows are normalised separately so that their total magnitudes both remain $16~\rm{\upmu L/min}$.
    The boundary-averaged mean pressure is $\overline{p_{mean}}=16~\rm{mmHg}$.}
    \label{surface_averages_pressures_aperture}
\end{figure}

Figure$~$\ref{surface_averages_pressures_aperture}(a) and (b) show the dependence of $\langle p_{cap}\rangle$ and $\langle p_t\rangle$, respectively, on the aperture parameter $\beta_{oc}$.
The grey solid and black dashed curves show the results for uniform terminal flows with $R_t^2\gamma_0^2\approx118.6$ and $R_t^2\gamma_0^2\approx1186$, respectively, whereas the blue dotted curves show the results for lognormally perturbed terminal-flow magnitudes at $R_t^2\gamma_0^2\approx118.6$.
The grey solid and black dashed curves overlap because the contribution involving $\langle f_{exch}\rangle$ vanishes when the total arteriolar and venular flows are balanced.
Consequently, changing $R^2_t\gamma^2_0$ does not affect the surface-averaged pressures for balanced total arteriolar and venular source flows.
The blue dotted curves show that applying independent lognormal noise to $\vert q_n\vert$ slightly shifts both surface-averaged pressures but preserves their weak dependence on $\beta_{oc}$.
Because the arteriolar and venular flows are normalised separately, the compatibility condition $\sum^{N_{cap}}_{n=1}q_n=0$ remains satisfied, and the difference from the uniform-flow case arises from weighting the source-position-dependent $\langle f_{mean}(r_n)\rangle$ by nonuniform $q_n$.
The surface averages $\langle p_{cap}\rangle$ and $\langle p_t\rangle$ remain close to $32~\rm{mmHg}$ and $16~\rm{mmHg}$, respectively, and their difference remains equal to the prescribed osmotic pressure difference of $16~\rm{mmHg}$ with $\alpha_r=1$.
Both surface-averaged pressures increase slightly with $\beta_{oc}$, with changes of less than $0.02~\rm{mmHg}$ over the range shown.

\section{Numerical methods}
\label{NumericalMethods}
\subsection{Finite volume method for the 1D flow in arterioles and venules}
The 1D model is solved using the finite volume method (FVM).
For each vessel, the domain $0\leq s\leq l_v$ is discretised into $N_s+1$ spatial points $s_i=i\Delta s$, where $i=0,\dots,N_s$ and $\Delta s=l_v/N_s$.
The cells are $C_i=[s_{i-1/2},s_{i+1/2}]=[(i-1/2)\Delta s,(i+1/2)\Delta s]$, whose centres are $s_i$ with $1\leq i\leq N_s-1$, and the boundary points are thus $s_0=0$ and $s_{N_s}=l_v$.
Time is discretised into steps $t_n=n\Delta t$, where $0\leq n\leq N_t$.

The FVM is implemented for the variables at the interior points $s_i$ for each time step $t_{n+1}$, including the cross-sectional area $A(s_i,t_{n+1})$ and flow rate $q(s_i,t_{n+1})$, where $1\leq i\leq N_s-1$.
To simplify the notation, the governing equations of the 1D model are written in the conservative form:
\begin{eqnarray}
    \frac{\partial {\bm U}}{\partial t}+\frac{\partial {\bm F}}{\partial s}={\bm S},
    \label{conservative_form}
\end{eqnarray}
where the state vector ${\bm U}$, flux vector ${\bm F}$ and source vector ${\bm S}$ are given by
\begin{eqnarray}
    &&{\bm U}=(A,q)^T, \\
    &&{\bm F}=\left(q,\frac{4}{3}\frac{q^2}{A}+\frac{K}{\rho}\left(\frac{\beta_1}{\beta_1+1}\frac{A^{\beta_1+1}}{A^{\beta_1}_0}-\frac{\beta_2}{\beta_2+1}\frac{A^{\beta_2+1}}{A^{\beta_2}_0}\right)\right)^T, \\
    &&{\bm S}=\left(0,C_f\right)^T,
\end{eqnarray}
respectively.
Integrating equation$~$(\ref{conservative_form}) over the control volume $C_i\times[t_n,t_{n+1}]$ yields
\begin{eqnarray}
    \int^{t_{n+1}}_{t_n}\int_{C_i}{\bm S}(s,t)dsdt=\int_{C_i}{\bm U}(s,t)\vert^{t=t_{n+1}}_{t=t_n}ds+\int^{t_{n+1}}_{t_n}{\bm F}(s,t)\vert^{s=s_{i+1/2}}_{s=s_{i-1/2}}dt.
    \label{FVM_integral}
\end{eqnarray}
The following approximations are then used for the integrals in equation$~$(\ref{FVM_integral}):
\begin{eqnarray}
    \int^{s_{i+1/2}}_{s_{i-1/2}}{\bm U}(s,t_{n+1})ds\approx\Delta s{\bm U}^{n+1}_i,\\
    \int^{t_{n+1}}_{t_n}{\bm F}(s_{i+1/2},t)dt\approx\Delta t{\bm F}^{n+1/2}_{i+1/2},\\
    \int^{t_{n+1}}_{t_n}\int^{s_{i+1/2}}_{s_{i-1/2}}{\bm S}(s,t)dsdt\approx\frac{\Delta s\Delta t}{2}\left({\bm S}^{n+1/2}_{i+1/2}+{\bm S}^{n+1/2}_{i-1/2}\right),
\end{eqnarray}
where ${\bm U}^{n+1}_i={\bm U}(s_i,t_{n+1})$, and similar notations are used for flux and source vectors.
With these approximations, the FVM equation$~$(\ref{FVM_integral}) becomes
\begin{eqnarray}
    {\bm U}^{n+1}_{i}={\bm U}^n_i-\frac{\Delta t}{\Delta s}\left({\bm F}^{n+1/2}_{i+1/2}-{\bm F}^{n+1/2}_{i-1/2}\right)+\frac{\Delta t}{2}\left({\bm S}^{n+1/2}_{i+1/2}+{\bm S}^{n+1/2}_{i-1/2}\right),
    \label{fvm_flow_rate_area}
\end{eqnarray}
where ${\bm F}^{n+1/2}_{i+1/2}$, ${\bm F}^{n+1/2}_{i-1/2}$, ${\bm S}^{n+1/2}_{i+1/2}$ and ${\bm S}^{n+1/2}_{i-1/2}$ are the interface variables, depending on the unknown variables ${\bm U}^{n+1/2}_{i-1/2}$ and ${\bm U}^{n+1/2}_{i+1/2}$.
These two unknown variables are determined by the Lax-Wendroff scheme \citep{kolachalama_predictive_2007}:
\begin{eqnarray}
    {\bm U}^{n+1/2}_{i-1/2}=\frac{{\bm U}^n_i+{\bm U}^n_{i-1}}{2}+\frac{\Delta t}{2}\left(-\frac{{\bm F}^n_i-{\bm F}^n_{i-1}}{\Delta s}+\frac{{\bm S}^n_i+{\bm S}^n_{i-1}}{2}\right), \label{lax_wendroff_left}\\
    {\bm U}^{n+1/2}_{i+1/2}=\frac{{\bm U}^n_{i+1}+{\bm U}^n_i}{2}+\frac{\Delta t}{2}\left(-\frac{{\bm F}^n_{i+1}-{\bm F}^n_i}{\Delta s}+\frac{{\bm S}^n_{i+1}+{\bm S}^n_i}{2}\right).
    \label{lax_wendroff_right}
\end{eqnarray}
For steady flow, state vector ${\bm U}$, flux vector ${\bm F}$ and source vector ${\bm S}$ are independent of the time $t$, and the vectors at the interface $s_{i+1/2}$ are thus denoted as ${\bm U}_{i+1/2}$, ${\bm F}_{i+1/2}$ and ${\bm S}_{i+1/2}$.
The equation$~$(\ref{fvm_flow_rate_area}) becomes
\begin{eqnarray}
    {\bm F}_{i+1/2}-{\bm F}_{i-1/2}=\frac{\Delta s}{2}({\bm S}_{i+1/2}+{\bm S}_{i-1/2}).
\end{eqnarray}
The first component of this equation yields
\begin{eqnarray}
    q_{i+1/2}=q_{i-1/2},
\end{eqnarray}
which implies that the flow rate is constant along each vessel segment.
The second component is solved implicitly for the cross-sectional area at cell interfaces $s_{i-1/2}$ and $s_{i+1/2}$, with the boundary and multiscale coupling conditions described in the following sections.
\subsection{The implementation of boundary conditions for the vasculature and multiscale coupling condition}
The following implementation is formulated for pulsatile flow.
For steady-flow computations, the same boundary and coupling conditions are used, with all discrete time-difference terms set to zero.
\subsubsection{Boundary conditions for the vasculature}
At the CRA inlet, the area $A^{n+1}_{CRA,0}$ is determined by solving the equation
\begin{eqnarray}
   p_{in,CRA}(t_{n+1})-p_{ext}=K\left[\left(\frac{A^{n+1}_{CRA,0}}{A_0}\right)^{\beta_1}-\left(\frac{A^{n+1}_{CRA,0}}{A_0}\right)^{\beta_2}\right],
   \label{inlet_boundary_area_CRA}
\end{eqnarray}
where $p_{in,CRA}(t_{n+1})$ is the prescribed CRA inlet pressure at time $t_{n+1}$, while the flow rate $q^{n+1}_{CRA,0}$ is computed by the equation
\begin{eqnarray}
    q^{n+1}_{CRA,0}=q^{n+1}_{CRA,1}+\frac{\Delta s_{CRA}}{2\Delta t}\left[(A^{n+1}_{CRA,0}+A^{n+1}_{CRA,1})-(A^n_{CRA,0}+A^n_{CRA,1})\right],
    \label{inlet_boundary_flow_rate_CRA_inlet}
\end{eqnarray}
which is obtained by discretising the mass-conservation equation \citep{he_one-dimensional_2004}:
\begin{eqnarray}
    \frac{(A^{n+1}_{CRA,0}+A^{n+1}_{CRA,1})-(A^n_{CRA,0}+A^n_{CRA,1})}{2\Delta t}+\frac{q^{n+1}_{CRA,1}-q^{n+1}_{CRA,0}}{\Delta s_{CRA}}=0,
\end{eqnarray}
The area at the CRV outlet $A^{n+1}_{CRV,0}$ is similarly computed from the equation
\begin{eqnarray}
    p_{out,CRV}-p_{ext}=K\left[\left(\frac{A^{n+1}_{CRV,0}}{A_0}\right)^{\beta_1}-\left(\frac{A^{n+1}_{CRV,0}}{A_0}\right)^{\beta_2}\right],
    \label{outlet_boundary_area_CRV}
\end{eqnarray}
where $p_{out,CRV}$ is the CRV outlet pressure, while the flow rate $q^{n+1}_{CRV,0}$ is determined by
\begin{eqnarray}
    q^{n+1}_{CRV,0}=q^{n+1}_{CRV,1},
    \label{outlet_boundary_flow_rate_CRV}
\end{eqnarray}
implied by the constant pressure condition at the CRV outlet.
The unknown variables at the bifurcation boundary are
\begin{eqnarray}
    {\bm x}_b=[A^{n+1}_{M,N_s},q^{n+1}_{M,N_s},A^{n+1}_{d_1,0},q^{n+1}_{d_1,0},A^{n+1}_{d_2,0},q^{n+1}_{d_2,0}]^T.
\end{eqnarray}
The mass-conservation equations in the system$~$(\ref{bifurct_boundary_conditions}) are discretised, which yields the following system at $t=t_{n+1}$ \citep{he_one-dimensional_2004,koppl_dimension_2023}:
\begin{eqnarray}
    {\bm Y}({\bm x}_b)=[Y_1({\bm x}_b),Y_2({\bm x}_b),\dots,Y_6({\bm x}_b)]^T={\bm 0},
    \label{bifurcation_boundary_area_flow_rate}
\end{eqnarray}
where
\begin{eqnarray}
    &Y_1=q^{n+1}_{M,N_s}-q^{n+1}_{d_1,0}-q^{n+1}_{d_2,0}, \\
    &Y_2=p^{n+1}_{M,N_s}(A^{n+1}_{M,N_s})-p^{n+1}_{d_1,0}(A^{n+1}_{d_1,0}), \\
    &Y_3=p^{n+1}_{M,N_s}(A^{n+1}_{M,N_s})-p^{n+1}_{d_2,0}(A^{n+1}_{d_2,0}), \\
    &Y_4=\frac{(A^{n+1}_{M,N_s}+A^{n+1}_{M,N_s-1})-(A^n_{M,N_s}+A^n_{M,N_s-1})}{2\Delta t}+\frac{q^{n+1}_{M,N_s}-q^{n+1}_{M,N_s-1}}{\Delta s_M}, \\
    &Y_5=\frac{(A^{n+1}_{d_1,0}+A^{n+1}_{d_1,1})-(A^n_{d_1,0}+A^n_{d_1,1})}{2\Delta t}+\frac{q^{n+1}_{d_1,1}-q^{n+1}_{d_1,0}}{\Delta s_{d_1}},\\
    &Y_6=\frac{(A^{n+1}_{d_2,0}+A^{n+1}_{d_2,1})-(A^n_{d_2,0}+A^n_{d_2,1})}{2\Delta t}+\frac{q^{n+1}_{d_2,1}-q^{n+1}_{d_2,0}}{\Delta s_{d_2}}.
\end{eqnarray}
The pressures $p^{n+1}_{M,N_s}(A^{n+1}_{M,N_s})$, $p^{n+1}_{d_1,0}(A^{n+1}_{d_1,0})$ and $p^{n+1}_{d_2,0}(A^{n+1}_{d_2,0})$ are described by the elastic model$~$(\ref{elastic_model}), while the interior variables of each vessel, such as $A^{n+1}_{M,N_s-1}$ and $q^{n+1}_{M,N_s-1}$, are obtained by integration along each vessel.
This system is solved at each time step to determine the bifurcation boundary variables.

For steady flow, all discrete time derivative terms at the inlet, bifurcation boundary and outlet are set to zero, such as those in equations$~$(\ref{inlet_boundary_flow_rate_CRA_inlet}) and (\ref{outlet_boundary_flow_rate_CRV}).
The areas are then solved implicitly together with the FVM equations and multiscale coupling conditions whose implementation is described in the following section.
\subsubsection{Multiscale coupling condition}
We next establish pressure-flow relations between the capillary bed and the termini of the vasculature.
At time $t=t_{n+1}$, we define the capillary pressure vector ${\bm p}^{n+1}_{cap}$ and source flow vector ${\bm q}^{n+1}$ as
\begin{eqnarray}
    &&{\bm p}^{n+1}_{cap}=[p^{n+1}_{cap}({\bm x}_1),\dots,p^{n+1}_{cap}({\bm x}_{N_{cap}})]^T, \\
    &&{\bm q}^{n+1}=[q^{n+1}({\bm x}_1),\dots,q^{n+1}({\bm x}_{N_{cap}})]^T,
\end{eqnarray}
where $p^{n+1}_{cap}({\bm x}_i)$ and $q^{n+1}({\bm x}_i)$ are the pressure and the flow rate at ${\bm x}={\bm x}_i$, respectively.
Using the analytic solution for capillary pressure, ${\bm p}^{n+1}_{cap}$ can be written in terms of the source flow rates as
\begin{eqnarray}
    {\bm p}^{n+1}_{cap}={\bm M}^{cap}{\bm q}^{n+1}+(\overline{p_{mean}}+\alpha_r\Delta\pi){\bm 1},
    \label{pressure_flow_capillary_source}
\end{eqnarray}
where ${\bm M}^{cap}$ is the resistance matrix with entries
\begin{eqnarray}
    M^{cap}_{iw}=-G_{cap}(r_i,\theta_i;r_w,\theta_w)/(\kappa_{cap}+\kappa_t).
\end{eqnarray}
The pressure at source points should be equal to the pressure at the termini of arteriolar/venular trees, which yields
\begin{eqnarray}
    p^{n+1}_{cap}({\bm x}_i)=p^{n+1}_{i,N_s}, \quad i=1,\dots,N_{cap},
    \label{superficial_termini_sources}
\end{eqnarray}
where $p^{n+1}_{i,N_s}$ is the pressure at the terminus of vessel $i$.
This equation couples the capillary bed and arterioles/venules.
In order to write the pressure-flow relation in a compact form, we define the terminal pressure vector
\begin{eqnarray}
    {\bm p}^{n+1}=[p^{n+1}_{1,N_s},\dots,p^{n+1}_{N_{cap},N_s}]^T.
\end{eqnarray}
Then the pressure-flow relation is given by
\begin{eqnarray}
    {\bm p}^{n+1}={\bm M}^{cap}{\bm q}^{n+1}+(\overline{p_{mean}}+\alpha_r\Delta\pi){\bm 1}.
    \label{capillary_poiseuille_pressure_flow}
\end{eqnarray}

Solving equation$~$(\ref{capillary_poiseuille_pressure_flow}) for the flow rate vector ${\bm q}^{n+1}$ yields
\begin{eqnarray}
    {\bm q}^{n+1}=({\bm M}^{cap})^{-1}{\bm p}^{n+1}-(\overline{p_{mean}}+\alpha_r\Delta\pi)({\bm M}^{cap})^{-1}{\bm 1}.
\end{eqnarray}
Then, imposing conservation of mass leads to
\begin{eqnarray*}
    0={\bm 1}^T({\bm M}^{cap})^{-1}{\bm p}^{n+1}-(\overline{p_{mean}}+\alpha_r\Delta\pi){\bm 1}^T({\bm M}^{cap})^{-1}{\bm 1},
\end{eqnarray*}
which implies
\begin{eqnarray}
    \overline{p_{mean}}=\frac{1}{s_{cap}}{\bm 1}^T({\bm M}^{cap})^{-1}{\bm p}^{n+1}-\alpha_r\Delta\pi,
    \label{boundary_average}
\end{eqnarray}
with $s_{cap}={\bm 1}^T({\bm M}^{cap})^{-1}{\bm 1}$.
Substituting the expression for $\overline{p_{mean}}$ into the equation$~$(\ref{capillary_poiseuille_pressure_flow}), we obtain
\begin{eqnarray}
    {\bm p}^{n+1}={\bm M}^{cap}{\bm q}^{n+1}+\frac{1}{s_{cap}}\left[{\bm 1}^T({\bm M}^{cap})^{-1}{\bm p}^{n+1}\right]{\bm 1}.
    \label{pressure_flow_source}
\end{eqnarray}
This equation couples the flow rates and pressures of all termini in the vasculature.
The mass-conservation equation is discretised as that for CRA inlet and bifurcation boundary such that for each terminus,
\begin{eqnarray}
    q^{n+1}_{k,N_s}=q^{n+1}_{k,N_s-1}-\frac{\Delta s_k}{2\Delta t}\left[(A^{n+1}_{k,N_s-1}+A^{n+1}_{k,N_s})-(A^n_{k,N_s}+A^n_{k,N_s-1})\right],
\end{eqnarray}
where $k=1,\dots,N_{cap}$.
This enables us to express ${\bm q}^{n+1}$ as
\begin{eqnarray}
    {\bm q}^{n+1}=-{\bm B}\left[({\bm A}^{n+1}_{N_s}+{\bm A}^{n+1}_{N_s-1})-({\bm A}^n_{N_s}+{\bm A}^n_{N_s-1})\right]+{\bm f}^{n+1},
    \label{capillary_flow_rate_vector}
\end{eqnarray}
where area vectors are given by
\begin{eqnarray}
    {\bm A}^{n+1}_{N_s}&=&[A^{n+1}_{1,N_s},\dots,A^{n+1}_{N_{cap},N_s}]^T, \\
    {\bm A}^{n+1}_{N_s-1}&=&[A^{n+1}_{1,N_s-1},\dots,A^{n+1}_{N_{cap},N_s-1}]^T, \\
    {\bm A}^n_{N_s}&=&[A^n_{1,N_s},\dots,A^n_{N_{cap},N_s}]^T, \\
    {\bm A}^n_{N_s-1}&=&[A^n_{1,N_s-1},\dots,A^n_{N_{cap},N_s-1}]^T,
\end{eqnarray}
the matrix ${\bm B}$ is given by
\begin{eqnarray}
    {\bm B}=
    \begin{bmatrix}
        \Delta s_1/(2\Delta t) & &  \\
        & \ddots &\\
        & & \Delta s_{N_{cap}}/(2\Delta t)
    \end{bmatrix},
\end{eqnarray}
and ${\bm f}^{n+1}$ is given by
\begin{eqnarray}
    {\bm f}^{n+1}=[q^{n+1}_{1,N_s-1},\dots,q^{n+1}_{N_{cap},N_s-1}]^T.
\end{eqnarray}
The numerical implementation of the coupling condition is thus given by
\begin{eqnarray}
    \left\{
    \begin{aligned}
        &{\bm p}^{n+1}={\bm M}^{cap}{\bm q}^{n+1}+\frac{1}{s_{cap}}\left[{\bm 1}^T({\bm M}^{cap})^{-1}{\bm p}^{n+1}\right]{\bm 1}, \\
        &{\bm q}^{n+1}=-{\bm B}\left[({\bm A}^{n+1}_{N_s}+{\bm A}^{n+1}_{N_s-1})-({\bm A}^n_{N_s}+{\bm A}^n_{N_s-1})\right]+{\bm f}^{n+1} \\
        &p^{n+1}_{k,N_s}-p_{ext}=K\left[\left(\frac{A^{n+1}_{k,N_s}}{A_0}\right)^{\beta_1}-\left(\frac{A^{n+1}_{k,N_s}}{A_0}\right)^{\beta_2}\right],\quad k=1,\dots,N_{cap}.
    \end{aligned}\right.
    \label{capillary_tissue_coupling_condition}
\end{eqnarray}
For steady flow, the time-difference term is equal to zero, which yields $q_{k,N_s}=q_{k,N_s-1}$ for each terminal vessel.
Consequently, the second equation becomes ${\bm q}={\bm f}$.
\subsubsection{Numerical treatments of the infinite summation and the singularity in Green's functions}
The Green's functions $f_{mean}$ and $f_{exch}$ are singular at ${\bm x}={\bm x}_n$.
For the Green's function of mean pressure, the singular part is
\begin{eqnarray}
    f^s_{mean}=\frac{1}{4\pi}\log\vert {\bm z}-{\bm z}_n\vert^2,
\end{eqnarray}
while the singular part of exchange Green's function is
\begin{eqnarray}
    f^s_{exch}=\frac{1}{4\sin(\pi\nu)}P^0_\nu\left(-w_nw-\sqrt{1-w^2_n}\sqrt{1-w^2}\cos(\theta-\theta_n)\right),
\end{eqnarray}
These singularities are regularised by replacing the pointwise values of the Green's functions at the sources with their averages over small spherical vessel patches centred at the source points.
Since the vessel patches are physiological objects on the retinal surface, the averaging is performed on the surface of the spherical cap rather than on the projected domain.

Let $\epsilon_n({\bm x})$ denote the geodesic angle between a field point ${\bm x}$ and the source point ${\bm x}_n$.
We introduce spherical polar coordinates $(\epsilon_n,\varphi)$, where $\epsilon_n$ is the polar angle and $\varphi$ is the azimuthal angle.
In the coordinates $(w,\theta)$, the angle satisfies
\begin{eqnarray}
    \cos(\epsilon_n)=ww_n+\sqrt{1-w^2}\sqrt{1-w^2_n}\cos(\theta-\theta_n).
    \label{geodesic_angle}
\end{eqnarray}
We define the spherical vessel patch by
\begin{eqnarray}
    \Omega^{\mathcal{S}}_n=\left\{{\bm x} \in \mathcal{S} \vert 0\leq \epsilon_n({\bm x})\leq \epsilon_{v,n}\right\},
\end{eqnarray}
where $\epsilon_{v,n}$ is the angular radius of the vessel patch and the corresponding geodesic radius is
\begin{eqnarray}
    \ell_{v,n}=R_t\epsilon_{v,n}.
\end{eqnarray}
We assume that each vessel patch $\Omega^{\mathcal{S}}_n$ lies entirely within the spherical-cap domain.
The area of the spherical vessel patch is
\begin{eqnarray}
    A^{\mathcal{S}}_n=2\pi R^2_t(1-\cos\epsilon_{v,n}).
\end{eqnarray}
The averages of the singular parts are given by
\begin{eqnarray}
    \langle f^s_{mean}\rangle_{\Omega^{\mathcal{S}}_n}=\frac{1}{A^{\mathcal{S}}_n}\int_{\Omega^{\mathcal{S}}_n}\frac{1}{4\pi}\log\vert {\bm z}-{\bm z}_n\vert^2dA, \\
    \langle f^s_{exch}\rangle_{\Omega^{\mathcal{S}}_n}=\frac{1}{A^{\mathcal{S}}_n}\int_{\Omega^{\mathcal{S}}_n}\frac{1}{4\sin(\pi\nu)}P^0_\nu\left(-w_nw-\sqrt{1-w^2_n}\sqrt{1-w^2}\cos(\theta-\theta_n)\right)dA,
\end{eqnarray}
where $dA$ is the area element in the spherical polar coordinates centred at ${\bm x}_n$ given by
\begin{eqnarray}
    dA=R^2_t\sin\epsilon_nd\epsilon_nd\varphi.
\end{eqnarray}
The evaluation of these averages is provided in Appendix$~$\ref{AveragesSingularitiesGreensFunction}.

For computation, the infinite series in $f_{exch}$ is truncated.
We denote the series in $f_{exch}$ as
\begin{eqnarray}
    s=\sum^\infty_{m=0}c_m(\nu,\beta_{oc})P^m_\nu(w_n)P^m_\nu(w)\cos(m(\theta-\theta_n)),
\end{eqnarray}
and its truncation is given by
\begin{eqnarray}
    s_M=\sum^M_{m=0}c_m(\nu,\beta_{oc})P^m_\nu(w_n)P^m_\nu(w)\cos(m(\theta-\theta_n)),
\end{eqnarray}
where $M$ is the truncation order.
Define the truncation error as
\begin{eqnarray}
    \mathcal{E}_M=\sum^\infty_{m=M+1}c_m(\nu,\beta_{oc})P^m_\nu(w_n)P^m_\nu(w)\cos(m(\theta-\theta_n)).
\end{eqnarray}
We next estimate the truncation error using the following ratio between terms of orders $m$ and $m+1$:
\begin{eqnarray}
    \mathcal{E}_{m,m+1}=\frac{T_{m+1}}{T_m},
\end{eqnarray}
where
\begin{eqnarray}
    T_m=\left\vert\frac{\Gamma(\nu-m+1)}{\Gamma(\nu+ m+1)}\frac{1}{2\sin(\pi\nu)}\frac{P^{m\prime}_\nu(\beta_{oc})}{P^{m\prime}_\nu(-\beta_{oc})}P^m_\nu(w_n)P^m_\nu(w)\right\vert.
\end{eqnarray}
For fixed degree $\nu$ and fixed $w\in(-\beta_{oc},1)$, the large-order asymptotic expansions of the associated Legendre functions imply \citep[\S14.15,\S14.10]{olver_nist_2010-1}
\begin{eqnarray}
    \vert P^m_\nu(w)\vert \simeq \frac{\vert\sin(\pi\nu)\vert}{\pi}\Gamma(m)\left(\frac{1-w}{1+w}\right)^{m/2}, \quad P^{m\prime}_\nu(w)\simeq -\frac{m}{1-w^2}P^m_\nu(w),
\end{eqnarray}
while the large-order expansion and reflection formula for the gamma function imply \citep[\S5.11,\S5.5]{olver_nist_2010-1}
\begin{eqnarray}
    \left\vert \frac{\Gamma(\nu-m+1)}{\Gamma(\nu+m+1)}\right\vert \simeq\frac{\pi}{\vert \sin(\pi \nu)\vert m\Gamma(m)^2}.
\end{eqnarray}
Therefore,
\begin{eqnarray}
    T_m\simeq \frac{1}{2\pi m}\left(\frac{1-\beta_{oc}}{1+\beta_{oc}}\left(\frac{1-w}{1+w}\right)^{1/2}\left(\frac{1-w_n}{1+w_n}\right)^{1/2}\right)^m,
    \label{radial_component_relative_ratio}
\end{eqnarray}
and hence the relative ratio satisfies
\begin{eqnarray}
    \mathcal{E}_{m,m+1}\simeq \frac{m}{m+1}\frac{1-\beta_{oc}}{1+\beta_{oc}}\left(\frac{1-w}{1+w}\right)^{1/2}\left(\frac{1-w_n}{1+w_n}\right)^{1/2},
    \label{estimate_relative_ratio}
\end{eqnarray}

\begin{figure}
    \centering
    \includegraphics[width=\textwidth]{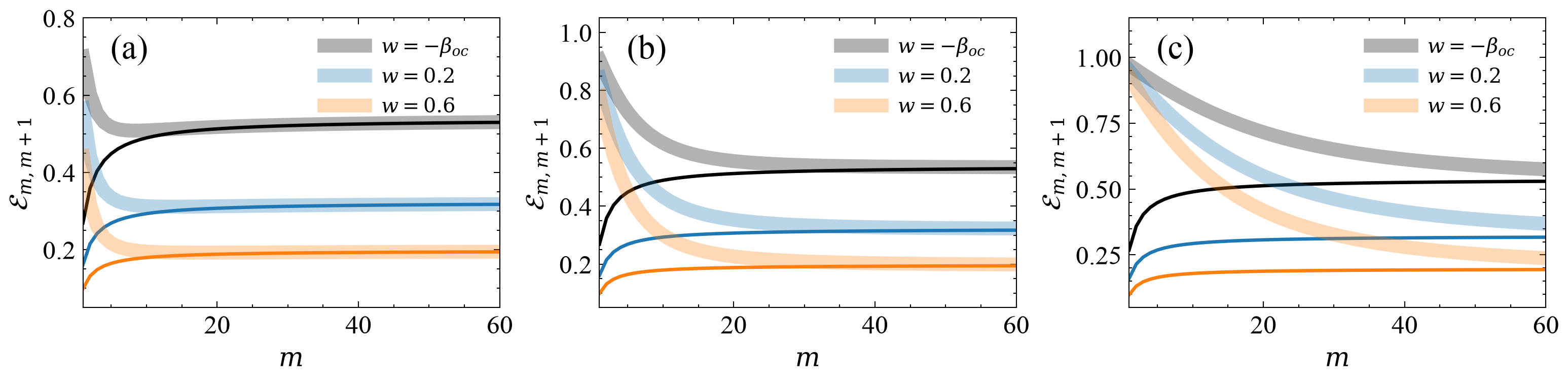}
    \caption{The relative ratio for the spherical exchange Green's function. Panels (a), (b) and (c) correspond to $R^2_t\gamma^2_0\approx11.86$, $118.6$ and $1186$, respectively. The black, blue and orange curves show $\mathcal{E}_{m,m+1}$ for $w=-\beta_{oc}$, $w=0.2$ and $w=0.6$, with $w_n=0.3$ and $\beta_{oc}=0.3$. The wide transparent curves denote the exact values, while the thin curves denote the asymptotic estimates.}
    \label{relative_ratio_spherical_exchange}
\end{figure}

Figure$~$\ref{relative_ratio_spherical_exchange} compares the exact value of the relative ratio with the asymptotic estimate$~$(\ref{estimate_relative_ratio}), where the wide transparent and the thin curves denote the exact values and the asymptotic estimates, respectively.
For all three values of $R^2_t\gamma^2_0$, the asymptotic estimate closely follows the exact value for all $m$.
The relative ratio is largest near the aperture boundary $w=-\beta_{oc}$, and decreases for larger $w$, which suggests faster decay of $\mathcal{E}_{m,m+1}$ away from the aperture boundary.
The agreement improves as $m$ increases, which indicates that the asymptotic expression provides an accurate estimate of the decay ratio between terms of orders $m$ and $m+1$.

Using the triangle inequality and the estimate for the relative ratio$~$(\ref{estimate_relative_ratio}), we estimate an upper bound on the truncation error $\mathcal{E}_M$.
For fixed $\nu$, $\beta_{oc}$ and interior source and field positions, as $M\to\infty$,
\begin{eqnarray}
    \vert \mathcal{E}_M\vert \leq \sum^\infty_{m=M+1}T_m\simeq T_{M+1}\sum^\infty_{m=0} \left(\frac{1-\beta_{oc}}{1+\beta_{oc}}\left(\frac{1-w}{1+w}\right)^{1/2}\left(\frac{1-w_n}{1+w_n}\right)^{1/2}\right)^m.
\end{eqnarray}
Using equation$~$(\ref{radial_component_relative_ratio}) and evaluating the geometric series gives the following asymptotic estimate of this upper bound:
\begin{eqnarray}
    \sum^\infty_{m=M+1}T_m \simeq \frac{1}{2\pi(M+1)}\frac{\left(\frac{1-\beta_{oc}}{1+\beta_{oc}}\left(\frac{1-w}{1+w}\right)^{1/2}\left(\frac{1-w_n}{1+w_n}\right)^{1/2}\right)^{M+1}}{1-\frac{1-\beta_{oc}}{1+\beta_{oc}}\left(\frac{1-w}{1+w}\right)^{1/2}\left(\frac{1-w_n}{1+w_n}\right)^{1/2}}.
\end{eqnarray}
Since the function $((1-w)/(1+w))^{1/2}$ is monotonically decreasing in $w$, for interior source and field points satisfying $-\beta_{oc}<w_n<1$ and $-\beta_{oc}<w<1$, we obtain
\begin{eqnarray}
    \frac{1-\beta_{oc}}{1+\beta_{oc}}\left(\frac{1-w}{1+w}\right)^{1/2}\left(\frac{1-w_n}{1+w_n}\right)^{1/2}<\frac{1-\beta_{oc}}{1+\beta_{oc}}\frac{1+\beta_{oc}}{1-\beta_{oc}}=1.
\end{eqnarray}
Thus, for source and field points inside the aperture boundary, the upper bound on the truncation error decays geometrically with $M$, up to an algebraic factor $1/(M+1)$.
\subsection{Summary of the numerical methods}
In summary, the numerical solution of the multiscale model is obtained as follows.
The 1D model for arteriolar and venular trees is solved using the FVM, with the Lax-Wendroff scheme used for time integration.
The multiscale coupling between the vascular trees and the capillary-tissue system is imposed through the resistance matrix derived from the analytic solution on the spherical-cap domain, while the singularities in the Green's functions are regularised by averaging over spherical vessel patches.
The resulting system is solved implicitly at each time step.
Steady and pulsatile flows are handled within the same framework.
\begin{table}
    \centering
    \resizebox{0.52\textwidth}{!}{%
    \begin{tabular}{l l l}
         Parameter& Value& References\\
         \hline
         \textit{Viscosity model}\\
         $\eta_p$ & $1.2~\rm{cP}$& \cite{pries_resistance_1994} \\
         & & \cite{secomb_blood_2013} \\
         & & \cite{bappoo_viscosity_2017} \\
         \hline
         \textit{Elastic model}\\
         $E$& $10^6~\rm{g/cm/s^2}$& \cite{julien_one-dimensional_2023}\\
         & & \cite{ferrara_biomechanical_2021} \\
         $\rm{WLR}$& $0.16$ (arteriole)& \cite{arichika_effects_2015} \\
          & $0.12$ (venule) & \cite{ertop_evaluation_2023} \\
         IOP & $16~\rm{mmHg}$ & \cite{guidoboni_intraocular_2014} \\
         \hline
         \textit{Vasculature model} \\
         CRA diameter & $200~\rm{\upmu m}$& \cite{dorner_calculation_2002} \\
         & & \cite{brown_physics-informed_2024} \\
         CRV diameter & $180~\rm{\upmu m}$& \cite{riva_blood_1985}\\
         Minimum angle & $\pi/3$& \cite{hernandez_linking_2024} \\
         $\beta$ & $2.3$ & \cite{hernandez_linking_2024} \\
         $N_{term}$ & $60$ & Estimated \\
         \hline
         \textit{Capillary-tissue coupled system}\\
         $R_t$& $1.26~\rm{cm}$& \cite{causin_blood_2016} \\
         $k_{cap}$ & $0.03~\rm{\upmu m}^2$ & \cite{dziubek_effect_2016}\\
         $k_t$ & $0.0002~\rm{\upmu m}^2$ & \cite{ruffini_mathematical_2024} \\
         & & \cite{antcliff_hydraulic_2001} \\
         & & \cite{fatt_flow_1971} \\
         $\mu_{cap}$ & computed by the viscosity model& \cite{pries_resistance_1994} \\
         $\mu_t$ & $0.7~\rm{cP}$& Estimated\\
         $\alpha_{exch}$ & $0.0002~\rm{\upmu m \cdot s/g}$ & \cite{bradbury_blood-brain_1990} \\
         $\alpha_r$ & $1$ & \cite{truskey_transport_2009} \\
         $\Delta\pi$ & $16~\rm{mmHg}$ & \cite{morissette_colloid_1977} \\
         \hline
         \textit{Boundary conditions for the vasculature}\\
         $p_{in,CRA}$ (steady) &$62~\rm{mmHg}$ & \cite{guidoboni_intraocular_2014} \\
          $p_{in,CRA}$ (pulsatile) &$45-77~\rm{mmHg}$ & \cite{rebhan_computational_2019} \\
          & & \cite{guidoboni_intraocular_2014} \\
         $p_{out,CRV}$ & $16~\rm{mmHg}$ & \cite{rebhan_computational_2019} \\
         & & \cite{causin_blood_2016} \\
         & & \cite{guidoboni_effect_2014}
    \end{tabular}%
    }
    \caption{Parameter values for the computation of the multiscale model.}
    \label{parameters_multiscale_model}
\end{table}
\section{Validation of the multiscale model}
\label{ValidationMultiscaleModel}
In this section, we assess the numerical reliability and physiological relevance of the multiscale model.
We first examine convergence with respect to the truncation order and consistency between the pulsatile and steady schemes for the 1D model.
To determine whether the model produces physiologically reasonable values, we then compute the steady flow under physiological boundary conditions and compare the predicted flow rates with experimental data.
We also compare the spherical-cap analytic solution with the circular-disc analytic solution for the capillary-tissue system, using the same spherical-cap vasculature shown in figure$~$\ref{spherical_fundus}(c).
Finally, pulsatile flow is computed under a prescribed pulsatile CRA inlet pressure.
The CRV outlet pressure is set to $p_{out,CRV}=16~\rm{mmHg}$ in all simulations, while the remaining parameters are summarised in table$~$\ref{parameters_multiscale_model}.
\begin{figure}
    \centering
    \includegraphics[width=0.76\textwidth]{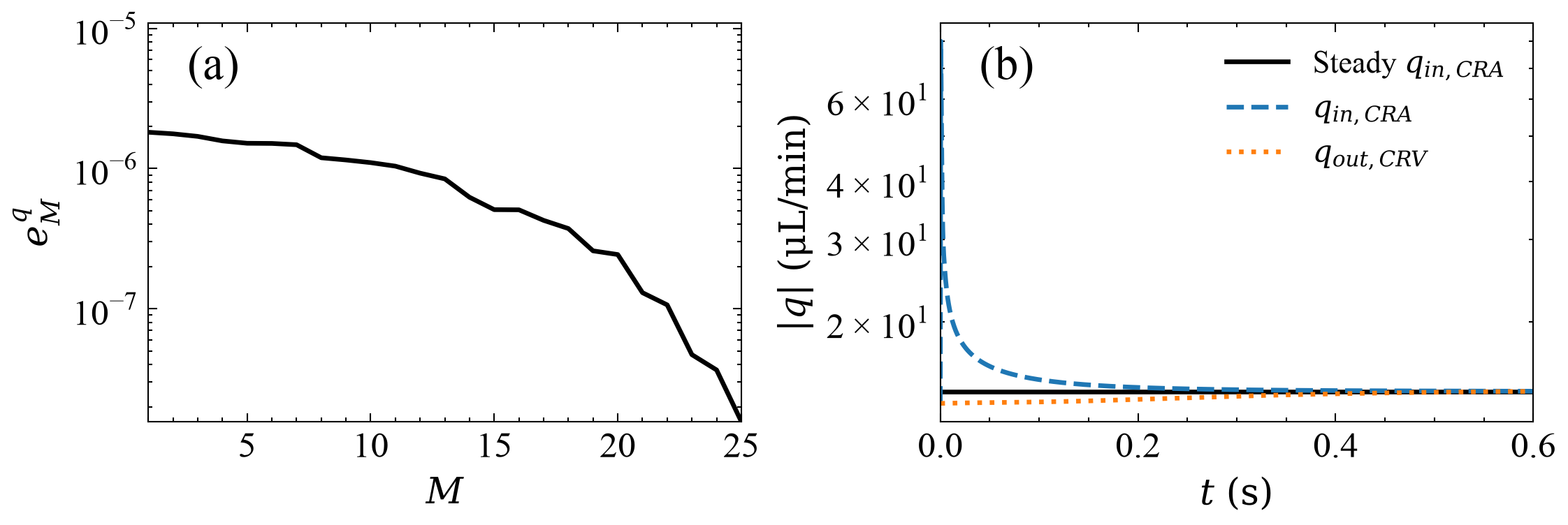}
    \caption{Numerical convergence and consistency of the steady and pulsatile schemes for the 1D model. (a) Relative error in the CRA inlet flow, given by equation$~$(\ref{relative_error_q_in_CRA}), as a function of the truncation order $M$. (b) CRA inlet flow rate (blue dashed curve) and CRV outlet flow rate (orange dotted curve) computed by the pulsatile scheme under a constant CRA inlet pressure. The black solid line denotes the CRA inlet flow computed by the steady scheme.}
    \label{truncation_CRA_inlet_flow}
\end{figure}
\subsection{Numerical convergence and consistency of pulsatile and steady schemes for the 1D model}
Since the exchange Green's function is represented by an infinite series, the truncation determines the accuracy of the analytic resistor formulation.
Figure$~$\ref{truncation_CRA_inlet_flow}(a) shows the relative error $e^q_M$ in the CRA inlet flow rate as a function of the truncation order $M$, where the relative error is given by
\begin{eqnarray}
    e^q_M=\frac{\vert \hat{q}_{in,CRA}(M)-\hat{q}_{in,CRA}(M_{max})\vert}{\vert \hat{q}_{in,CRA}(M_{max})\vert},
    \label{relative_error_q_in_CRA}
\end{eqnarray}
where $\hat{q}_{in,CRA}(M)$ is the CRA inlet flow rate computed using truncation order $M$, with $M_{max}=26$.
The error decreases from its maximum value of $1.82\times 10^{-6}$ at $M=1$ to its minimum value of $1.57\times 10^{-8}$ at $M=25$.
This demonstrates convergence of the computed CRA inlet flow rate with truncation order $M$, and thus a small truncation order $M$ is sufficient to obtain accurate results.
In the following computations under constant CRA inlet pressure, the truncation order $M$ is set to $M=6$.
\begin{figure}
    \centering
    \includegraphics[width=0.76\textwidth]{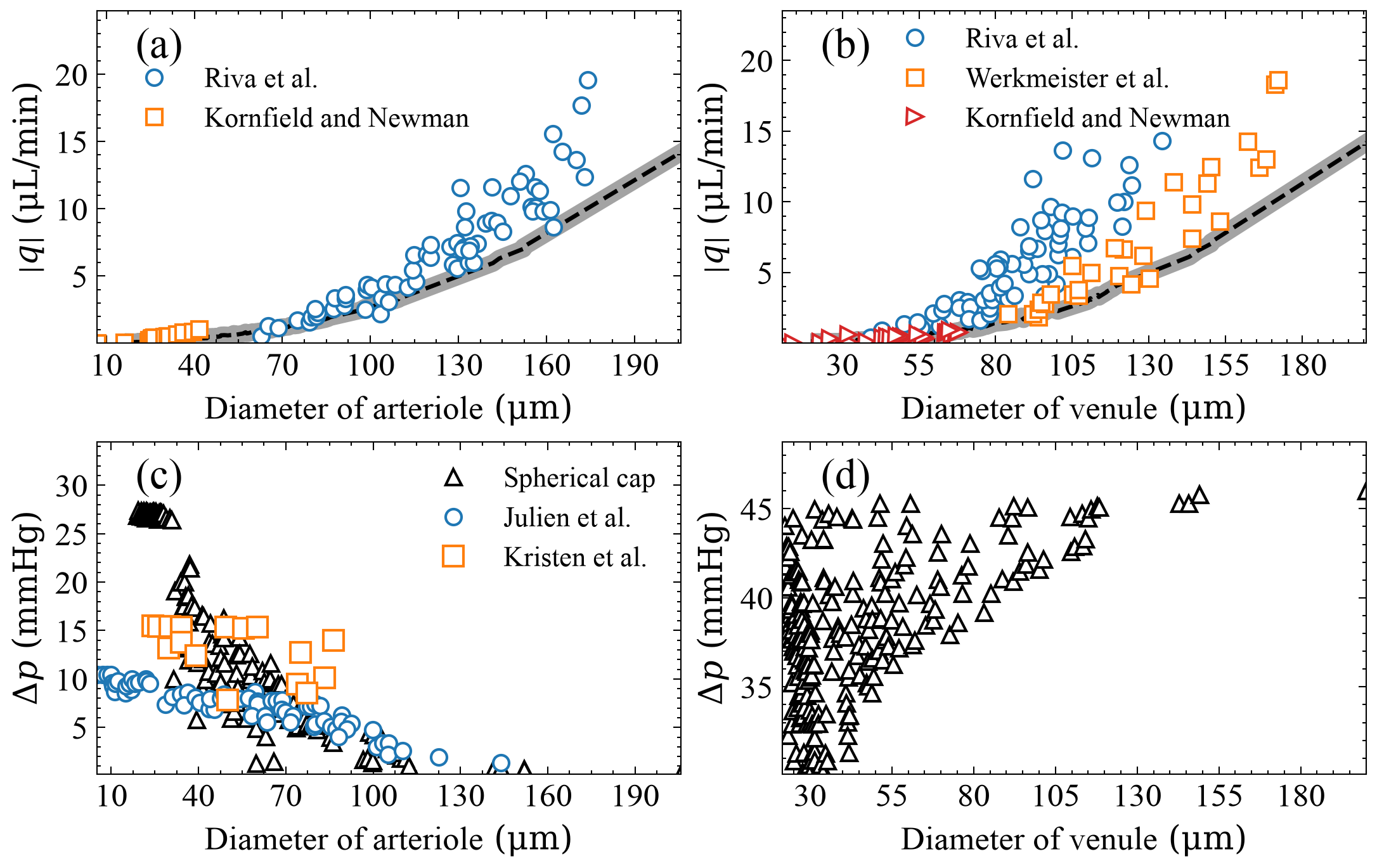}
    \caption{Flow rate and pressure difference, measured from the CRA inlet to each vessel outlet, as functions of vessel diameter.
    (a) Flow rate in the arteriolar tree.
    (b) Flow rate in the venular tree.
    (c) Pressure difference $\Delta p$ in the arteriolar tree.
    (d) Pressure difference $\Delta p$ in the venular tree.
    In panels$~$(a) and (b), the wide transparent grey solid curves denote the spherical-cap formulation, while the black dashed curves denote the circular-disc formulation. The blue circles denote the measurements of \citet{riva_blood_1985}; the orange squares denote those of \citet{kornfield_measurement_2015} in panel$~$(a) and \citet{werkmeister_measurement_2012} in panel$~$(b); and the red right-pointing triangles in panel$~$(b) denote those of \citet{kornfield_measurement_2015}.
    The black triangles in panels$~$(c) and (d) show the results from the spherical-cap formulation for each individual vessel, while the blue circles and orange squares in panel$~$(c) show predictions from other models \citep{julien_one-dimensional_2023,joldes_fundus_2016}.}
    \label{steady_1d_arterioles_venules_circular_spherical}
\end{figure}

Figure$~$\ref{truncation_CRA_inlet_flow}(b) compares the transient flow rates predicted by the pulsatile scheme under a constant CRA inlet pressure with the CRA inlet flow rate computed by the steady scheme.
The blue dashed and orange dotted curves denote $q_{in,CRA}$ and $q_{out,CRV}$ computed by the pulsatile scheme, while the black solid line denotes $q_{in,CRA}$ computed by the steady scheme.
The pulsatile computation is initialised using the vessel flow rates and cross-sectional areas obtained from the steady scheme with $p_{in,CRA}=45~\rm{mmHg}$ to improve numerical stability, after which a constant CRA inlet pressure of $62~\rm{mmHg}$ is imposed.
The CRA inlet flow decreases from its initial value, whereas the CRV outlet flow progressively increases, and both flow rates converge to the steady CRA inlet flow.
Thus, the pulsatile scheme converges to the result obtained from the steady scheme when a constant inlet pressure is prescribed, and we use the steady scheme for the following computations under constant $p_{in,CRA}$.

\subsection{Steady flow}
We next validate the multiscale model against clinical observations and results from other models.
Figure$~$\ref{steady_1d_arterioles_venules_circular_spherical}(a) and (b) show the flow-diameter relation in the arteriolar and venular trees, respectively, where the grey solid curves and black dashed curves correspond to the spherical-cap and circular-disc formulations.
In both formulations, the same hemispherical vasculature is used, and only the terminal arteriolar and venular sources are projected onto the circular disc for the circular-disc formulation.
The blue circles denote the measurements of \citet{riva_blood_1985}; the orange squares denote those of \citet{kornfield_measurement_2015} for the arterioles and \citet{werkmeister_measurement_2012} for the venules; and the red right-pointing triangles denote the venular measurements of \citet{kornfield_measurement_2015}.
For both arterioles and venules, the predicted flow rate increases with vessel diameter, and the curves from the two formulations are almost identical over the full diameter range.
Specifically, the results for the arterioles are close to the lower range of the experimental data, whereas those for the venules lie within the observed range.
This indicates that the spherical-cap analytic solution agrees with the circular-disc analytic solution when the same vasculature is used, and both formulations produce physiologically reasonable flow rates.
Because the measurements are drawn from different studies and are not matched to the modelled vascular geometry, this comparison assesses agreement with population-level ranges rather than patient-specific predictive accuracy.
The results therefore support the physiological plausibility of the model but do not constitute subject-specific validation.

Figure$~$\ref{steady_1d_arterioles_venules_circular_spherical}(c) and (d) show the relation between vessel diameter and pressure difference in the arteriolar and venular trees, respectively.
For the vessel $i$, the pressure difference is defined relative to the CRA inlet pressure as
\begin{eqnarray}
    \Delta p_i=p_{in,CRA}-p_{i,outlet},
\end{eqnarray}
where $p_{i,outlet}$ is the pressure at the vessel outlet.
The black triangle markers in panels$~$(c) and (d) denote the predictions from the spherical-cap formulation.
In the arterial tree, the pressure difference generally decreases as vessel diameter increases because larger arterioles are closer to the CRA inlet and thus have outlet pressures closer to $p_{in,CRA}$.
In the venular tree, the pressure difference generally increases with vessel diameter, as the blood moves from the smaller venules toward the CRV.
The variation among vessels of similar diameter reflects differences in their locations and paths within the vasculature.
Moreover, the arteriolar results are generally consistent with predictions from other established models \citep{julien_one-dimensional_2023,joldes_fundus_2016}, denoted by blue circles and orange squares in panel (c).
\begin{figure}
    \centering
    \includegraphics[width=\textwidth]{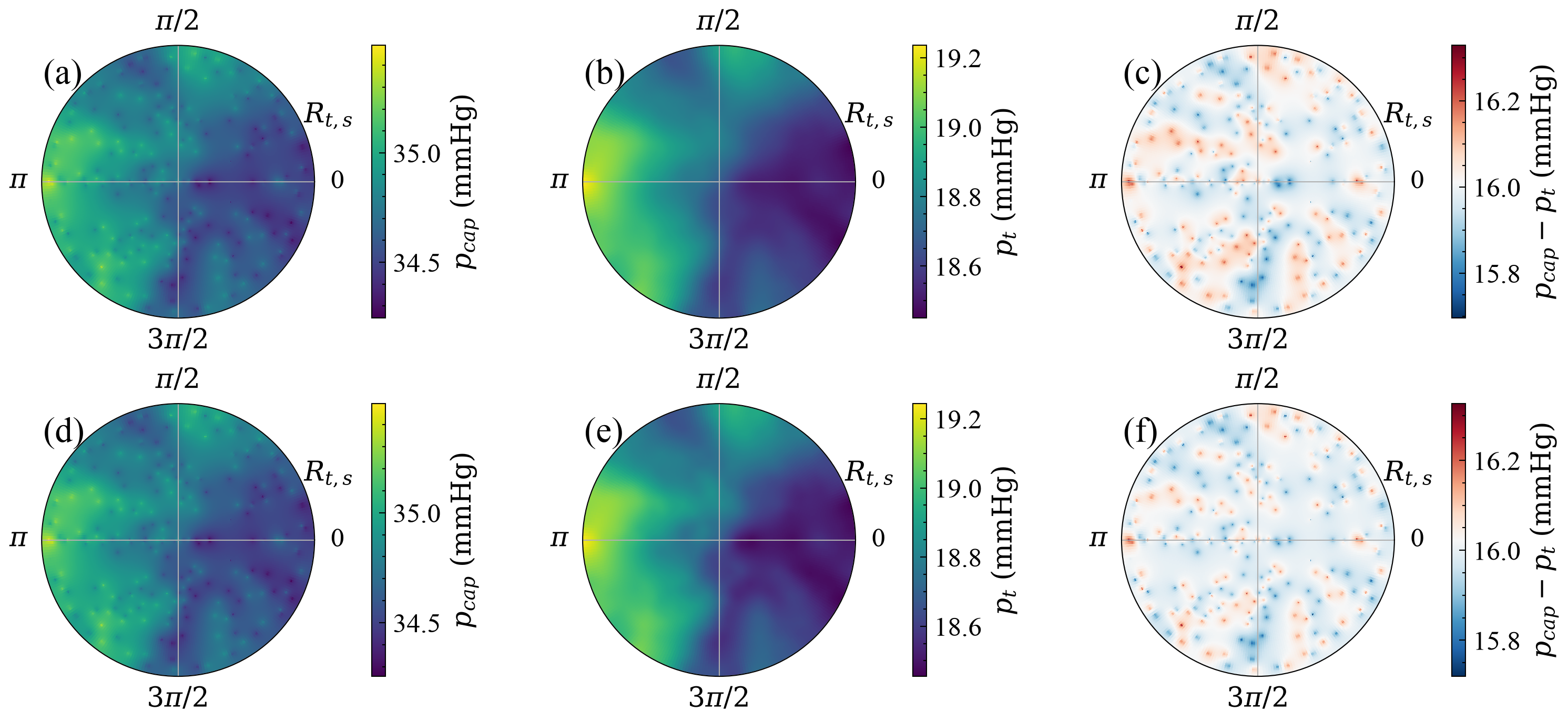}
    \caption{Capillary pressure, tissue pressure and their difference for the circular-disc and spherical-cap formulations.
    Panels (a)--(c) show the circular-disc results, and panels (d)--(f) show the spherical-cap results.
    The first, second and third columns show $p_{cap}$, $p_t$ and $p_{cap}-p_t$, respectively.}
    \label{steady_1d_capillary_tissue_pressure_circular_spherical_projected}
\end{figure}

Figure$~$\ref{steady_1d_capillary_tissue_pressure_circular_spherical_projected} compares the pressures obtained from the circular-disc and spherical-cap formulations using the same spherical-cap vasculature.
Both formulations produce capillary pressure between $34.24~\rm{mmHg}$ and $35.50~\rm{mmHg}$ and tissue pressure between $18.45~\rm{mmHg}$ and $19.24~\rm{mmHg}$.
Moreover, the capillary and tissue pressures from the two formulations exhibit slight differences over the retinal surface, for example, near $\theta=\pi$ and $r=R_{t,s}/2$.
The pressure differences in figure$~$\ref{steady_1d_capillary_tissue_pressure_circular_spherical_projected}(c) and (f) are close to the prescribed osmotic pressure and show variations due to the distribution of terminal arteriolar sources and venular sinks.
\begin{figure}
    \centering
    \includegraphics[width=\textwidth]{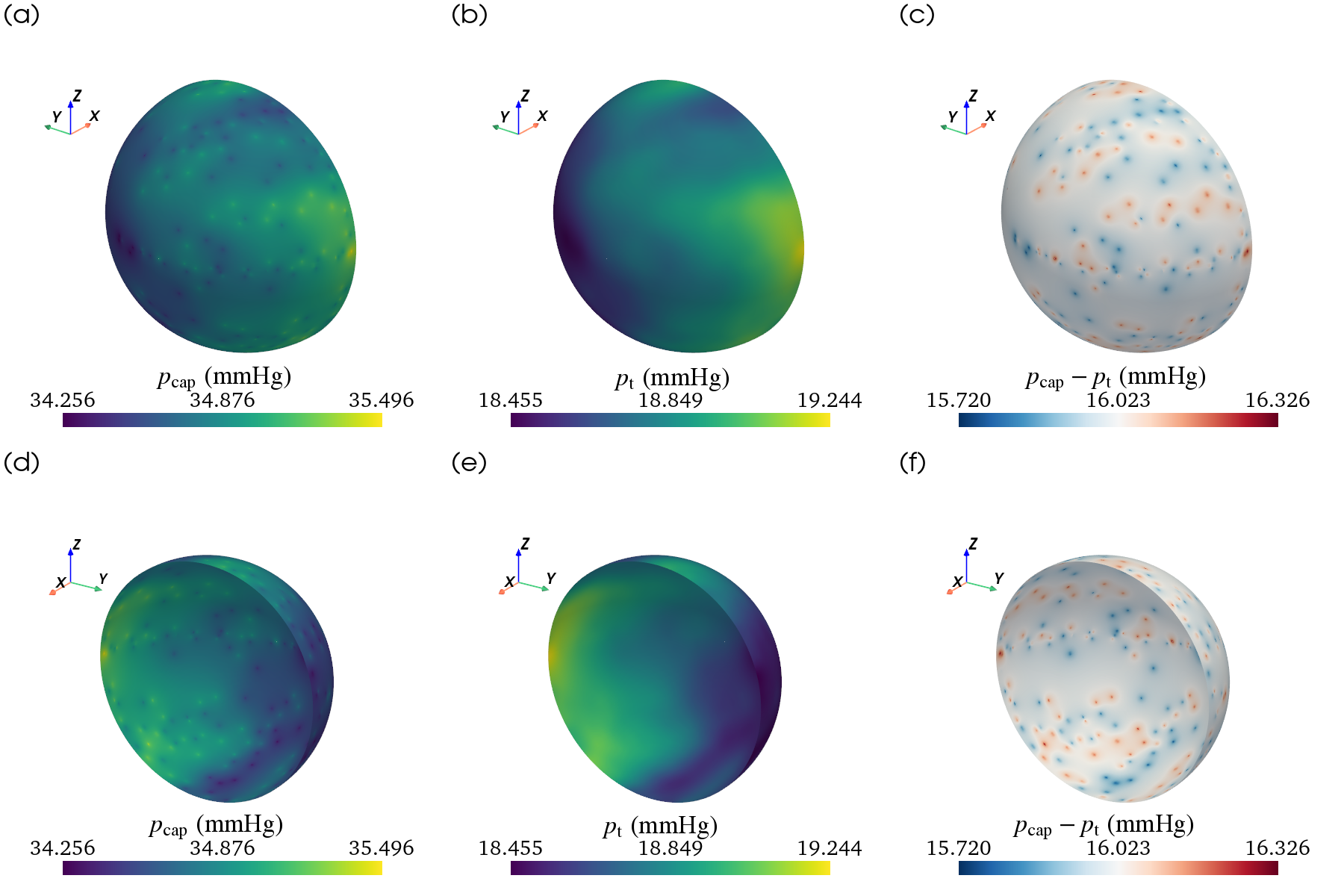}
    \caption{Capillary and tissue pressures and their difference rendered on the spherical-cap domain. Panels$~$(a), (b) and (c) show the front view of the spherical cap for $p_{cap}$, $p_t$ and $p_{cap}-p_t$, respectively, while panels (d), (e) and (f) show the corresponding opposite views.}
    \label{steady_1d_spherical_cap_pressure_spherical}
\end{figure}

Figure$~$\ref{steady_1d_spherical_cap_pressure_spherical} shows the pressures from the spherical-cap formulation on the retinal surface, where the first and second rows correspond to the front and opposite views, respectively.
The capillary pressure and pressure difference exhibit variations near the sources and sinks, while the tissue pressure varies more smoothly over the retinal surface.

\begin{figure}
    \centering
    \includegraphics[width=0.86\textwidth]{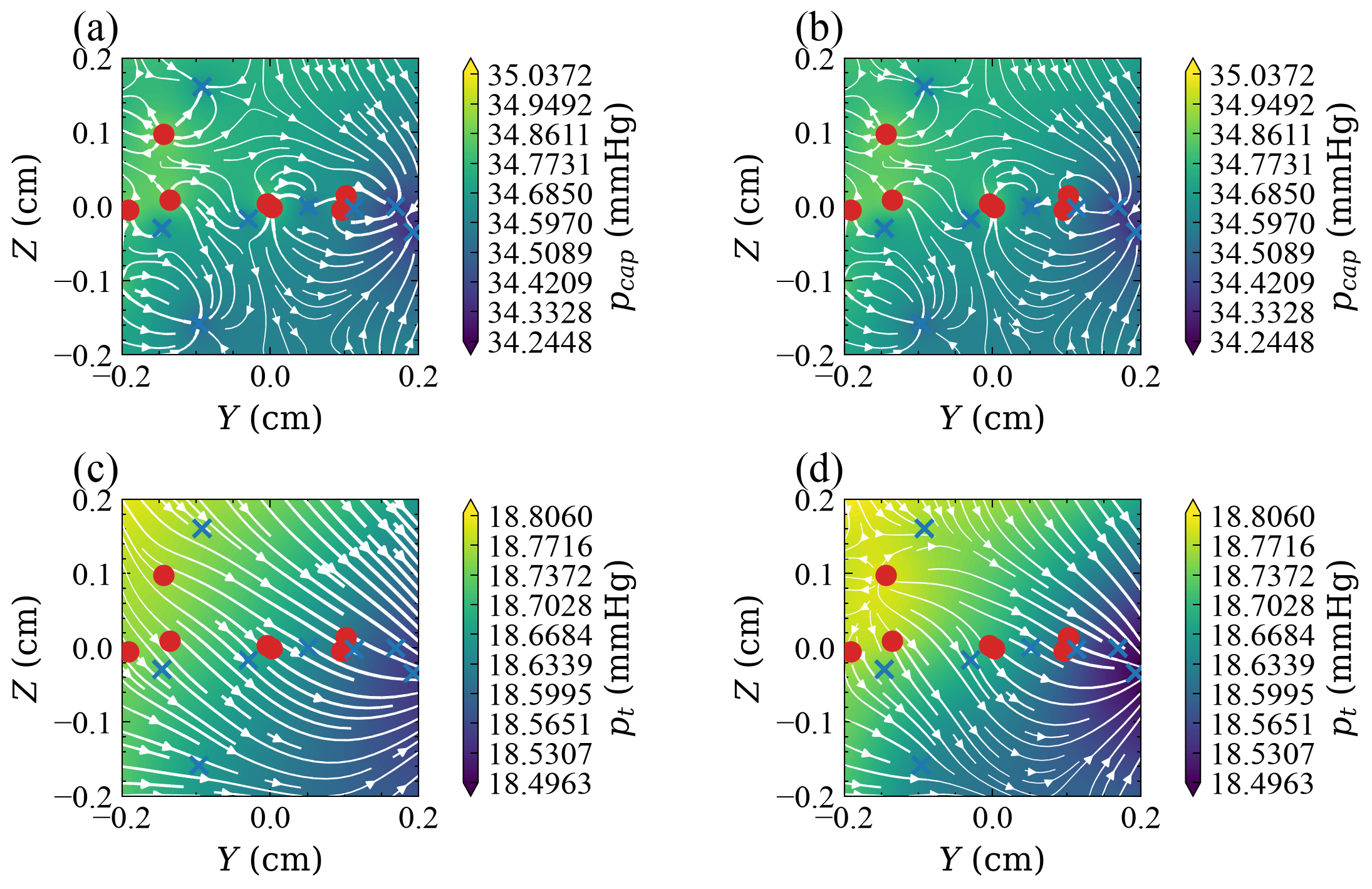}
    \caption{Local capillary and tissue pressure fields and the associated Darcy fluxes near the centre of the projected domain.
    Panels$~$(a) and (c) show the circular-disc formulation, while panels$~$(b) and (d) show the spherical-cap formulation.
    Panels$~$(a) and (b) show the capillary pressure $p_{cap}$, and panels$~$(c) and (d) show the tissue pressure $p_t$, with the same colour scale used within each row.
    The white curves denote the streamlines of the corresponding Darcy fluxes, with line width proportional to the local flux magnitude and normalised consistently within each row, while the red circles and blue crosses denote terminal arteriolar sources and venular sinks, respectively.
    The displayed region is $Y,Z\in[-0.2,0.2]~\rm{cm}$.}
    \label{steady_1d_capillary_tissue_pressure_velocity_circular_spherical_closeup}
\end{figure}

Figure$~$\ref{steady_1d_capillary_tissue_pressure_velocity_circular_spherical_closeup} provides a local comparison of the pressure fields and Darcy fluxes near the centre of the projected retinal domain.
The capillary pressure exhibits pronounced local variations around the terminal arteriolar sources and venular sinks, whereas the tissue pressure varies more smoothly because the terminal vessels couple directly to the capillary compartment.
Within each row, thicker streamlines indicate regions of larger local Darcy-flux magnitude. The capillary flux varies strongly around individual terminal sources and sinks, whereas the tissue flux exhibits a smoother spatial distribution.
Although the two formulations produce similar pressure ranges and overall patterns, differences in the orientation and curvature of the tissue Darcy-flux streamlines around the same terminal configuration demonstrate that the surface geometry affects the pathways of fluid flow.
\subsection{Pulsatile flow}
\begin{figure}
    \centering
    \includegraphics[width=0.76\textwidth]{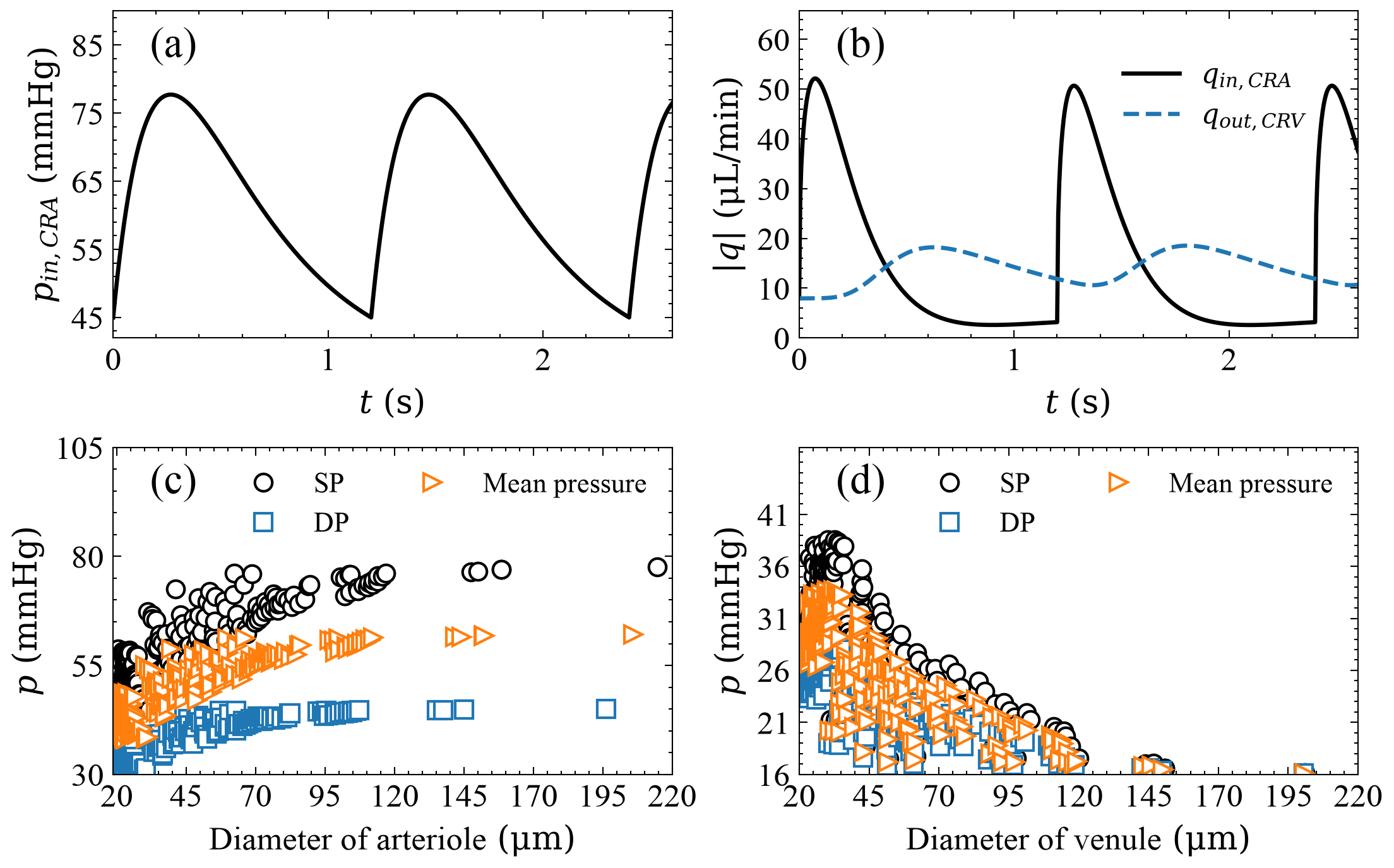}
    \caption{The CRA inlet pressure and pulsatile flow predicted by the spherical-cap formulation. (a) Prescribed CRA inlet pressure. (b) Flow rates at the CRA inlet and CRV outlet. (c) Pressures in the arteriolar tree. (d) Pressures in the venular tree. In panels$~$(c) and (d), black circles, blue squares and orange triangles correspond to systolic pressure (SP), diastolic pressure (DP) and mean pressure, respectively.}
    \label{spherical_pulsatile_flow}
\end{figure}

We next explore retinal haemodynamics under a pulsatile arteriolar pressure imposed at the CRA inlet.
The inlet pressure waveform is constructed using a periodic exponential pulse $t\exp(-3.6t)$ \citep{song_multi-domain_2024}, with a peak at $t_p=0.268~\rm{s}$.
Specifically, for a cardiac cycle period of $T=1.2~\rm{s}$ \citep{rebhan_computational_2019}, the CRA inlet pressure is prescribed as
\begin{eqnarray}
    p_{in,CRA}(t)=45+332(\tilde{t}\exp(-3.6\tilde{t})-\tilde{t}\exp(-3.6T)),
\end{eqnarray}
where $\tilde{t}=t \mod T$, and the second term ensures continuity of the waveform over consecutive cardiac cycles.
The scaling factor and offset are set to produce physiological CRA inlet pressures, with a mean pressure of approximately $62.19~\rm{mmHg}$, a systolic pressure of $77.72~\rm{mmHg}$ and a diastolic pressure of $45~\rm{mmHg}$.
The CRA inlet pressure is shown in figure$~$\ref{spherical_pulsatile_flow}(a).

Figure$~$\ref{spherical_pulsatile_flow}(b) shows the flow rates at the CRA inlet and CRV outlet, where the black solid and blue dashed curves correspond to $q_{in,CRA}$ and $q_{out,CRV}$, respectively.
The CRA inlet flow exhibits significant periodic peaks, whereas the CRV outlet flow varies more smoothly, with a smaller amplitude and a temporal delay.
This attenuation and delay result from vascular resistance and compliance, together with flow through the capillary-tissue system.
Although the instantaneous inlet and outlet flow rates differ markedly, their averages over a cardiac cycle are nearly identical.

Figure$~$\ref{spherical_pulsatile_flow}(c) and (d) show the systolic pressure (SP), diastolic pressure (DP) and mean pressure as functions of vessel diameter in the arteriolar and venular trees, respectively.
The black, blue and orange markers denote SP, DP and mean pressure, respectively.
In arterioles, the pressures generally increase with vessel diameter, reflecting the higher pressure toward the CRA inlet.
The pressures in venules generally decrease with vessel diameter.
The variation at a specific diameter reflects differences in vessel location and branching level within the vasculature.
Moreover, the pressures in the venular tree are more closely grouped than those in the arteriolar tree, consistent with the damping of pressure pulsations in arterioles and the capillary bed.
\section{Effects of the aperture}
In this section, we explore the effects of the aperture on retinal haemodynamics.
These effects comprise changes in the boundary of the retinal surface and changes in the vasculature grown on that surface.
The CRA inlet flow rate $q_{in,CRA}$ is used as an indicator of vascular flow, while the surface-averaged capillary and tissue pressures characterise the capillary bed and surrounding tissue, respectively.
Constant pressures of $p_{in,CRA}=62~\rm{mmHg}$ and $p_{out,CRV}=16~\rm{mmHg}$ are imposed at the CRA inlet and CRV outlet, respectively, and the steady scheme is used for the 1D elastic vascular segments.
As shown in figure$~$\ref{surface_averages_pressures_aperture}, for prescribed terminal source flows, the surface-averaged pressures depend only weakly on $\beta_{oc}$; however, changing the aperture can also affect the geometry of the constructed vasculature and thereby retinal haemodynamics.
Thus, we investigate these effects separately.
We first vary $\beta_{oc}$ while retaining the same hemispherical vasculature to isolate the direct effect of the aperture on the capillary-tissue domain, and then use aperture-dependent vasculature to assess the additional effect of changes in vascular construction.

\begin{figure}
    \centering
    \includegraphics[width=\textwidth]{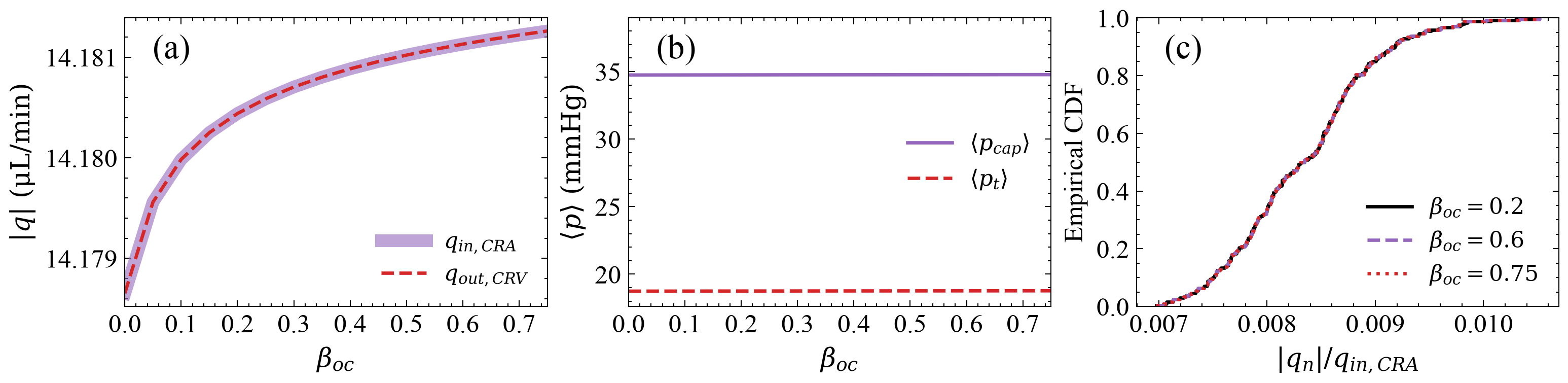}
    \caption{Dependence of retinal haemodynamics on the aperture parameter $\beta_{oc}$ while preserving the geometry of the fixed hemispherical vasculature.
    (a) Magnitudes of the CRA inlet and CRV outlet flow rates.
    (b) Surface-averaged capillary and tissue pressures.
    (c) Empirical cumulative distributions of the terminal-flow magnitudes normalised by the CRA inlet flow rate.
    In panel$~$(a), the wide transparent purple solid curve and red dashed curve denote $q_{in,CRA}$ and $q_{out,CRV}$, respectively.
    In panel$~$(b), the purple solid and red dashed curves denote $\langle p_{cap}\rangle$ and $\langle p_t\rangle$, respectively.
    In panel$~$(c), the black solid, purple dashed and red dotted curves correspond to $\beta_{oc}=0.2$, $\beta_{oc}=0.6$ and $\beta_{oc}=0.75$, respectively.}
    \label{aperture_sensitivity_fixed_vasculature}
\end{figure}

Figure$~$\ref{aperture_sensitivity_fixed_vasculature}(a) and (b) show the dependence of the CRA inlet and CRV outlet flow-rate magnitudes and the surface-averaged capillary and tissue pressures, respectively, on $\beta_{oc}$ using the fixed hemispherical vasculature.
Because the vascular geometry is preserved, these results isolate the effect of changing the spherical-cap domain.
The CRA inlet flow rate increases by only approximately $0.00261~\rm{\upmu L/min}$ over the range shown, corresponding to approximately $0.0184~\%$ of its value for the hemisphere ($\beta_{oc}=0$).
The surface averages $\langle p_{cap}\rangle$ and $\langle p_t\rangle$ each vary by approximately $0.0264~\rm{mmHg}$ and are therefore also insensitive to the aperture parameter.
Figure$~$\ref{aperture_sensitivity_fixed_vasculature}(c) shows the empirical cumulative distribution functions of the normalised terminal-flow magnitudes $\vert q_n\vert/q_{in,CRA}$, rather than their absolute magnitudes, for $\beta_{oc}=0.2$, $\beta_{oc}=0.6$ and $\beta_{oc}=0.75$.
The three curves are nearly identical, demonstrating that the relative distribution of terminal flows remains essentially unchanged as $\beta_{oc}$ varies when the vasculature is fixed.
\begin{figure}
    \centering
    \includegraphics[width=\textwidth]{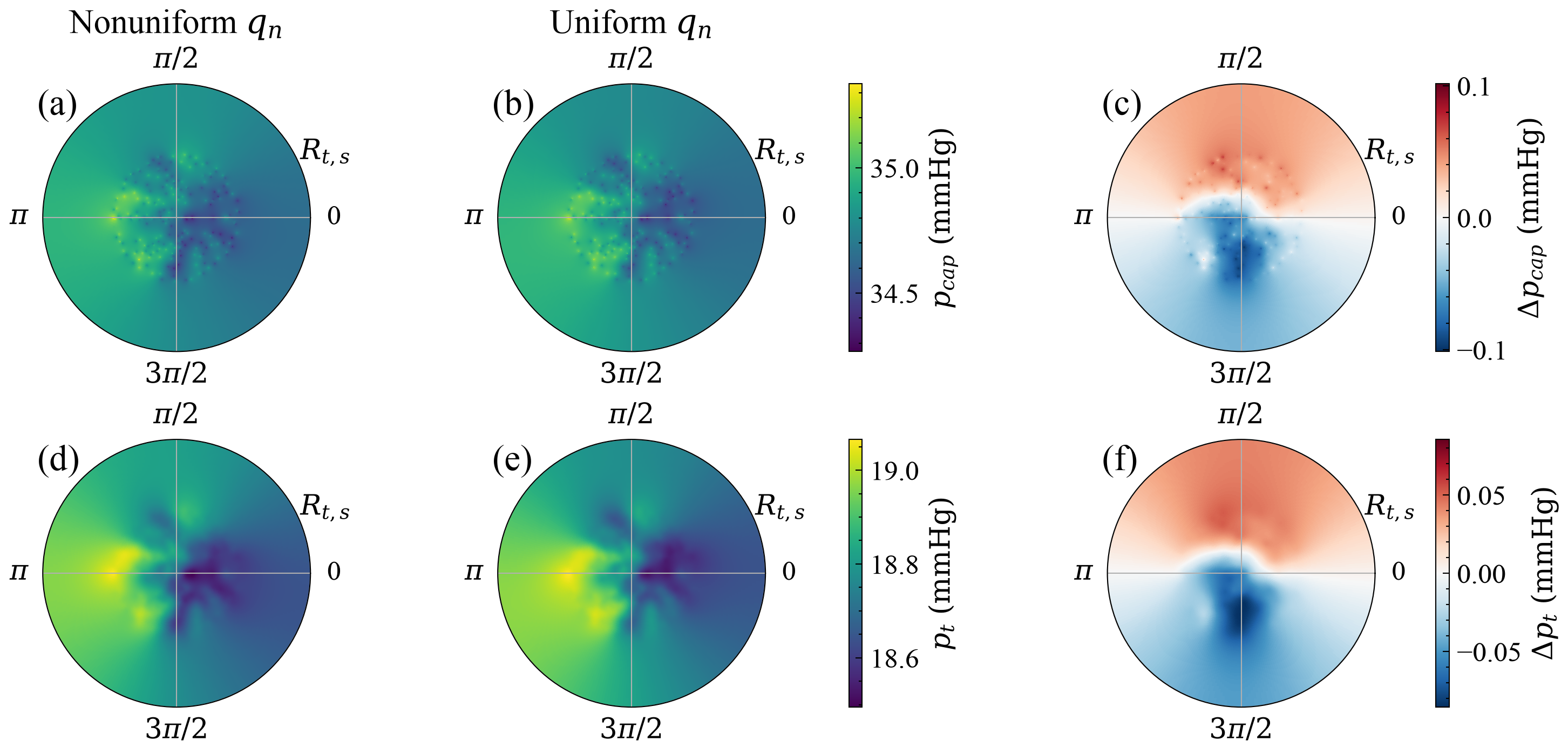}
    \caption{Effect of terminal-flow allocation on the pressure fields for the fixed hemispherical vasculature at $\beta_{oc}=0.6$.
    Panels (a)--(c) show the capillary pressure for the nonuniform-flow and uniform-flow cases and their difference (nonuniform minus uniform), respectively.
    Panels (d)--(f) show the corresponding tissue pressures and their difference in the same order.
    In the uniform-flow case, $q_n=q_{in,CRA}/120$ for arteriolar sources and $q_n=-q_{in,CRA}/120$ for venular sources, with the same total CRA inlet flow as in the nonuniform-flow case.}
    \label{pressure_terminal_flow_allocation_beta_06}
\end{figure}

To explore the effect of terminal-flow allocation on the pressure fields, figure$~$\ref{pressure_terminal_flow_allocation_beta_06} compares the nonuniform terminal flows obtained from the steady 1D model with terminal flows of uniform magnitude at $\beta_{oc}=0.6$.
The total CRA inlet flow rate is identical in both cases, and the total arteriolar source flow is balanced by the magnitude of the total venular sink flow.
Figures$~$\ref{pressure_terminal_flow_allocation_beta_06}(a), (b) and (c) show that the two terminal-flow allocations produce similar capillary pressure distributions, although local differences occur near the terminal source positions.
The difference in capillary pressure ranges from approximately $-0.1014~\rm{mmHg}$ to $0.0716~\rm{mmHg}$.
Figures$~$\ref{pressure_terminal_flow_allocation_beta_06}(d), (e) and (f) show that the overall tissue pressure distributions are similar, with differences from approximately $-0.0855~\rm{mmHg}$ to $0.0522~\rm{mmHg}$.
Thus, for a fixed total flow, replacing the nonuniform terminal flows with uniform values preserves the overall pressure distributions but smooths the local pressure variations associated with individual terminal vessels.
\begin{figure}
    \centering
    \includegraphics[width=\textwidth]{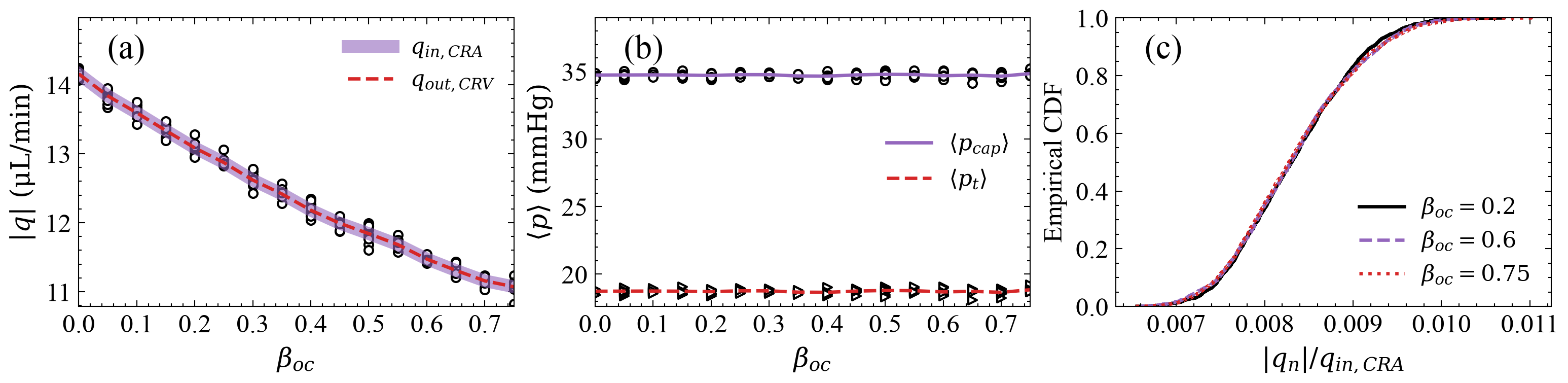}
    \caption{Dependence of retinal haemodynamics on the aperture parameter $\beta_{oc}$ using aperture-dependent vasculature.
    (a) Magnitudes of the CRA inlet and CRV outlet flow rates.
    (b) Surface-averaged capillary and tissue pressures.
    (c) Empirical cumulative distributions of the terminal-flow magnitudes normalised by the CRA inlet flow rate and pooled over $10$ realisations.
    In panel$~$(a), the wide transparent purple solid and red dashed curves denote the mean values of $q_{in,CRA}$ and $q_{out,CRV}$, respectively, while the black open circles denote $q_{out,CRV}$ for the individual realisations.
    In panel$~$(b), the purple solid and red dashed curves denote the mean values of $\langle p_{cap}\rangle$ and $\langle p_t\rangle$, respectively, while the black open circles and triangles denote their values for the individual realisations.
    In panel$~$(c), the black solid, purple dashed and red dotted curves correspond to $\beta_{oc}=0.2$, $\beta_{oc}=0.6$ and $\beta_{oc}=0.75$, respectively.}
    \label{aperture_sensitivity_varying_vasculature}
\end{figure}

We next explore the haemodynamics for different values of $\beta_{oc}$ using aperture-dependent vasculature.
For each value of $\beta_{oc}$, new arteriolar and venular trees are constructed on the corresponding spherical-cap domain, and the results are averaged over $10$ realisations to account for the stochasticity of vascular growth.
The curves in figure$~$\ref{aperture_sensitivity_varying_vasculature}(a) and (b) denote the means over the $10$ realisations, while the markers show the individual realisations.
Figure$~$\ref{aperture_sensitivity_varying_vasculature}(a) shows that the mean CRA inlet flow rate decreases by approximately $3.08~\rm{\upmu L/min}$ as $\beta_{oc}$ increases over the range shown.
This decrease is consistent with an increase in vessel lengths and hence in the hydraulic resistance of the constructed vasculature as the retinal surface is extended.
In figure$~$\ref{aperture_sensitivity_varying_vasculature}(b), both surface-averaged pressures vary by approximately $0.191~\rm{mmHg}$ over the range shown and therefore remain comparatively insensitive to $\beta_{oc}$ despite the marked change in total flow.
Figure$~$\ref{aperture_sensitivity_varying_vasculature}(c) shows that the distributions of $\vert q_n\vert/q_{in,CRA}$ are close but not identical for the three aperture values.
Comparison of figures$~$\ref{aperture_sensitivity_fixed_vasculature} and~\ref{aperture_sensitivity_varying_vasculature} indicates that changing the aperture alone has a negligible effect on retinal flow.
The substantial decrease in total flow observed with aperture-dependent vasculature therefore arises primarily from changes in the constructed vascular geometry and its hydraulic resistance.
In contrast, the surface-averaged pressures and normalised allocation of flow among the terminal vessels remain comparatively insensitive to the aperture parameter.
\begin{figure}
    \centering
    \includegraphics[width=\textwidth]{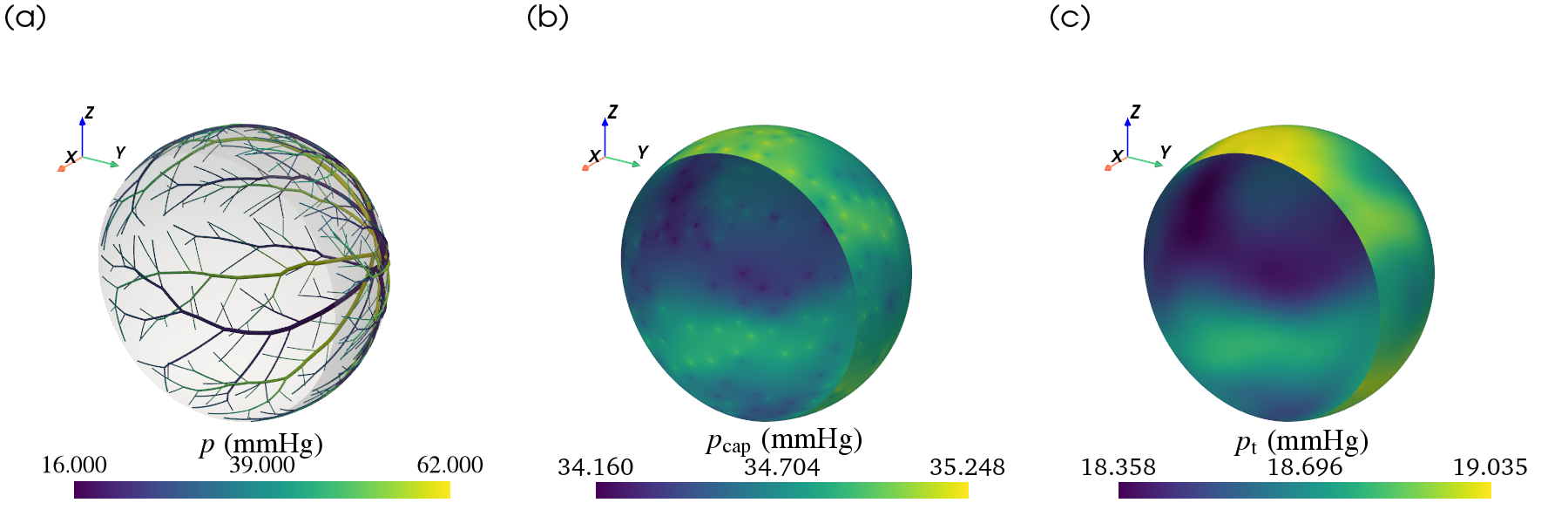}
    \caption{Pressure distributions for one realisation of the aperture-dependent vasculature at $\beta_{oc}=0.3$.
    (a) Pressures in the arteriolar and venular trees.
    (b) Capillary pressure $p_{cap}$.
    (c) Tissue pressure $p_t$.}
    \label{pressure_varying_vasculature_beta_03}
\end{figure}

Figure$~$\ref{pressure_varying_vasculature_beta_03} shows the vascular, capillary and tissue pressure distributions for one of the aperture-dependent vascular realisations at $\beta_{oc}=0.3$.
Panel$~$(a) shows the pressure variation along the arteriolar and venular trees.
The capillary pressure in panel$~$(b) exhibits local variations associated with the discrete terminal sources and sinks, whereas the tissue pressure in panel$~$(c) varies more smoothly over the spherical-cap domain.
\section{Discussion and conclusion}
In a recent study \citep{lin_novel_2026}, we proposed a multiscale model that combines three key components: (i) 1D elastic flow in the arteriolar and venular trees; (ii) coupled Darcy flow in the capillary bed and interstitial tissue, treated as interacting porous media; and (iii) point source coupling between the vascular trees and the capillary-tissue continuum.
In this work, we extend our previous work in several directions.
The retinal surface is modelled as a spherical-cap domain with an aperture, and the tree-based descriptions for the arterioles and venules are integrated with a coupled porous medium description for the capillary bed and surrounding tissue.
The current model is closer to physiological reality but mathematically more complicated.
The central theoretical contribution is an analytic solution for the coupled capillary-tissue system on the spherical-cap domain.
It compresses the system into an effective resistor while retaining local fluid exchange between the two components.
Obtained via stereographic projection and a decoupling transformation, this analytic solution provides a more physiologically realistic description of flow on the retinal surface than previous models developed on the circular disc.

Comparison of the proposed model with experimental data shows that it produces physiologically reasonable flow rates and pressures.
Moreover, the close agreement between the spherical-cap and circular-disc formulations suggests that retinal flow is determined mainly by the vasculature. Thus, although geometrically simplified, the planar-disc multiscale model may be sufficient for predicting retinal flow in a fixed vasculature.
However, the spherical-cap multiscale model is needed when the vasculature is constructed directly on the curved surface.
We then explore the effects of the aperture on retinal flow.
Computations with a varying aperture parameter and fixed hemispherical vasculature reveal that retinal haemodynamics is largely insensitive to the aperture parameter.
Under the balanced source condition, the surface-averaged pressures depend only weakly on the aperture parameter.
Furthermore, computations with aperture-dependent vasculature demonstrate that the aperture affects retinal flow mainly through its effect on the vasculature itself, underscoring the crucial role of vascular trees in determining perfusion.

Our model has several limitations.
First, the conductivities and the exchange rate are assumed constant.
This simplification enables the analytic solution; however, such parameters are expected to vary spatially in the retina.
A natural extension of this work is to consider spatially varying conductivities and exchange coefficients, which would provide a more realistic representation of the properties of the capillary bed and surrounding tissue.
Second, both the capillary bed and surrounding tissue are treated as two-dimensional porous media, ignoring the finite thickness and variations through the thickness \citep{causin_blood_2016}.
Further studies could employ a similar method to solve the capillary-tissue system on more general geometric domains.
Third, the osmotic pressure difference is treated as a constant, which is reasonable under normal physiological conditions but may not hold in pathological conditions, such as diabetic retinopathy or macular oedema, where protein leakage and osmotic gradients can vary significantly.
Coupling this model with transport models \citep{song_multi-domain_2024,zhu_bidomain_2019,zhu_membranes_2021,xiao_potassium_2025,xiao_glymphatic_2025,zhu_tridomain_2021,zhu_optic_2021} would capture these variations and thereby enable a more accurate description of the oxygen and nutrient delivery in the retina.
Additionally, this framework can be adapted for patient-specific computation by combining individual retinal geometries and vasculature from clinical imaging.

Despite the necessary simplifications, the proposed model produces physiologically reasonable results, permits efficient computation and provides a more realistic geometric representation of the retinal surface.
The derived analytic solution contributes to the modelling of flow in curved porous media, and the overall framework can provide further insights into retinal physiology and support future patient-specific computations.

\noindent \textbf{Acknowledgement.} This work is supported in part by NSFC (Project number 12231004).

\noindent \textbf{Declaration of interests.} The authors report no conflict of interest.

\appendix
\section{Stereographic projection and conformal factor}
\label{StereographicProjectionConformalFactor}
Using the spherical coordinates$~$(\ref{spherical_coordinates_spherical_cap}), the coordinates $(y,z)$ can be written as
\begin{eqnarray}
    y=\frac{R^2_t\sin\phi\cos\theta}{R_t(1+\cos\phi)}=\frac{R_t\sin\phi}{1+\cos\phi}\cos\theta \\
    z=\frac{R^2_t\sin\phi\sin\theta}{R_t(1+\cos\phi)}=\frac{R_t\sin\phi}{1+\cos\phi}\sin\theta.
\end{eqnarray}
Using the identities $\sin\phi=2\sin(\phi/2)\cos(\phi/2)$ and $1+\cos\phi=2\cos^2(\phi/2)$, we obtain
\begin{eqnarray}
    y=R_t\tan\frac{\phi}{2}\cos\theta, \\
    z=R_t\tan\frac{\phi}{2}\sin\theta,
\end{eqnarray}
and the radial coordinate in the projected domain is
\begin{eqnarray}
    r=R_t\tan\frac{\phi}{2}.
    \label{radial_coordinate_theta}
\end{eqnarray}
From equation$~$(\ref{radial_coordinate_theta}), $d\phi$ can be written as
\begin{eqnarray}
    d\phi=\left(\frac{dr}{d\phi}\right)^{-1}dr=\frac{2R_t}{r^2+R^2_t}dr.
\end{eqnarray}
Moreover,
\begin{eqnarray}
    \sin^2\frac{\phi}{2}=\frac{r^2}{r^2+R^2_t}, \quad \cos^2\frac{\phi}{2}=\frac{R^2_t}{r^2+R^2_t},
\end{eqnarray}
so that $\sin\phi$ in the metric can be written as
\begin{eqnarray}
    \sin\phi=2\sin\frac{\phi}{2}\cos\frac{\phi}{2}=\frac{2R_tr}{r^2+R^2_t}.
\end{eqnarray}
Substituting these expressions into the metric$~$(\ref{spherical_cap_domain_metric}) yields
\begin{eqnarray}
    ds^2_{\mathcal{S}}=R^2_t\left(\frac{4R^2_t}{(r^2+R^2_t)^2}dr^2+\frac{4R^2_tr^2}{(r^2+R^2_t)^2}d\theta^2\right)=R_t^2\frac{4R^2_t}{(R^2_t+r^2)^2}(dr^2+r^2d\theta^2).
\end{eqnarray}
Since $dr^2+r^2d\theta^2=dy^2+dz^2$, we obtain
\begin{eqnarray}
    ds^2_{\mathcal{S}}=\frac{4R^4_t}{(R^2_t+r^2)^2}(dy^2+dz^2).
\end{eqnarray}
Thus, the stereographic projection from the spherical cap to the circular disc is conformal, and the conformal factor $\lambda(y,z)$ is given by
\begin{eqnarray}
    \lambda(y,z)=\frac{2R^2_t}{R^2_t+y^2+z^2}.
\end{eqnarray}
\section{Solution for the radial component in the Green's function of exchange pressure and its simplification}
\label{SolutionRadialComponentGreensFunctionExchangePressureSimplification}
The solution $f^m_1$ can be constructed as
\begin{eqnarray}
    f^m_1(w)=P^{\vert m\vert}_\nu(w)+A_mQ^{\vert m\vert}_\nu(w),
\end{eqnarray}
where $P^{\vert m\vert}_\nu(w)$ and $Q^{\vert m\vert}_\nu(w)$ are associated Legendre functions of first and second kind, respectively, and $A_m$ is a coefficient to be determined.
The function $f^m_1$ must satisfy $df^m_1/dw=0$ at $w=-\beta_{oc}$, which yields
\begin{eqnarray}
    A_m=-\frac{P^{\vert m\vert\prime}_\nu(-\beta_{oc})}{Q^{\vert m\vert\prime}_\nu(-\beta_{oc})},
\end{eqnarray}
where $P^{\vert m\vert\prime}_\nu(w)$ and $Q^{\vert m\vert\prime}_\nu(w)$ are the derivatives of $P^{\vert m\vert}_\nu(w)$ and $Q^{\vert m\vert}_\nu(w)$, respectively, and hence $f^m_1$ is
\begin{eqnarray}
    f^m_1(w)=P^{\vert m\vert}_\nu(w)-\frac{P^{\vert m\vert\prime}_\nu(-\beta_{oc})}{Q^{\vert m\vert\prime}_\nu(-\beta_{oc})}Q^{\vert m\vert}_\nu(w).
\end{eqnarray}
The solution $f^m_2$ is given by
\begin{eqnarray}
    f^m_2(w)=P^{\vert m\vert}_\nu(w),
\end{eqnarray}
since $Q^{\vert m\vert}_\nu(w)$ is singular at $w=1$.
Then, integrating the equation over a small interval around $w=w_n$ yields
\begin{eqnarray}
    \int^{w_n+\epsilon}_{w_n-\epsilon}\frac{d}{dw}\left((1-w^2)\frac{df^m_{exch}}{dw}\right)dw-\int^{w_n+\epsilon}_{w_n-\epsilon}\left(R^2_t\gamma^2_0+\frac{m^2}{1-w^2}\right)f^m_{exch}dw=\frac{1}{2\pi}.
\end{eqnarray}
The nonsingular integral vanishes as $\epsilon\to 0$; thus the jump condition is
\begin{eqnarray}
    (1-w^2_n)\left(\frac{df^m_{exch}}{dw}\vert_{w=w^+_n}-\frac{df^m_{exch}}{dw}\vert_{w=w^-_n}\right)=\frac{1}{2\pi}.
\end{eqnarray}
Substituting the expression$~$(\ref{f_m_exch_small_large}) implies
\begin{eqnarray}
    C_m(1-w^2_n)\left(f^m_1(w_n)f^\prime_2(w_n)-f^m_2(w_n)f^\prime_1(w_n)\right)=\frac{1}{2\pi}.
\end{eqnarray}
Thus,
\begin{eqnarray}
    C_m&=&\frac{1}{2\pi(1-w^2_n)\left(f^m_1(w_n)f^\prime_2(w_n)-f^m_2(w_n)f^\prime_1(w_n)\right)}, \\
    &=&-\frac{1}{2\pi A_m(1-w^2_n)(P^{\vert m\vert}_\nu(w_n)Q^{\vert m\vert\prime}_\nu(w_n)-Q^{\vert m\vert}_\nu(w_n)P^{\vert m\vert\prime}_\nu(w_n))}.
\end{eqnarray}
We employ the following identity:
\begin{eqnarray}
    P^{\vert m\vert}_\nu(w)Q^{\vert m\vert\prime}_\nu(w)-Q^{\vert m\vert}_\nu(w)P^{\vert m\vert\prime}_\nu(w)=\frac{\Gamma(\nu+\vert m\vert+1)}{\Gamma(\nu-\vert m\vert+1)}\frac{1}{1-w^2}.
\end{eqnarray}
Substituting the expression for $A_m$ and employing this identity, the coefficient $C_m$ can be written as
\begin{eqnarray}
    C_m=\frac{Q^{\vert m\vert\prime}_\nu(-\beta_{oc})}{P^{\vert m\vert\prime}_\nu(-\beta_{oc})}\frac{\Gamma(\nu-\vert m\vert+1)}{2\pi\Gamma(\nu+\vert m\vert+1)}.
\end{eqnarray}
Substituting the expressions for $f^m_1$, $f^m_2$ and the coefficient $C_m$ into equation$~$(\ref{f_m_exch_small_large}) yields the solution$~$(\ref{f_m_exch_Q_P}) for the radial component $f^m_{exch}$.

We next simplify the Green's function $f_{exch}$ using a connection formula for $Q^{\vert m\vert}_\nu(w)$ and the Legendre addition theorem.
For $-\beta_{oc}\leq w \leq 1$, $m\in \mathbb{Z}$ and $\nu\notin \mathbb{Z}$, the function $Q^{\vert m\vert}_\nu(w)$ can be expressed as follows \citep[\S14.9]{olver_nist_2010-1}:
\begin{eqnarray}
    Q^{\vert m\vert}_\nu(w)=\frac{\pi}{2\sin(\pi\nu)}\left[\cos(\pi\nu)P^{\vert m\vert}_\nu(w)-(-1)^mP^{\vert m\vert}_\nu(-w)\right].
\end{eqnarray}
Thus, the Green's function $f_{exch}$ can be expressed only in terms of the associated Legendre function of the first kind $P^{\vert m\vert}_\nu(w)$.
The terms in $f_{exch}$ including the function $Q^{\vert m\vert}_\nu(w)$ can thus be written as
\begin{eqnarray}
    \frac{Q^{\vert m\vert\prime}_\nu(-\beta_{oc})}{P^{\vert m\vert\prime}_\nu(-\beta_{oc})}=\frac{\pi}{2\sin(\pi\nu)}\left[\cos(\pi\nu)+(-1)^m\frac{P^{\vert m\vert\prime}_\nu(\beta_{oc})}{P^{\vert m\vert\prime}_\nu(-\beta_{oc})}\right], \\
    Q^{\vert m\vert}_\nu(w_<)=\frac{\pi}{2\sin(\pi\nu)}\left[\cos(\pi\nu)P^{\vert m\vert}_\nu(w_<)-(-1)^mP^{\vert m\vert}_\nu(-w_<)\right],
\end{eqnarray}
and substituting these expressions into the equation$~$(\ref{f_m_exch_Q_P}) implies
\begin{eqnarray}
    f_{exch}&=&\sum^\infty_{m=-\infty}\frac{\Gamma(\nu-\vert m\vert+1)}{2\pi\Gamma(\nu+\vert m\vert+1)}\frac{\pi\cos(\pi\nu)}{2\sin(\pi\nu)}P^{\vert m\vert}_\nu(w_<)P^{\vert m\vert}_\nu(w_>)\exp[im(\theta-\theta_n)] \nonumber\\
    &&+\sum^\infty_{m=-\infty}\frac{\Gamma(\nu-\vert m\vert+1)}{2\pi\Gamma(\nu+\vert m\vert+1)}\frac{(-1)^m\pi}{2\sin(\pi\nu)}\frac{P^{\vert m\vert\prime}_\nu(\beta_{oc})}{P^{\vert m\vert\prime}_\nu(-\beta_{oc})}P^{\vert m\vert}_\nu(w_<)P^{\vert m\vert}_\nu(w_>)\exp[im(\theta-\theta_n)] \nonumber\\
    &&-\sum^\infty_{m=-\infty}\frac{\Gamma(\nu-\vert m\vert+1)}{2\pi\Gamma(\nu+\vert m\vert+1)}\frac{\pi\cos(\pi\nu)}{2\sin(\pi\nu)}P^{\vert m\vert}_\nu(w_<)P^{\vert m\vert}_\nu(w_>)\exp[im(\theta-\theta_n)] \nonumber \\
    &&+\sum^\infty_{m=-\infty}\frac{\Gamma(\nu-\vert m\vert+1)}{2\pi\Gamma(\nu+\vert m\vert+1)}\frac{(-1)^m\pi}{2\sin(\pi\nu)}P^{\vert m\vert}_\nu(-w_<)P^{\vert m\vert}_\nu(w_>)\exp[im(\theta-\theta_n)], \\
    &=&\sum^\infty_{m=-\infty}\frac{\Gamma(\nu-\vert m\vert+1)}{2\pi\Gamma(\nu+\vert m\vert+1)}\frac{(-1)^m\pi}{2\sin(\pi\nu)}\frac{P^{\vert m\vert\prime}_\nu(\beta_{oc})}{P^{\vert m\vert\prime}_\nu(-\beta_{oc})}P^{\vert m\vert}_\nu(w_<)P^{\vert m\vert}_\nu(w_>)\exp[im(\theta-\theta_n)] \nonumber\\
    &&+\sum^\infty_{m=-\infty}\frac{\Gamma(\nu-\vert m\vert+1)}{2\pi\Gamma(\nu+\vert m\vert+1)}\frac{(-1)^m\pi}{2\sin(\pi\nu)}P^{\vert m\vert}_\nu(-w_<)P^{\vert m\vert}_\nu(w_>)\exp[im(\theta-\theta_n)].
\end{eqnarray}
We denote the two series by $s_1$ and $s_2$:
\begin{eqnarray}
    s_1=\sum^\infty_{m=-\infty}\frac{\Gamma(\nu-\vert m\vert+1)}{2\pi\Gamma(\nu+\vert m\vert+1)}\frac{(-1)^m\pi}{2\sin(\pi\nu)}\frac{P^{\vert m\vert\prime}_\nu(\beta_{oc})}{P^{\vert m\vert\prime}_\nu(-\beta_{oc})}P^{\vert m\vert}_\nu(w_<)P^{\vert m\vert}_\nu(w_>)\exp[im(\theta-\theta_n)], \\
    s_2=\sum^\infty_{m=-\infty}\frac{\Gamma(\nu-\vert m\vert+1)}{2\pi\Gamma(\nu+\vert m\vert+1)}\frac{(-1)^m\pi}{2\sin(\pi\nu)}P^{\vert m\vert}_\nu(-w_<)P^{\vert m\vert}_\nu(w_>)\exp[im(\theta-\theta_n)],
\end{eqnarray}
and hence
\begin{eqnarray}
    f_{exch}=s_1+s_2.
\end{eqnarray}
Combining the terms paired with $m$ and $-m$, $s_1$ can be written as
\begin{eqnarray}
    s_1&=&\frac{1}{4\sin(\pi\nu)}\frac{P^{0\prime}_\nu(\beta_{oc})}{P^{0\prime}_\nu(-\beta_{oc})}P^0_\nu(w_<)P^0_\nu(w_>) \nonumber\\
    &&+\sum^\infty_{m=1}\frac{\Gamma(\nu-m+1)}{\Gamma(\nu+ m+1)}\frac{(-1)^m}{2\sin(\pi\nu)}\frac{P^{ m\prime}_\nu(\beta_{oc})}{P^{m\prime}_\nu(-\beta_{oc})}P^m_\nu(w_<)P^m_\nu(w_>)\cos(m(\theta-\theta_n)).
\end{eqnarray}
The series $s_2$ can be written as
\begin{eqnarray}
    s_2&=&\frac{1}{4\sin(\pi\nu)}P^0_\nu(-w_<)P^0_\nu(w_>) \nonumber \\
    &&+2\sum^\infty_{m=1}\frac{\Gamma(\nu-\vert m\vert+1)}{2\pi\Gamma(\nu+\vert m\vert+1)}\frac{(-1)^m\pi}{2\sin(\pi\nu)}P^m_\nu(-w_<)P^m_\nu(w_>)\cos(m(\theta-\theta_n)).
\end{eqnarray}
The series $s_2$ can be summed using the Legendre addition formula \citep[\S14.18(ii)]{olver_nist_2010-1}:
\begin{eqnarray}
    &&P^0_\nu(\cos z_1\cos z_2+\sin z_1\sin z_2\cos \varphi) \nonumber \\
    &&=P^0_\nu(\cos z_1)P^0_\nu(\cos z_2)+2\sum^\infty_{m=1}(-1)^mP^{-m}_\nu(\cos z_1)P^m_\nu(\cos z_2)\cos(m \varphi),
    \label{legendre_addition}
\end{eqnarray}
where $z_1,z_2,z_1+z_2\in[0,\pi)$ and $\varphi$ is real-valued.
Setting $\cos z_1=-w_<$ and $\cos z_2=w_>$ gives
\begin{eqnarray}
    z_1+z_2=\cos^{-1}(-w_<)+\cos^{-1}w_>=\pi-\cos^{-1}w_<+\cos^{-1}w_>,
\end{eqnarray}
Since $w_<\leq w_>$, we obtain $\cos^{-1}w_<\geq \cos^{-1}w_>$ and $0\leq z_1+z_2\leq \pi$.
Then, if $w_< <w_>$, the conditions for equation$~$(\ref{legendre_addition}) are satisfied.
For $m\in \mathbb{N}$, the following identity holds:
\begin{eqnarray}
    P^{-m}_\nu(w)=(-1)^m\frac{\Gamma(\nu-m+1)}{\Gamma(\nu+m+1)}P^m_\nu(w).
\end{eqnarray}
Setting $\varphi=\pi-(\theta-\theta_n)$ and using this identity, the series $s_2$ simplifies to
\begin{eqnarray}
    s_2=\frac{1}{4\sin(\pi\nu)}P^0_\nu\left(-w_<w_>-\sqrt{1-w^2_<}\sqrt{1-w^2_>}\cos(\theta-\theta_n)\right).
\end{eqnarray}
When $w_<=w_>$, we obtain $z_1+z_2=\pi$ and the same identity for $s_2$ should be understood in the limit $w\to w_n$.
The Green's function $f_{exch}$ can then be written as
\begin{eqnarray}
    f_{exch}&=&\frac{1}{4\sin(\pi\nu)}\frac{P^{0\prime}_\nu(\beta_{oc})}{P^{0\prime}_\nu(-\beta_{oc})}P^0_\nu(w_<)P^0_\nu(w_>)+\frac{1}{4\sin(\pi\nu)}P^0_\nu\left(-w_<w_>-\sqrt{1-w^2_<}\sqrt{1-w^2_>}\cos(\theta-\theta_n)\right) \nonumber\\
    &&+\sum^\infty_{m=1}\frac{\Gamma(\nu-m+1)}{\Gamma(\nu+ m+1)}\frac{(-1)^m}{2\sin(\pi\nu)}\frac{P^{ m\prime}_\nu(\beta_{oc})}{P^{m\prime}_\nu(-\beta_{oc})}P^m_\nu(w_<)P^m_\nu(w_>)\cos(m(\theta-\theta_n)).
\end{eqnarray}
which is expressed entirely in terms of the associated Legendre functions of the first kind.
Employing the identity that $P^0_\nu(w_<)P^0_\nu(w_>)=P^0_\nu(w_n)P^0_\nu(w)$ yields the equation$~$(\ref{f_exch_solution}).
\section{Analytic solution for the capillary-tissue system on a circular disc}
\label{CircularDiscCapillaryTissueSolution}
Consider a planar circular disc $\mathcal{D}_c=\{(r,\theta)\mid 0\leq r\leq R_d,\ 0\leq\theta<2\pi\}$, where $R_d$ is the disc radius and the source ${\bm z}_n$ has polar coordinates $(r_n,\theta_n)$.
The constant inverse screening length is
\begin{eqnarray}
    \gamma_0=\sqrt{\alpha_{exch}\left(\frac{1}{\kappa_{cap}}+\frac{1}{\kappa_t}\right)}.
\end{eqnarray}
The mean-pressure Green's function is given by \citep{lin_novel_2026}
\begin{eqnarray}
    f^c_{mean}(r,\theta;r_n,\theta_n)&=&\frac{1}{4\pi}\log\left(r_n^2+r^2-2r_nr\cos(\theta_n-\theta)\right) \nonumber\\
    &&+\frac{1}{4\pi}\log\left(r_n^2+\frac{R_d^4}{r^2}-2r_n\frac{R_d^2}{r}\cos(\theta_n-\theta)\right),
\end{eqnarray}
and the exchange-pressure Green's function is
\begin{eqnarray}
    f^c_{exch}(r,\theta;r_n,\theta_n)=\frac{1}{2\pi}\sum^{\infty}_{m=-\infty}I_m(\gamma_0r^n_<)\left[\frac{K_m^\prime(\gamma_0R_d)}{I_m^\prime(\gamma_0R_d)}I_m(\gamma_0r^n_>)-K_m(\gamma_0r^n_>)\right]\exp[im(\theta-\theta_n)],
\end{eqnarray}
where $r^n_<=\min(r,r_n)$, $r^n_>=\max(r,r_n)$, and $I_m$ and $K_m$ are the modified Bessel functions of order $m$.
The capillary- and tissue-pressure Green's functions are
\begin{eqnarray}
    G^c_{cap}(r,\theta;r_n,\theta_n)=f^c_{mean}(r,\theta;r_n,\theta_n)+\frac{\kappa_t}{\kappa_{cap}}f^c_{exch}(r,\theta;r_n,\theta_n), \\
    G^c_t(r,\theta;r_n,\theta_n)=f^c_{mean}(r,\theta;r_n,\theta_n)-f^c_{exch}(r,\theta;r_n,\theta_n),
\end{eqnarray}
and the pressures follow from
\begin{eqnarray}
    p_{cap}({\bm z})=-\frac{1}{\kappa_{cap}+\kappa_t}\sum^{N_{cap}}_{n=1}q_nG^c_{cap}({\bm z},{\bm z}_n)+\overline{p_{mean}}+\alpha_r\Delta\pi, \\
    p_t({\bm z})=-\frac{1}{\kappa_{cap}+\kappa_t}\sum^{N_{cap}}_{n=1}q_nG^c_t({\bm z},{\bm z}_n)+\overline{p_{mean}}.
\end{eqnarray}
\section{Surface averages of the Green's functions}
\label{SurfaceBoundaryAveragesGreensFunction}
The surface area is computed using the equation$~$(\ref{conformal_factor_radial_coordinate}) for $\lambda(r)$ as follows:
\begin{eqnarray}
    \vert\mathcal{S}\vert=\int_{\mathcal{D}}\lambda^2({\bm z})d{\bm z}=\int^{2\pi}_0\int^{R_{t,s}}_0\left(\frac{2R^2_t}{R^2_t+r^2}\right)^2rdrd\theta=2\pi R^2_t(1+\beta_{oc}).
\end{eqnarray}
Integrating the governing equation for $f_{exch}$ in system$~$(\ref{f_exch_equation_boundary_condition}) over ${\mathcal D}$ yields
\begin{eqnarray}
    \int_{\mathcal{D}}\left(\nabla^2 f_{exch}-\gamma^2({\bm z})f_{exch}\right)d{\bm z}=\int_{\mathcal{D}}\delta({\bm z}-{\bm z}_n)d{\bm z}=1,
\end{eqnarray}
while the boundary condition gives
\begin{eqnarray}
    \int_{\mathcal{D}}\nabla^2 f_{exch}d{\bm z}=\int_{\partial \mathcal{D}}\nabla f_{exch}\cdot {\bm n}d\ell_{\mathcal D}=0.
\end{eqnarray}
Thus, using $\gamma^2({\bm z})=\gamma^2_0\lambda^2({\bm z})$, we obtain
\begin{eqnarray}
    \langle f_{exch}\rangle=-\frac{1}{\gamma^2_0\vert\mathcal{S}\vert}=-\frac{1}{2\pi\gamma^2_0 R^2_t(1+\beta_{oc})},
\end{eqnarray}
which is independent of the source position ${\bm z}_n$.
The surface average $\langle f_{mean}\rangle$ can be written as
\begin{eqnarray}
    \langle f_{mean}\rangle=\frac{1}{\vert \mathcal{S}\vert}\int^{R_{t,s}}_0\lambda^2(r)\int^{2\pi}_0f_{mean}(r,\theta;r_n,\theta_n)d\theta rdr.
\end{eqnarray}
To compute the angular integral, we use the identity
\begin{eqnarray}
    \frac{1}{2\pi}\int^{2\pi}_0\log(A^2+B^2-2AB\cos(\theta-\theta^\prime))d\theta=2\log(\max\{A,B\}),
    \label{logarithm_integral_maximum}
\end{eqnarray}
which holds for $A>0$, $B>0$ and $\theta^\prime\in\mathbb{R}$. It follows from the identity \citep[4.224.9]{gradshtein_table_2015}
\begin{eqnarray}
    \int^{\pi}_0\log(a+b\cos \theta)d\theta=\pi\log\frac{a+\sqrt{a^2-b^2}}{2}, \quad a\geq \vert b\vert>0,
    \label{logarithmic_integral}
\end{eqnarray}
by setting $a=A^2+B^2$ and $b=-2AB$, and using
\begin{eqnarray}
    \int^{2\pi}_0\log(a+b\cos (\theta-\theta^\prime))d\theta=2\int^{\pi}_0\log(a+b\cos \theta)d\theta.
\end{eqnarray}
Using this identity, the angular integral becomes
\begin{eqnarray}
    \int^{2\pi}_0f_{mean}(r,\theta;r_n,\theta_n)d\theta=\log(\max\{r,r_n\})+\log \frac{R^2_{t,s}}{r}.
\end{eqnarray}
The integral is split at $r=r_n$ because
\begin{eqnarray}
    \log(\max\{r,r_n\})=\left\{
    \begin{aligned}
        &\log r_n, \quad 0\leq r\leq r_n, \\
        &\log r, \quad r_n\leq r\leq R_{t,s}.
    \end{aligned}\right.
\end{eqnarray}
Thus,
\begin{eqnarray}
    \langle f_{mean}\rangle=\frac{1}{\vert \mathcal{S}\vert}\left(\int^{r_n}_0\lambda^2(r)r\log \frac{R^2_{t,s}r_n}{r}dr+\int^{R_{t,s}}_{r_n}\lambda^2(r)r\log R^2_{t,s}dr\right).
\end{eqnarray}
Using $\lambda(r)=2R^2_t/(R^2_t+r^2)$, $\vert \mathcal{S}\vert=2\pi R^2_t(1+\beta_{oc})$ and $w_n=(R^2_t-r^2_n)/(R^2_t+r^2_n)$, we obtain
\begin{eqnarray}
    \langle f_{mean}\rangle=\frac{1}{2\pi(1+\beta_{oc})}\left(2(1+\beta_{oc})\log R_{t,s}+\log\frac{2}{1+w_n}\right).
\end{eqnarray}
Since it depends on the source position $r_n$ through $w_n$, we denote it as $\langle f_{mean}\rangle=\langle f_{mean}(r_n)\rangle$.
\section{Averages for the singularities of Green's function}
\label{AveragesSingularitiesGreensFunction}
The stereographic distance $\vert {\bm z}-{\bm z}_n\vert^2$ is given by
\begin{eqnarray}
    \vert {\bm z}-{\bm z}_n\vert^2=r^2+r^2_n-2rr_n\cos(\theta-\theta_n).
\end{eqnarray}
Combining the stereographic relation
\begin{eqnarray}
    r^2=\frac{1-w}{1+w}R^2_t, \quad r^2_n=\frac{1-w_n}{1+w_n}R^2_t,
\end{eqnarray}
with the geodesic relation$~$(\ref{geodesic_angle}), the distance can be expressed as
\begin{eqnarray}
    \vert {\bm z}-{\bm z}_n\vert^2=\frac{4R^2_t}{(1+w)(1+w_n)}\sin^2\left(\frac{\epsilon_n}{2}\right),
\end{eqnarray}
and
\begin{eqnarray}
    \log\vert {\bm z}-{\bm z}_n\vert^2=\log(4R^2_t)-\log(1+w_n)-\log(1+w)+2\log\sin\left(\frac{\epsilon_n}{2}\right).
\end{eqnarray}
For the patch average, we introduce the azimuthal angle $\varphi$ around ${\bm x}_n$.
In the source-centred coordinates,
\begin{eqnarray}
    w=w_n\cos\epsilon_n-\sqrt{1-w^2_n}\sin\epsilon_n\cos\varphi,
\end{eqnarray}
and thus,
\begin{eqnarray}
    1+w=1+w_n\cos\epsilon_n-\sqrt{1-w^2_n}\sin\epsilon_n\cos\varphi.
\end{eqnarray}
Using the identity$~$(\ref{logarithmic_integral}) with
\begin{eqnarray}
    a=1+w_n\cos\epsilon_n, \quad b=-\sqrt{1-w^2_n}\sin\epsilon_n,
\end{eqnarray}
Since $w_n+\cos\epsilon_n>0$ within each vessel patch, the logarithmic integral identity yields
\begin{eqnarray}
    \frac{1}{2\pi}\int^{2\pi}_0\log(1+w)d\varphi=\log(1+w_n)+2\log\cos\left(\frac{\epsilon_n}{2}\right),
\end{eqnarray}
which implies
\begin{eqnarray}
    \frac{1}{2\pi}\int^{2\pi}_0\log\vert {\bm z}-{\bm z}_n\vert^2d\varphi=\log\left[\frac{4R^2_t}{(1+w_n)^2}\tan^2\left(\frac{\epsilon_n}{2}\right)\right].
\end{eqnarray}
Therefore,
\begin{eqnarray}
    \langle f^s_{mean}({\bm z},{\bm z}_n)\rangle_{\Omega^{\mathcal{S}}_n}=\frac{1}{4\pi}\frac{1}{1-\cos\epsilon_{v,n}}\int^{\epsilon_{v,n}}_0\log\left[\frac{4R^2_t}{(1+w_n)^2}\tan^2\left(\frac{\epsilon_n}{2}\right)\right]\sin\epsilon_nd\epsilon_n.
\end{eqnarray}
Let $C_{v,n}=\cos{\epsilon_{v,n}}$.
This integral can be evaluated analytically as
\begin{eqnarray}
    \langle f^s_{mean}({\bm z},{\bm z}_n)\rangle_{\Omega^{\mathcal{S}}_n}=\frac{1}{4\pi}\Bigg[\log\left(\frac{4R^2_t}{(1+w_n)^2}\right)+\log(1-C_{v,n})+\frac{(1+C_{v,n})}{(1-C_{v,n})}\log(1+C_{v,n})-\frac{2\log 2}{(1-C_{v,n})}\Bigg].
\end{eqnarray}

For the Green's function of exchange pressure, the argument of the singular part is related to the geodesic angle $\epsilon_n$ by
\begin{eqnarray}
    \xi_n=-w_nw-\sqrt{1-w^2_n}\sqrt{1-w^2}\cos(\theta-\theta_n)=-\cos(\epsilon_n).
\end{eqnarray}
Thus, $\xi_n\to-1$ at the source point, which corresponds to the logarithmic singularity in $f_{exch}$.
Employing the same spherical vessel patch,
\begin{eqnarray}
    \langle P^0_\nu(\xi_n)\rangle_{\Omega^{\mathcal{S}}_n}
    =\frac{1}{A^{\mathcal{S}}_n}\int_{\Omega^\mathcal{S}_n}P^0_\nu(-\cos\epsilon_n)dA,
\end{eqnarray}
and substituting $z=-\cos\epsilon_n$ yields
\begin{eqnarray}
    \langle P^0_\nu(\xi_n)\rangle_{\Omega^{\mathcal{S}}_n}=\frac{1}{1-\cos\epsilon_{v,n}}\int^{-\cos\epsilon_{v,n}}_{-1}P^0_\nu(z)dz.
\end{eqnarray}
Since $P^0_\nu(z)$ satisfies the associated Legendre equation \citep[\S14.2]{olver_nist_2010-1},
\begin{eqnarray}
    \frac{d}{dz}((1-z^2)P^{0\prime}_\nu(z))+\nu(\nu+1)P^0_\nu(z)=0,
\end{eqnarray}
we obtain
\begin{eqnarray}
    \int^{-\cos\epsilon_{v,n}}_{-1}P^0_\nu(z)dz=-\int^{-\cos\epsilon_{v,n}}_{-1}\frac{\frac{d}{dz}((1-z^2)P^{0\prime}_\nu(z))}{\nu(\nu+1)}dz.
\end{eqnarray}
Thus, the average becomes
\begin{eqnarray}
    \langle P^0_\nu(\xi_n)\rangle_{\Omega^{\mathcal{S}}_n}=\frac{1}{1-\cos\epsilon_{v,n}}\frac{\lim_{z\to -1^+}(1-z^2)P^{0\prime}_\nu(z)-\sin^2\epsilon_{v,n}P^{0\prime}_\nu(-\cos\epsilon_{v,n})}{\nu(\nu+1)}.
\end{eqnarray}
Using the endpoint relation $\lim_{z\to -1^+}(1-z^2)P^{0\prime}_\nu(z)=2\sin(\pi\nu)/\pi$ \citep{olver_nist_2010-1}, the average is given by
\begin{eqnarray}
    \langle P^0_\nu(\xi_n)\rangle_{\Omega^{\mathcal{S}}_n}=\frac{2\sin(\pi\nu)-\pi\sin^2\epsilon_{v,n}P^{0\prime}_\nu(-\cos\epsilon_{v,n})}{\pi\nu(\nu+1)(1-\cos\epsilon_{v,n})}.
\end{eqnarray}
Substituting $\nu(\nu+1)=-R^2_t\gamma^2_0$, the averaged singular exchange contribution is
\begin{eqnarray}
    \langle f^s_{exch}\rangle_{\Omega^{\mathcal{S}}_n}=-\frac{2\sin(\pi\nu)-\pi\sin^2\epsilon_{v,n}P^{0\prime}_\nu(-\cos\epsilon_{v,n})}{4\pi\sin(\pi\nu)R^2_t\gamma^2_0(1-\cos\epsilon_{v,n})}.
\end{eqnarray}
\bibliographystyle{elsarticle-harv}
\bibliography{Ref}

\end{document}